\documentclass[a4paper,11pt]{article}
\pdfoutput=1 

\usepackage{jheppub} 

\usepackage{color}
\usepackage{graphicx}
\usepackage{amsmath,latexsym}
\usepackage{amssymb}
\usepackage{xspace}
\usepackage[small]{subfigure}
\usepackage{xcolor}
\usepackage{appendix}

\graphicspath{{figures/}}

\newlength{\fighskip} \fighskip=2pt
\newlength{\figvskip} \figvskip=3pt
\newcommand*{\figbox}[2]{{
\def\figscale{#1}
\def\arraystretch{0.8}
\arraycolsep=0pt
\begin{array}{c}
\vbox{\vskip\figscale\figvskip
\hbox{\hskip\figscale\fighskip
\includegraphics[scale=\figscale]{#2}}}
\end{array}}}

\title{Information retrieval from Hawking radiation in the non-isometric model of black hole interior: theory and quantum simulations}

\author[a,\dagger,*]{Ran Li,}
\author[b,\dagger,*]{Xuanhua Wang,\note[$\dagger$]{Equal contributions.}}
\author[c]{Kun Zhang,}
\author[d,*]{Jin Wang \note[*]{Corresponding authors.}}
\affiliation[a]{Department of Physics, Qufu Normal University, Qufu, Shandong 273165, China}
\affiliation[b]{Center for Theoretical Interdisciplinary Sciences, Wenzhou Institute, University of Chinese Academy of Sciences, Wenzhou, Zhejiang 325001, China}
\affiliation[c]{School of Physics, Northwest University, Xi'an, Shaanxi 710069, China}
\affiliation[d]{Department of Chemistry, and Department of Physics and Astronomy, The State University of New York at Stony Brook, Stony Brook, NY 11794, USA}

\emailAdd{liran@qfnu.edu.cn}
\emailAdd{wangxh@ucas.ac.cn}
\emailAdd{kunzhang@nwu.edu.cn}
\emailAdd{jin.wang.1@stonybrook.edu}

\abstract{The non-isometric holographic model of the black hole interior stands out as a potential resolution of the long-standing black hole information puzzle since it remedies the friction between the effective calculation and the microscopic description. In this study, combining the final-state projection model, the non-isometric model of black hole interior and Hayden-Preskill thought experiment, we investigate the information recovery from decoding Hawking radiation and demonstrate the emergence of the Page time in this setup. We incorporate the effective modes into the scrambling inside the horizon, which are usually disregarded in Hayden-Preskill protocols, and show that the Page time can be identified as the transition of information transmission channels from the EPR projection to the local projections. This offers a new perspective on the Page time. We compute the decoupling condition under which retrieving information is feasible and show that this model computes the black hole entropy consistent with the quantum extremal surface calculation. Assuming the full knowledge of the dynamics of the black hole interior, we show how Yoshida-Kitaev decoding strategy can be employed in the modified Hayden-Preskill protocol. Furthermore, we perform experimental tests of both probabilistic and Grover's search decoding strategies on the 7-qubit IBM quantum processors to validate our analytical findings and confirm the feasibility of retrieving information in the non-isometric model. This study would stimulate more interests to explore black hole information problem on the quantum processors.}

\begin{document}

\maketitle
\section{Introduction} 

The most important quantum behavior of the black holes is that from the effective field theory calculation, it can radiate particles in the form of thermal spectrum at the temperature proportional to the surface gravity of the event horizon \cite{Hawking:1975vcx}. If the collapsing matter that forms the black hole is initially in a pure state, the whole system will evolve into a mixed state after the black hole is completely evaporated according to Hawking's calculation. This is incisively inconsistent with the principle of unitarity in quantum mechanics and it results in the long-standing puzzle regarding the conservation of information in black holes \cite{Hawking:1976ra}. After the discovery of AdS/CFT correspondence \cite{Maldacena:1997re,Gubser:1998bc,Witten:1998qj}, it is generally believed that the dynamical process of black hole collapsing and evaporating is an unitary one satisfying the principles of quantum mechanics. During this process, the information inside the black hole is released through its radiation and the information conservation is guaranteed \cite{Page:1993df,Page:1993wv}. However, this poses the question of how the information contained in the infalling objects is released in the evaporating process and also how the information can be recovered by the observer outside of the black hole from collecting and decoding the Hawking radiation \cite{Hayden:2007cs,Yoshida:2017non}.

The biggest obstacle to answering these questions lies in that the dynamics of the black hole interior is not known for the outside observer due to the existence of causal boundary event horizon. Recently, motivated by the theory of the quantum error correction \cite{Almheiri:2014lwa,Pastawski:2015qua} and quantum computational complexity \cite{Harlow:2013tf,Brown:2019rox}, a holographic model of the black hole interior was proposed to resolve the black hole information puzzle \cite{Akers:2022qdl}. In this model, two descriptions of the black hole degrees of freedom, the effective field description and the fundamental description from the quantum gravity, are connected through a non-isometric map. Although a full quantum gravity theory is not completely constructed by now and the exact nature of the fundamental description of the black hole is unknown to us, from the central dogma of black hole physics \cite{Almheiri:2020cfm} we can treat the area of the event horizon as counting the fundamental quantum gravity degrees of freedom. From the effective description of the semiclassical gravity along with the evaporating process, the entangled pairs of the outside radiated modes and their inside partners are generated continuously and the number of the effective field theory modes inside the black hole eventually exceeds the number of the black hole degrees of freedom accounted by the horizon area from the fundamental description \cite{Mathur:2008wi}. In order to resolve this apparent contradiction, Akers, Engelhardt, Harlow, Penington, and Vardhan (AEHPV) proposed that there is a non-isometric holographic map from the effective description to the fundamental description \cite{Akers:2022qdl}. This means that a large number of ``null" states are annihilated by the non-isometric holographic map, which apparently violates the unitarity of the effective description. However, it is shown that on the average the deviation from the unitarity in the effective description is negligibly small in the entropy. Furthermore, the entanglement entropy of Hawking radiation in the fundamental description was shown to follow the quantum extremal surface formula \cite{Ryu:2006bv,Faulkner:2013ana,Engelhardt:2014gca}. Therefore, it is argued that the AEHPV model of the black hole interior can give a Hilbert space interpretation of the Page curve computation from the island rule \cite{Penington:2019npb,Almheiri:2019psf}. The non-isometric holographic model of encoding black hole interiors has inspired many interesting works \cite{Kar:2022qkf,Faulkner:2022ada,deBoer:2022zps,Kim:2022pfp,Basu:2022crn,Giddings:2022ipt,Gyongyosi:2023sue,Cao:2023gkw,DeWolfe:2023iuq}.

In the present work, based on the AEHPV model of the black hole interior and the final-state projection model, we explore the possibility of treating the non-isometric mapping as a dual non-unitary dynamics inside the black hole and investigate the viability of decoding Hawking radiation and information recovery from the black hole. By studying its corresponding Hayden-Preskill decoding strategy \cite{Hayden:2007cs}, we first address the problem of the decoupling condition under which the information swallowed by the black hole can be recovered by decoding the Hawking radiation. This amounts to estimating the operator distance between the reduced density matrix of black hole and reference system and the density matrix of their product state \cite{Nielsen,Hayden:2006,Hayden:2007}. In principle, when the decoupling condition is satisfied, the entanglement between the reference system and the black hole is transferred to the entanglement between the reference system and the Hawking radiation and that the information swallowed by the black hole can be recovered. Furthermore, under the assumption that the black hole interior dynamics is known to the outside observer, we discuss how the Yoshida-Kitaev decoding strategies \cite{Yoshida:2017non} can be used to decode the Hawking radiation in the modified version of the Hayden-Preskill protocol. For the probabilistic decoding strategy, the corresponding decoding probability and the fidelity on the average of the random unitary group are computed. Importantly, the probability of the EPR projection demonstrates a phase transition behavior so that we can identify the Page time with the phase transition of information channels \cite{Li:2023nfv}. We show that by including the scrambling of effective modes inside the horizon, a transition of information transmission channels naturally emerges around the Page time that switches the information transmission from through the EPR-projections to through the local projections. This offers a new perspective on the nature of the Page transition. Besides, the decoding strategy using the Grover's search algorithm \cite{Grover:1996rk} is performed to recover the initial quantum state of the system swallowed by the black hole. In addition, we test the decoding strategies on the 7-qubit IBM quantum processors. The experimental results validate our analytical findings and show the feasibility of the information recovery. Finally, inspired by the work of Kim and Preskill \cite{Kim:2022pfp}, where an infalling agent interacts with the radiation both outside and inside the black hole, we further study the effects caused by such interactions. We argue that the interaction of the infalling message system with the outside right-going Hawking radiation causes no additional effect in the modified Hayden-Preskill protocol.

This paper is arranged as follows. In Sec.~\ref{secI}, we briefly introduce the non-isometric holographic model of the black hole interior. In Sec.~\ref{secIII}, based on the non-isometric model of black hole interior, we propose a modified version of Hayden-Preskill thought experiment. In Sec.~\ref{Sec:island}, we prove that the black hole entropy calculated from this model is consistent with the quantum extremal surface calculation. In Sec.~\ref{secIV}, we discuss the decoupling condition to recovery the information swallowed by the black hole. In Sec.~\ref{secV}, we apply the Yoshida-Kitaev decoding scheme to our model to show that the information can be recovered from the Hawking radiation and the transition of information channels emerges around the Page time. Two types of decoding strategies are discussed. In Sec.~\ref{secVI}, the simulation experiments of the decoding Hawking radiation are implemented on the IBM quantum processors. In Sec.~\ref{secVII}, we comment on the interaction of the infalling message system with the outside radiation. The conclusion and discussion are presented in the last section.

\section{The non-isometric models of black hole interiors}\label{secII}

\subsection{Review of AEHPV model}\label{secI}

In this section, we give a brief review of the black hole interior model proposed by AEHPV \cite{Akers:2022qdl}.  There are two complementary descriptions for the dynamics of black hole interior, one is the effective field description and the other one is the fundamental description from the quantum gravity.

The fundamental description gives the viewpoint of the outside observer. In this description without other external matters, the Hilbert space can be partitioned into that of the black hole $B$ and that of the Hawking radiation $R$, viz.,
\begin{eqnarray}
    \mathcal{H}_B\otimes\mathcal{H}_{R} \;.
\end{eqnarray}
According to the central dogma of the black hole physics \cite{Almheiri:2020cfm}, the dimension of $\mathcal{H}_B$ is proportional to $e^{A_{\textrm{EH}}/4G}$ with $A_{\textrm{EH}}$ being the horizon area.

In contrast to the outside observer, the infalling observer will experience another picture. In a ``nice slice" \cite{Mathur:2008wi}, the infalling observer will find that there are the right-moving modes $R$ in the black hole exterior, and the left-moving modes $l$ and the right-moving modes $r$ in the black hole interior. Here, $R$ is again the degrees of freedom of the Hawking radiation, $l$ can be treated as the degrees of freedom that forms the black hole and $r$ denotes the interior partners of the Hawking radiation $R$. In this effective description of the infalling observer, the Hilbert space is given by
\begin{eqnarray}
\mathcal{H}_{l}\otimes\mathcal{H}_{f}\otimes\mathcal{H}_{r}\otimes\mathcal{H}_{R} \;,
\end{eqnarray}
where $f$ is an additional system that accounts the fixed degrees of freedom and does not play an essential role in the analysis. In addition, it is obvious that $r$ and $R$ are in the maximally entangled state from the effective field theory calculation.

As claimed in the introduction, there is an apparent contradiction for the outside observer and the infalling observer. As the black hole evaporates, the degrees of freedom of Hawking radiation $R$ increase monotonically while the degrees of freedom of black hole $B$ decrease. For the black hole at late time, the number of the degrees of freedom of Hawking radiation is larger than that of black hole, i.e. $|R|>|B|$, or $|r|>|B|$, where $|\cdot|$ denotes the Hilbert space dimension of the corresponding system. This will result in the contradiction that the entanglement entropy of Hawking radiation in the effective description exceeds the black hole entropy in the fundamental description.

In order to resolve this conflict, AEHPV proposed that there is a non-isometric holographic map from the effective description to the fundamental description 
\begin{eqnarray}
 V: \mathcal{H}_{l}\otimes\mathcal{H}_{r}\rightarrow  \mathcal{H}_{B} \;.
\end{eqnarray}
For the Hilbert space in the effective description $\mathcal{H}_{l}\otimes\mathcal{H}_{f}\otimes\mathcal{H}_{r}$, we consider its mapping into the fundamental Hilbert space by introducing an auxiliary system $P$ and scrambling unitary $U$ such that 
\begin{eqnarray}
\mathcal{H}_{l}\otimes\mathcal{H}_{f}\otimes\mathcal{H}_{r}\xrightarrow{V} \mathcal{H}_{B}\otimes\mathcal{H}_{P}\;.
\end{eqnarray}
The non-isometric map can be explicitly realized as 
\begin{eqnarray}
V=\sqrt{|P|}\langle 0|_P U|\psi_0\rangle_f=\sqrt{|P|} ~~~\figbox{0.2}{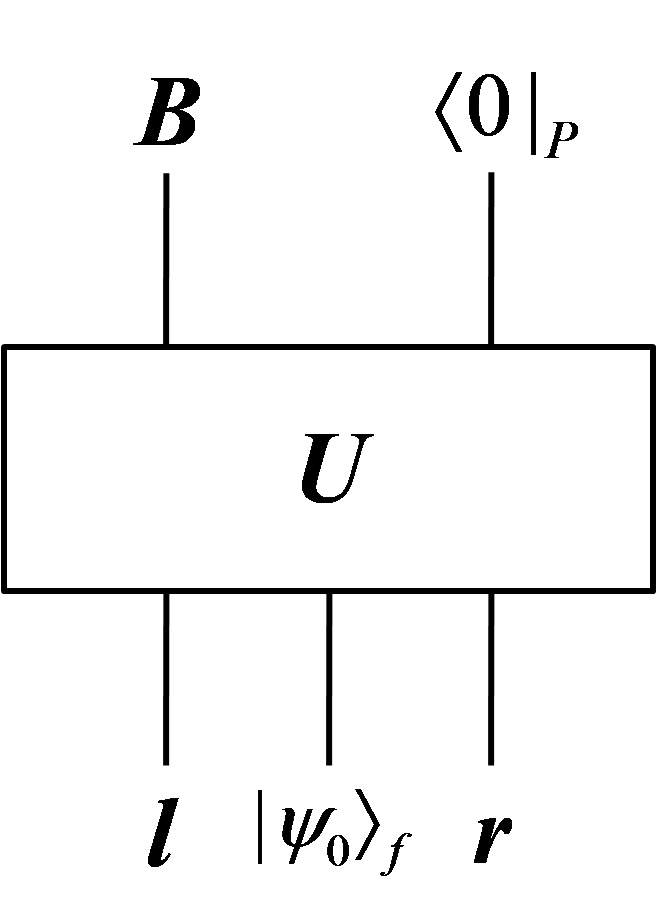}\;.
\label{Eq:map}
\end{eqnarray}
Here, $|\psi_0\rangle_f$ and $|0\rangle_P$ are the fixed states in $\mathcal{H}_f$ and $\mathcal{H}_{P}$, respectively. The prefactor $\sqrt{|P|}$ is introduced to preserve the normalization of the resulting state, which will be clarified in the next section. The graph gives the intuitive representation of the non-isometric map. The dynamics of the effective field theory degrees of freedom $l$, $r$ and $f$ in the black hole interior is modeled by a typical scrambling unitary operator $U$. The contradiction between the effective description and the fundamental description is resolved by post-selecting or projecting certain degrees of freedom on the auxilliary system $P$, resulting in a non-isometric mapping from the black hole interior of much larger degrees of freedom in the effective description to the black hole $B$ of much lower degrees of freedom in the fundamental description.

The post-selection or the projection in the holographic map is reminiscent of the final state proposal by Horowitz and Maldacena \cite{Horowitz:2003he} (see also \cite{Lloyd:2013bza}). In the final state proposal the post-selection comes from a modification of quantum mechanics and in AEHPV model the post-selection is a property of the non-isometric holographic map itself. In addition, in the final state model the post-selection is supposed to happen at the singularity while the post-selection in AEHPV model happens in the black hole interior. We should also mention another interesting proposal by Wang et.al in \cite{Wang:2023eyb} that is inspired by the Island rule. In \cite{Wang:2023eyb}, such projections or post-selected measurements occur on the horizon to avoid causal issues, and that the information is transferred to the outside once it enters the entanglement island.

Due to the post-selection or the projection in the black hole interior, the unitarity is apparently violated in the effective description. However, two important observations from the AEHPV model are: (1) averaged over the random unitary group, the deviation from the unitarity in the effective description is negligibly small in the entropy; (2) the entanglement entropy of the Hawking radiation in the fundamental description can be computed by the quantum extremal surface formula in the effective description. It is therefore argued that the non-isometric holographic model of the black hole interior gives a Hilbert space interpretation of the Page curve computation from the Island rule. This model has also been generalized to include the effects induced by the infalling agent interacting with the radiation both outside and inside the black hole horizon \cite{Kim:2022pfp}. It is tested that the unitarity of the S-matrix is guaranteed to a very high precision. Inspired by these works, a ``backwards-forwards" evolving model was also introduced to probe the non-trivial interactions between the infalling modes with the radiation modes outside and inside the horizon \cite{DeWolfe:2023iuq}.

\subsection{The non-unitary dynamical model and Hayden-Preskill protocol}\label{secIII}

\begin{figure}
  \centering
  \includegraphics[width=12cm]{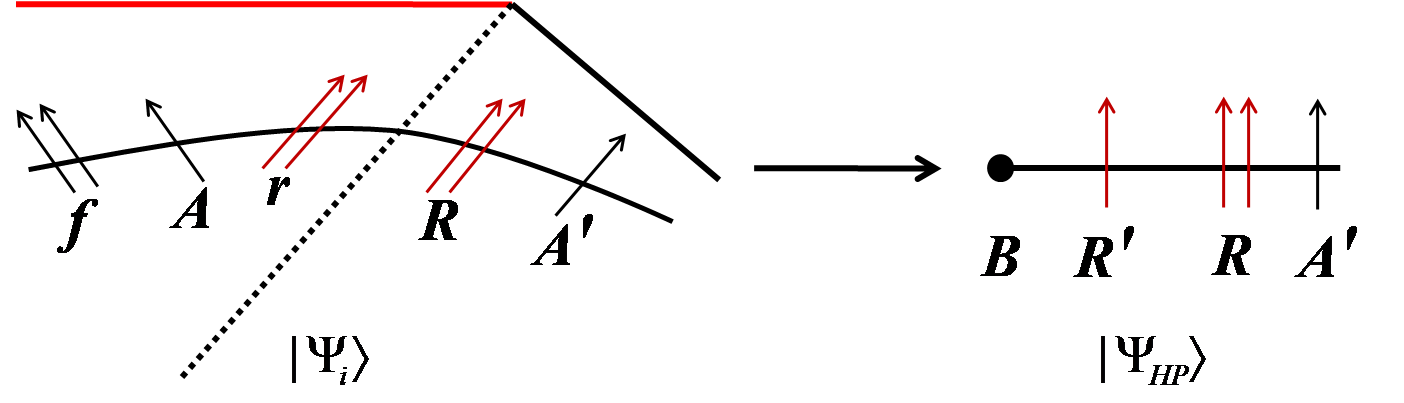}
  \caption{Left panel: An illustration of the systems $f$, $A$, $A'$, $r$ and $R$ that appear on the ``nice slice" Cauchy surface at the initial time. Nice slice is a Cauchy surface on which the semiclassical effective field theory calculations do not break down until a very late time.  $A$ is the message system thrown into the black hole and $A'$ is the reference system maximally entangled with $A$. The system $R$ denotes the outside Hawking radiation and $r$ is its entangled partner inside the black hole. $f$ is the matter system that forms the black hole. The corresponding state is given by Eq.(\ref{psi_i}). Right panel: An illustration of the systems $B$, $R'$, $R$ and $A'$ that observed by the outside decoder at a later time. The degrees of freedom inside the black hole are collectively denoted by $B$. The newly-generated Hawking radiation is denoted by $R'$. The quantum state is given by Eq.(\ref{psi_hp}).}
  \label{nice_slice_representation}
\end{figure}

In this section, we explore the consequences of a dual interpretation of treating the non-isometric mapping as an effective non-unitary dynamics on a set of given effective modes inside the black hole. We propose a non-unitary dynamic model of a radiating black hole and consider its Hayden-Preskill thought experiment. The initial state consists of a set of prearranged modes in the effective description, and the outcome of the mapping returns the fundamental modes on our interested time slice. In addition to the vacuum modes and their entangled partners shown in Eq.\eqref{Eq:map}, we introduce the information qudit $A$, which is thrown into the black hole, and its entangled partner qudit $A'$ which stays outside the black hole. The schematic illustration is provided on Figure~\ref{nice_slice_representation}. The effective modes at the initial time are shown by the left panel and the modes observed by an outside observer at a later time are represented on the right panel with the new-generated radiation $R'$.

As shown on the left of the Figure \ref{nice_slice_representation}, on the Cauchy surface at the initial time, there are the matter system $f$ that forms the black hole, the infalling message system $A$, the reference system $A'$ that is maximally entangled with $A$, the outside Hawking radiation $R$ and its interior partner $r$. In our model, $f$ denotes the matter system that collapses to the black hole. \footnote{Our notation is different from the AEHPV model, where $f$ denotes an auxilliary system in the fixed state while $l$ denotes the infalling matter system that forms the black hole.} For simplicity, we also set $f$ to be in the fixed state $|\psi_0\rangle_f$. The reference system $A'$ is introduced to purify the message system $A$. In addition, from the effective field viewpoint, $R$ and $r$ are maximally entangled. In this setup, the state of the total system at the initial time is given by 
\begin{eqnarray}\label{psi_i}
    |\Psi_{i}\rangle = |\textrm{EPR}\rangle_{A'A} \otimes |\psi_0\rangle_f \otimes |\textrm{EPR}\rangle_{rR}\;,
\end{eqnarray} 
where $\textrm{EPR}$ represents the maximally entangled state, for example, $|\textrm{EPR}\rangle_{A'A}=\frac{1}{\sqrt{|A|}}\sum_j |j,j\rangle$.

By the non-isometric holographic map, after some time the state of the whole system evolves into the the following modified Hayden-Preskill state, which is given by 
\begin{eqnarray}\label{psi_hp}
    |\Psi_{\textrm{HP}}\rangle &=& \sqrt{|P|} \langle 0|_P \left(I_{A'}\otimes U_{(Afr)(BPR')}\otimes I_{R}\right) |\Psi_{i}\rangle\nonumber\\
    &=&\sqrt{|P|} \langle 0|_P \left(I_{A'}\otimes U_{(Afr)(BPR')}\otimes I_{R}\right) |\textrm{EPR}\rangle_{A'A} \otimes |\psi_0\rangle_f \otimes |\textrm{EPR}\rangle_{rR}\;.
\end{eqnarray}
In this expression, the dynamical process of the information scrambling and the black hole radiating is typically represented by the random unitary operator $U$. In order to describe the black hole radiating, we introduce the newly generated Hawking radiation $R'$. The subscripts $(Afr)$ and $(BPR')$ denote the input systems and the output systems of the random unitary operator $U$. It is clear that the Hilbert space dimension of the input systems $Afr$ is equal to that of the output system $BPR'$, i.e., $|A||f||r|=|B||P||R'|=d$. Similar to the non-isometric model, after scrambling, certain degrees of freedom $P$ are projected onto the fixed state $\langle 0|_{P}$ in the black hole interior. The factor $\sqrt{|P|}$ is introduced to preserve the normalization. In this setup, $|\Psi_{\textrm{HP}}\rangle$ describes the state of the total system in the fundamental description including the remnant black hole $B$, the newly-generated Hawking radiation $R'$, the early Hawking radiation $R$ and the reference system $A'$, which are systematically depicted in the right panel of Figure \ref{nice_slice_representation}.

In this study, calculations are done using the graphical representation. The modified Hayden-Preskill state \eqref{psi_hp} can be graphically represented as 
\begin{eqnarray}
 |\Psi_{\textrm{HP}}\rangle =\sqrt{|P|}~~~\figbox{0.2}{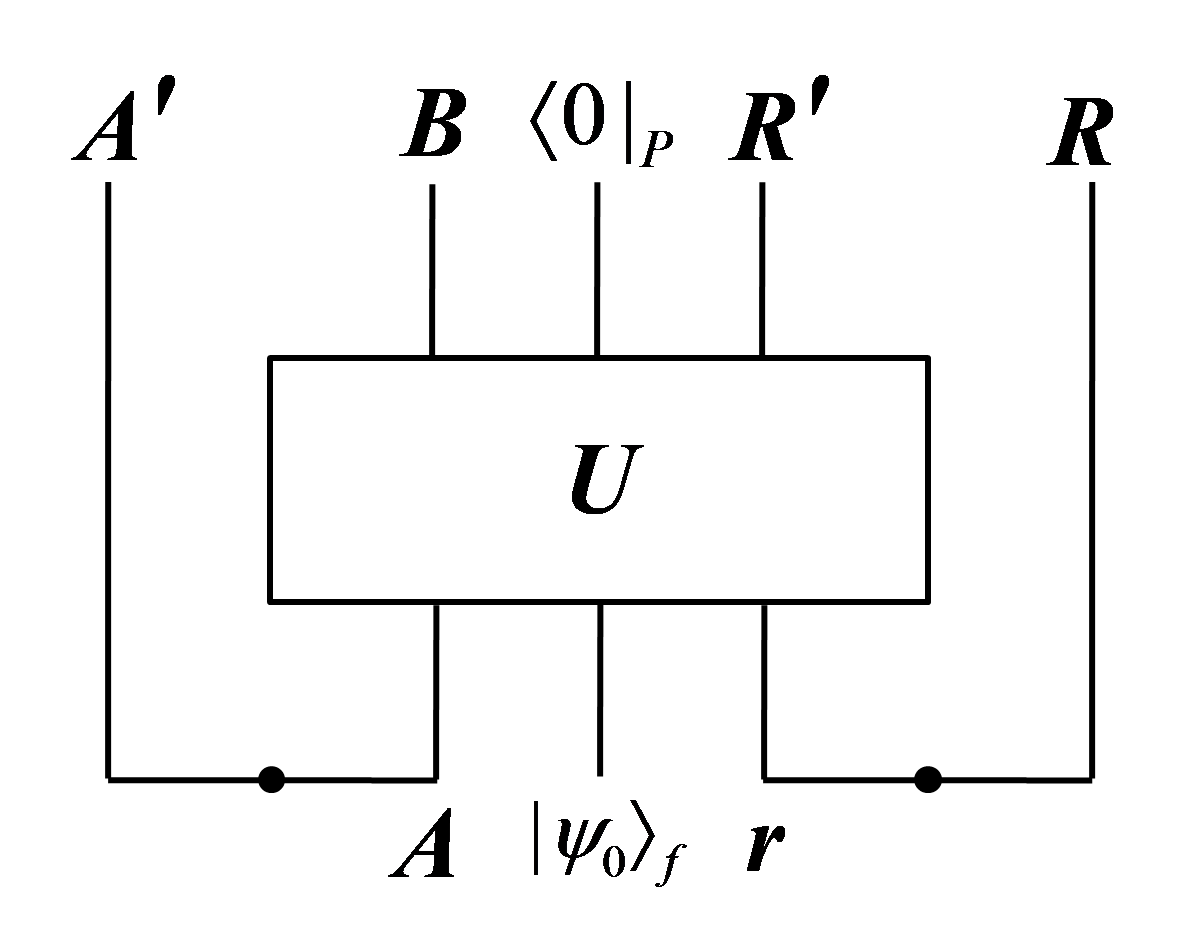}\;.
 \label{Fig:psi_hp}
\end{eqnarray}
In this graph, $\figbox{0.2}{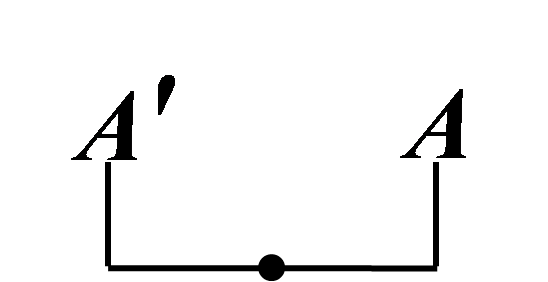}$ represents the $\textrm{EPR}$ state of $A$ and $A'$ and the black dot stands for the normalization factor $\frac{1}{\sqrt{|A|}}$. Similar rules applies to the system $r$ and $R$. As we have claimed, in the black hole interior, the infalling message system $A$, the matter system $f$ and the Hawking partner mode $r$ are scrambled by the random unitary operator $U$, resulting in the output system composed by the newly-generated Hawking radiation $R'$, the remnant black hole $B$ and an auxiliary system $P$. According to the non-isometric map, $P$ is postselected or projected onto the fixed state $|0\rangle_P$.  

With the same rules, we can represent the conjugate state $\langle \Psi_{\textrm{HP}}|$ as 
\begin{eqnarray}
 \langle\Psi_{\textrm{HP}}|=\sqrt{|P|}~~~\figbox{0.2}{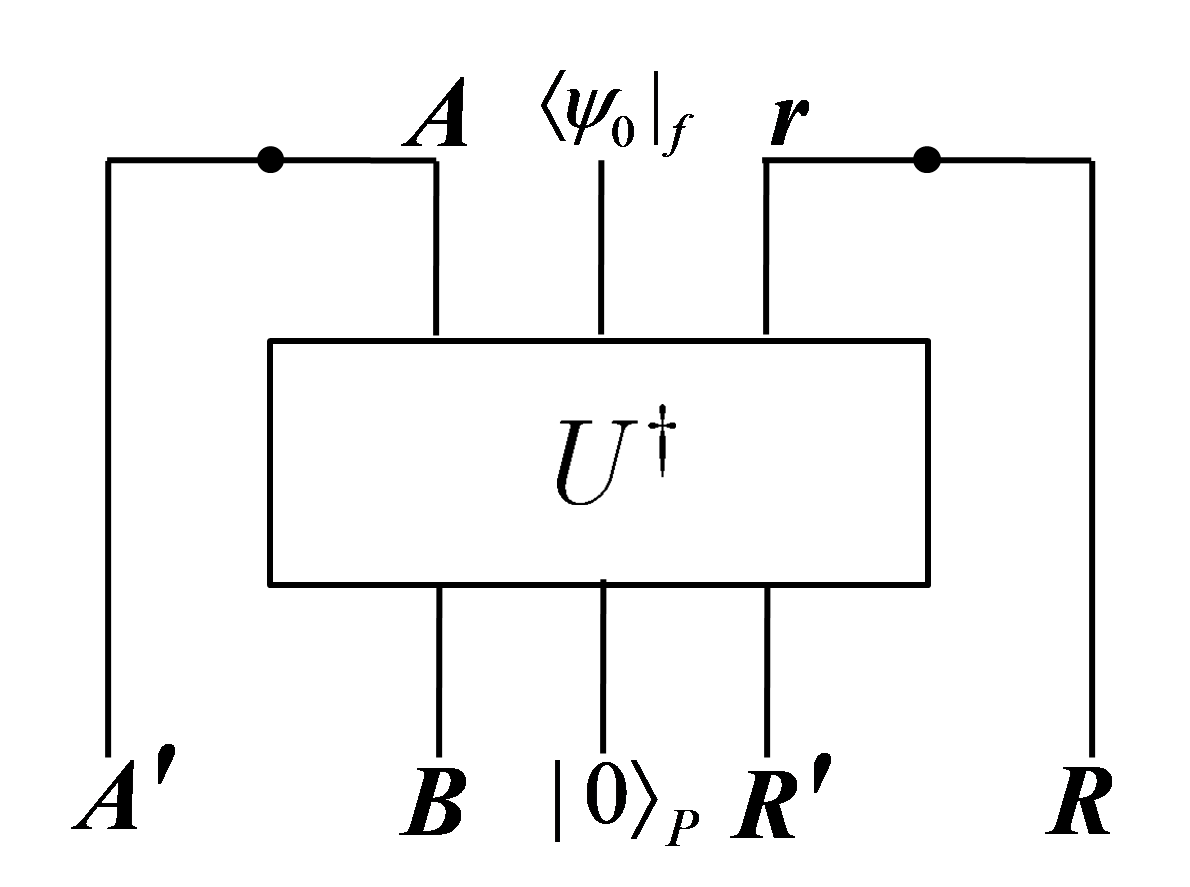}\;.
 \label{Fig:psi_hp_dagger}
\end{eqnarray}
This graph is obtained by flipping the graphical representation \eqref{Fig:psi_hp} of $|\Psi_{\textrm{HP}}\rangle$ and replacing the random unitary $U$ with $U^\dagger$. 

Because the holographic map from the effective description to the fundamental description is non-isometric, the modified Hayden-Preskill state is not normalized in general. One can see that the graphical representation of the inner product $\langle\Psi_{\textrm{HP}}|\Psi_{\textrm{HP}}\rangle$ can be obtained by connecting the open ends of the graphs in Eq.\eqref{Fig:psi_hp} and \eqref{Fig:psi_hp_dagger}
\begin{eqnarray}
 \langle\Psi_{\textrm{HP}}|\Psi_{\textrm{HP}}\rangle=|P|~~~\figbox{0.2}{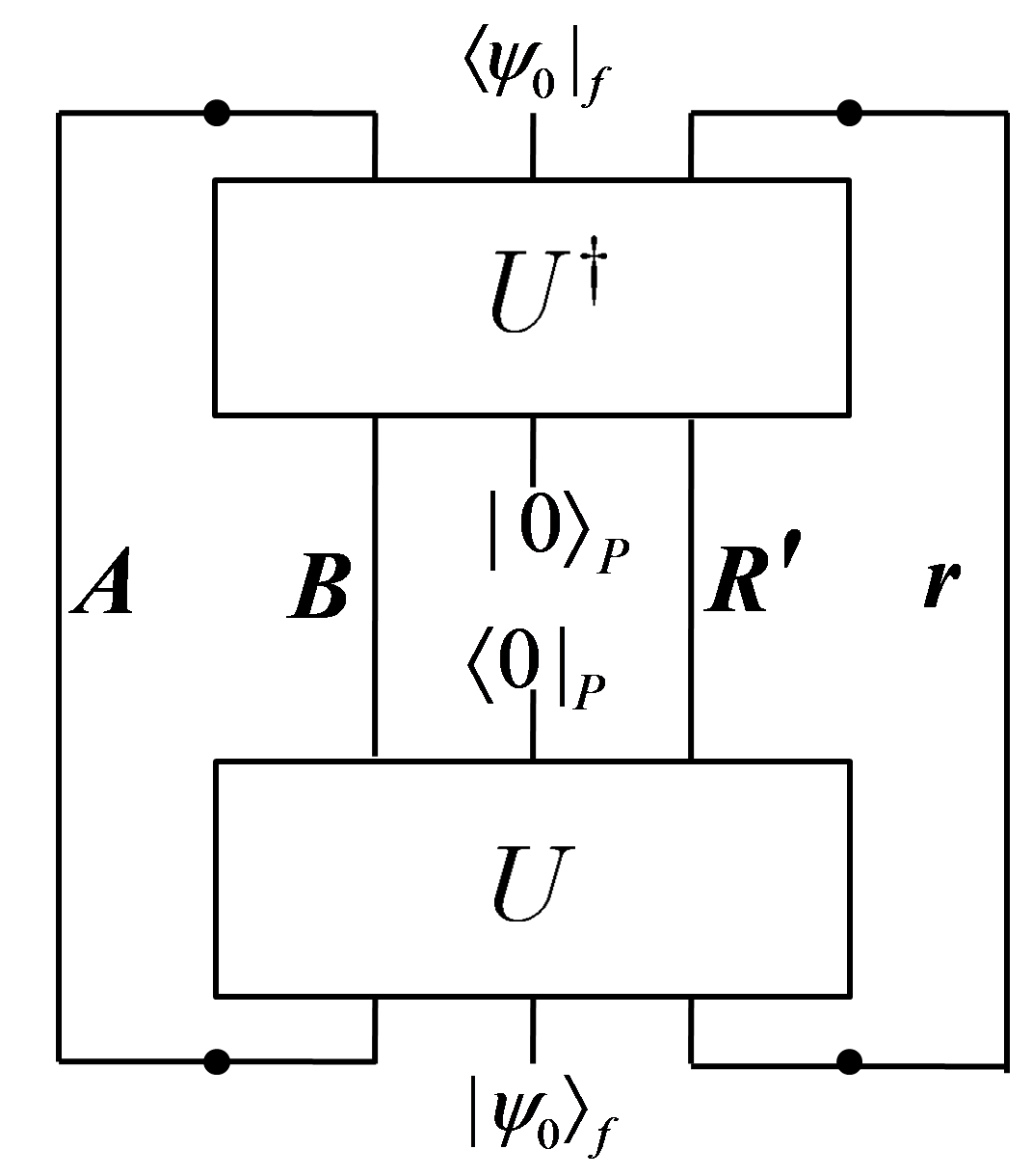}\;.
 \label{HP_inner_product}
\end{eqnarray}
Note that due to the existence of the postselection on the fixed state $|0\rangle_P$, the successive action of $U$ and $U^\dagger$ can not be treated as the identity operator $I$. In this sense, in general $\langle\Psi_{\textrm{HP}}|\Psi_{\textrm{HP}}\rangle\neq 1$. However, the normalization is preserved on the Haar average over the random unitary operator $U$. To realize this aim, we invoke the following integral formula
\begin{eqnarray}\label{2U_integral}
    \int dU U_{ij} U^\dagger_{j'i'}=\int dU U_{ij} U^*_{i'j'} =\frac{\delta_{ii'}\delta_{jj'} }{d}\;,
\end{eqnarray}
which can be graphically represented as \cite{Kim:2022pfp}
\begin{eqnarray}\label{2U_graph}
    \int dU \left(\figbox{0.2}{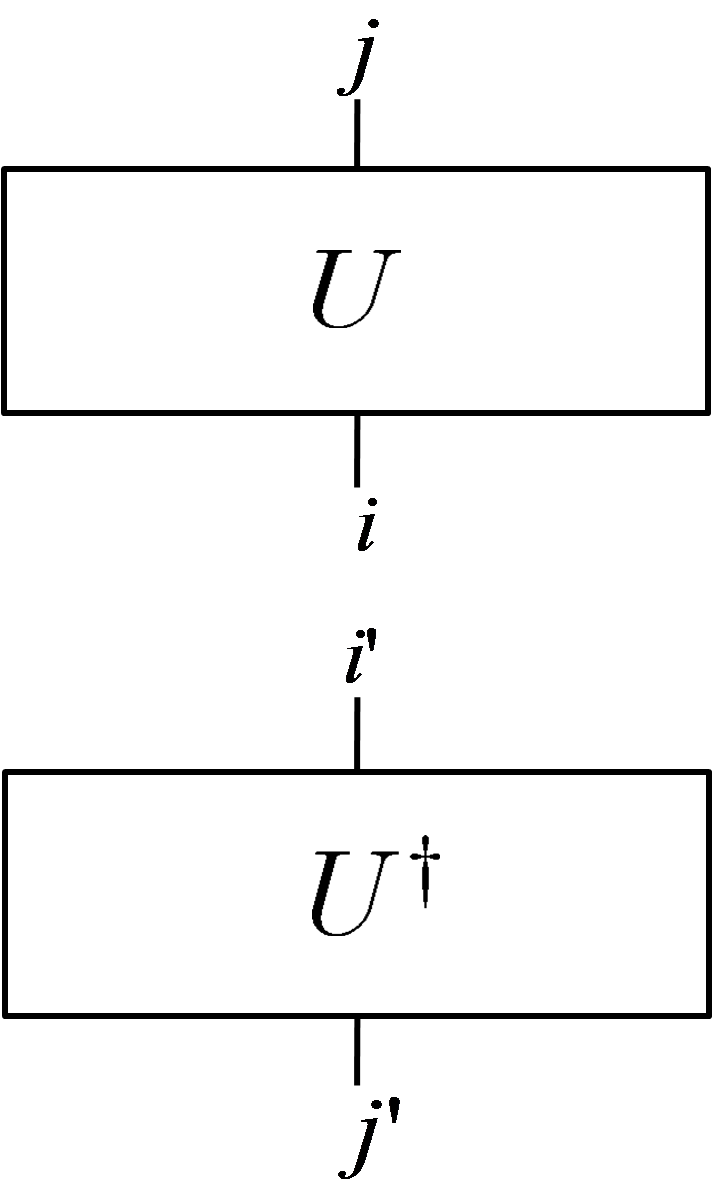}\right) =\frac{1}{d}\left(\figbox{0.2}{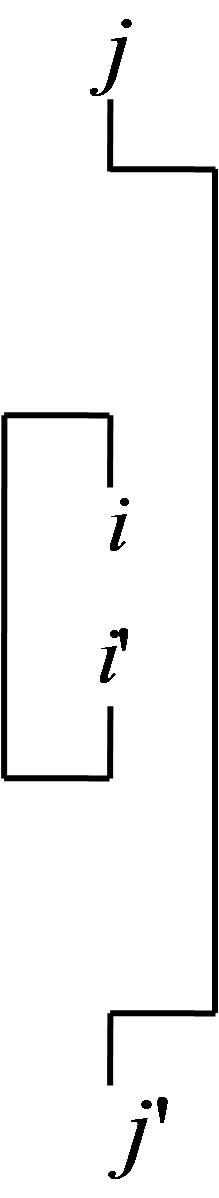}\right)\;.
\end{eqnarray}
Then the average of the inner product $\langle\Psi_{\textrm{HP}}|\Psi_{\textrm{HP}}\rangle$ of Eq.~\eqref{HP_inner_product} over the random unitary operator $U$ can be calculated as 
\begin{eqnarray}\label{HP_inner}
    \int dU \langle\Psi_{\textrm{HP}}|\Psi_{\textrm{HP}}\rangle &=& |P| \int dU
\left(\figbox{0.2}{HP_inner_product.png}\right)   \nonumber\\
&=&\frac{|P|}{d} 
\left(\figbox{0.2}{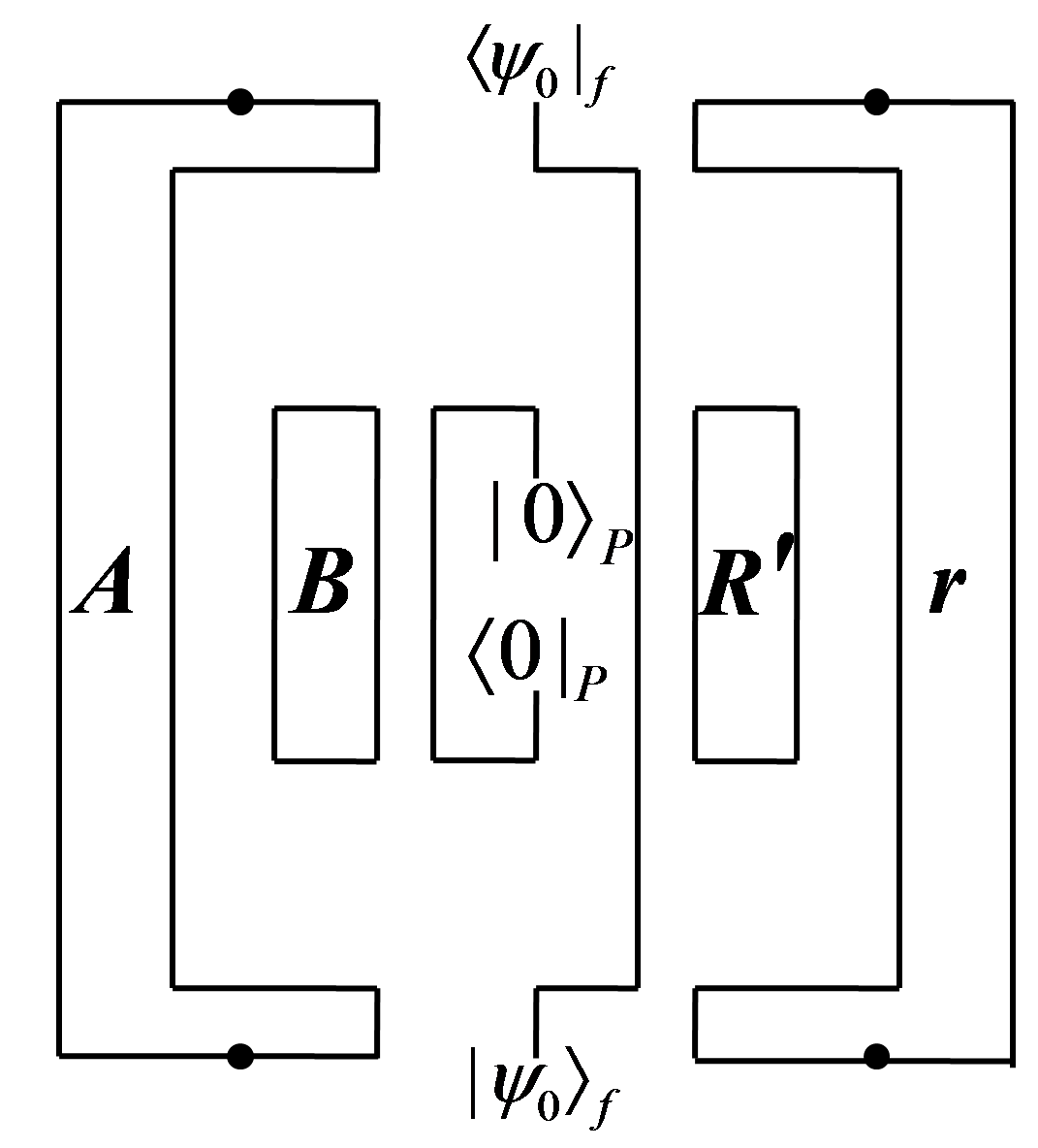}\right) \nonumber\\
&=& \frac{|P||B||R'|}{d}=1\;.
\end{eqnarray}
In deriving this result, we used the fact that the loop denotes the trace of identity operator over the Hilbert space of the corresponding system, which gives rise to the factor of its Hilbert space dimension. The loop with two dots is equal to unity. The line that connects the two fixed states (for example $|0\rangle_P$ and $\langle 0|_P$) represents the normalization condition of the fixed state and gives rise to the factor of unity.

The normalization condition of the modified Hayden-Preskill state is preserved on the average over the random unitary operator $U$. This also implies that the dynamical process is unitary on average for the observer in the effective description.

\subsection{Black hole entropy and Island formula}\label{Sec:island}
In this section, we show that the model gives the Island rule for the entanglement entropy of black holes \cite{Engelhardt:2014gca,Penington:2019npb,Almheiri:2019psf}. Using the graphic representation introduced in the previous section, we can compute the state of the black hole,
\begin{gather}
    \rho_B \equiv |\Psi_B\rangle \langle \Psi_B |=\frac{|P|}{|R||A|} \left(\figbox{0.6}{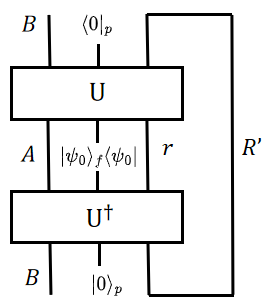}\right) \,.
\end{gather}
The $n$-th Renyi entropy is related to the density matrix by 
\begin{gather}
    S_n(B)=-\frac{1}{n-1} \log \mathrm{Tr}\left(\rho_B^n\right)\,.
\end{gather}
Invoking the integration formula for Haar random unitary matrices, we compute the second Renyi entropy,
\begin{eqnarray}
    \int dU e^{-S_2(B)} &=& \int dU \mathrm{Tr} \rho_B^2 =  \int dU \frac{|P|^2}{|R|^2|A|^2} \left(\figbox{0.5}{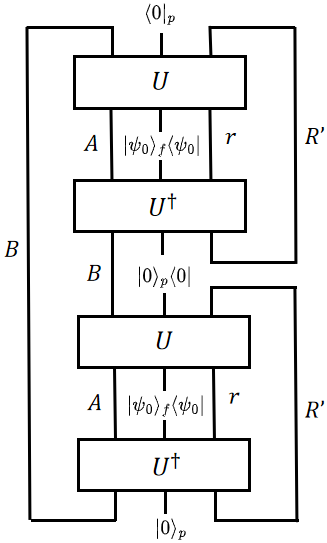}\right)\nonumber\\
    &\simeq& \frac{|P|^2}{|R|^2 |A|^2(d^2-1)}\left( |A|^2 |R|^2 |R'|^2 |B|+ |B|^2 |R'||A||R| \right)+\left(\mathrm{higher \, order\, terms}\right)\nonumber\\
    &=& \frac{d^2}{d^2-1} \frac{1}{|B|}+ \frac{d^2}{d^2-1} \frac{1}{|R'||A||R|}+\left(\mathrm{higher \, order\, terms}\right)\nonumber\\
    &\simeq& e^{-\bar S(B)}+e^{-\bar S(A'RR')}+\left(\mathrm{higher \, order\, terms}\right)\,,   
\end{eqnarray}
where $\bar S(B)$ represents the coarse-grained entropy of the remnant black hole. Therefore, for $|B|,|R|\gg 1$, the second Renyi entropy of the black hole $B$ is given by 
\begin{gather}
    S_2(B)\simeq \min \left(\bar S(B)\,,\bar S(A'RR')\right)\,,
\end{gather}
where $\bar S(A'RR')$ is the entropy of the exterior systems in the effective description. This is reminiscent of the island formula. To compute the von Neumann entropy of the black hole, we extend the calculation to the general $n$-th Renyi entropy, 
\begin{eqnarray}
    \int dU e^{-(n-1)S_n(B)} &=& \int dU \mathrm{Tr} \rho_B^n =  \int dU \frac{|P|^2}{|R|^2|A|^2} \left(\figbox{0.5}{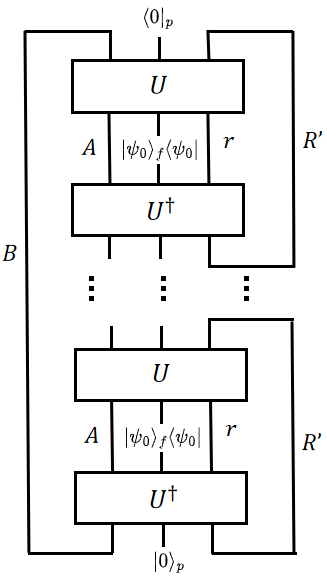}\right)\nonumber\,.
\end{eqnarray}
For this calculation, we make use of the Weingarten functions given in the Appendix B. The leading order contribution of the integral gives  
\begin{eqnarray}
    \int dU e^{-(n-1)S_n(B)} &\simeq& \frac{|P|^n}{|R|^n |A|^n(d^n)}\left( |A|^n |R|^n |R'|^n |B|+ |B|^n |R'||A||R| +\right)\nonumber\\
    &\simeq& e^{-(n-1)\bar S(B)}+e^{-(n-1) \bar S(A'RR')}\,.    
\end{eqnarray}
Hence, 
\begin{gather}
    S_n(B) \simeq \min \left(\bar S(B)\,,\bar S(A'RR')\right)
\end{gather}
for all $n$. The $n$-th Renyi entropy has a well-defined limit as $n\rightarrow 1$. This implies that the von Neumman entropy for the black hole satisfies $ S(B) \simeq \min \left(\bar S(B)\,,\bar S(A'RR')\right)$, which is the island formula of the black hole. This formula produces the expected Page curve for the entanglement entropy. To be specific, at the initial stage of the evaporation the black hole entropy is given by the coarse-grained entropy of the radiation and at the late times it is given by the coarse-grained entropy of  the black hole. 

Furthermore, we can consider the following situation where the black hole just radiated out the system $R'$ within a short time interval. We have the freedom to choose the cutoff surface such that the newly-generated radiation $R'$ near the horizon is inside the cutoff surface and the rest of the exterior systems are on the outside of it. Repeating the above calculation for the density matrix $\rho_{BR'}$ returns the von Neumann entropy of the systems inside (and outside) the cutoff surface
\begin{gather}
     S_{vN} \simeq \min \left(\log|B|+\bar S(R')\,,\bar S(A'R)\right)\,.
\end{gather}
This is precisely what one should expect from the quantum extremal surface calculation. The area term is represented by the coarse-grained entropy $\log|B|$ of the black hole and the entropy of the states between the cutoff surface and the quantum extremal surface is represented by $\bar S(R')$. The situation of null quantum extremal surface is given by the entropy of the systems outside the cutoff surface $\bar S(A'R)$.

\subsection{Decoupling condition of the modified Hayden-Preskill protocol}\label{secIV}

In this section, with the modified Hayden-Preskill state given in Eq.\eqref{Fig:psi_hp}, we now discuss whether the information contained in the message system $A$ can be recovered by the outside observer from collecting and decoding the early and the newly-generated Hawking radiation $R$ and $R'$. The condition that the aim can be achieved relies on the decoupling or the disentangling between the reference system $A'$ and the remnant black hole $B$. We refer to this condition as the decoupling condition. This is to say that the decoupling condition can be obtained by estimating the operator distance between the ``reduced density matrix" $\rho_{A'B}$ and the product state of $A'$ and $B$ averaged over the random unitary operator $U$.

The ``reduced density matrix" for the combined system of the reference $A'$ and the remnant black hole $B$ can be obtained from the density matrix of the modified Hayden-Preskill state by tracing out the early Hawking radiation $R$ and the newly-generated Hawking radiation $R'$, which can be graphically represented by  
\begin{eqnarray}
\rho_{A'B}=\textrm{Tr}_{RR'}|\Psi_{\textrm{HP}}\rangle\langle\Psi_{\textrm{HP}}|=\frac{|P|}{|r|}~~~ \figbox{0.2}{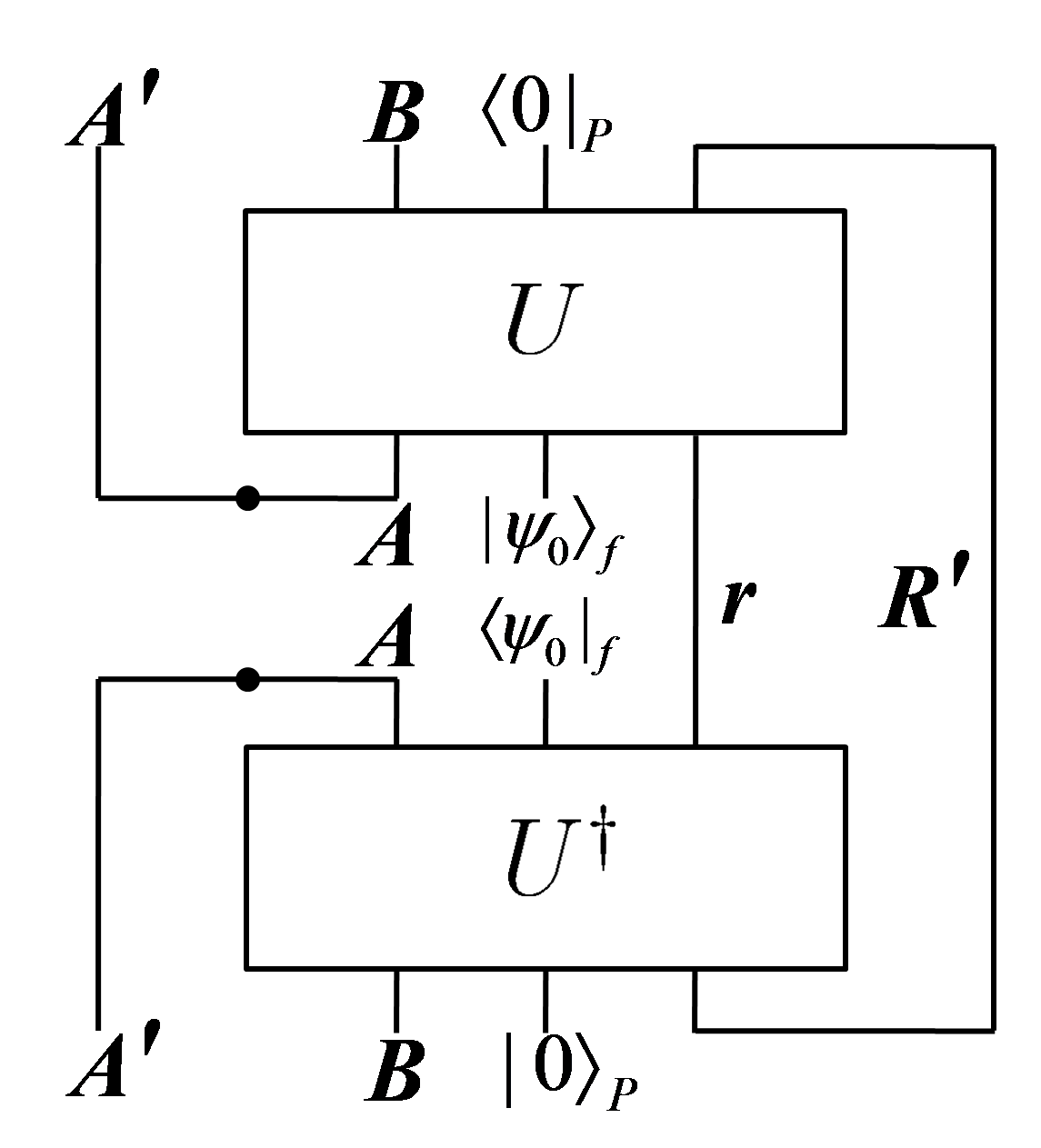}\;,
  \label{Eq:rho_a'b}
\end{eqnarray}
where the factor $\frac{1}{|r|}$ comes from the normalization factor of the $\textrm{EPR}$ state for the system $r$ and $R$. The above graph is obtained by juxtaposing the representation of $|\Psi_{\textrm{HP}}\rangle$ in Eq.~\eqref{Fig:psi_hp} with the representation of $\langle\Psi_{\textrm{HP}}|$ in Eq.~\eqref{Fig:psi_hp_dagger} and then connecting the same legs of the newly-generated radiation $R'$ and the early radiations $R$. Here, taking the trace over a specific system is simply realized by connecting the corresponding open ends in the graphical representation of the density matrix $|\Psi_{\textrm{HP}}\rangle\langle\Psi_{\textrm{HP}}|$.

Note that $\rho_{A'B}$ is not a real reduced density matrix in the usual sense, which can be observed by calculating its trace. Tracing out the remnant black hole $B$ and the reference $A'$ of Eq.~\eqref{Eq:rho_a'b} gives us
\begin{eqnarray}
  \textrm{Tr}\rho_{A'B}=\frac{|P|}{|A||r|} ~~~ \figbox{0.2}{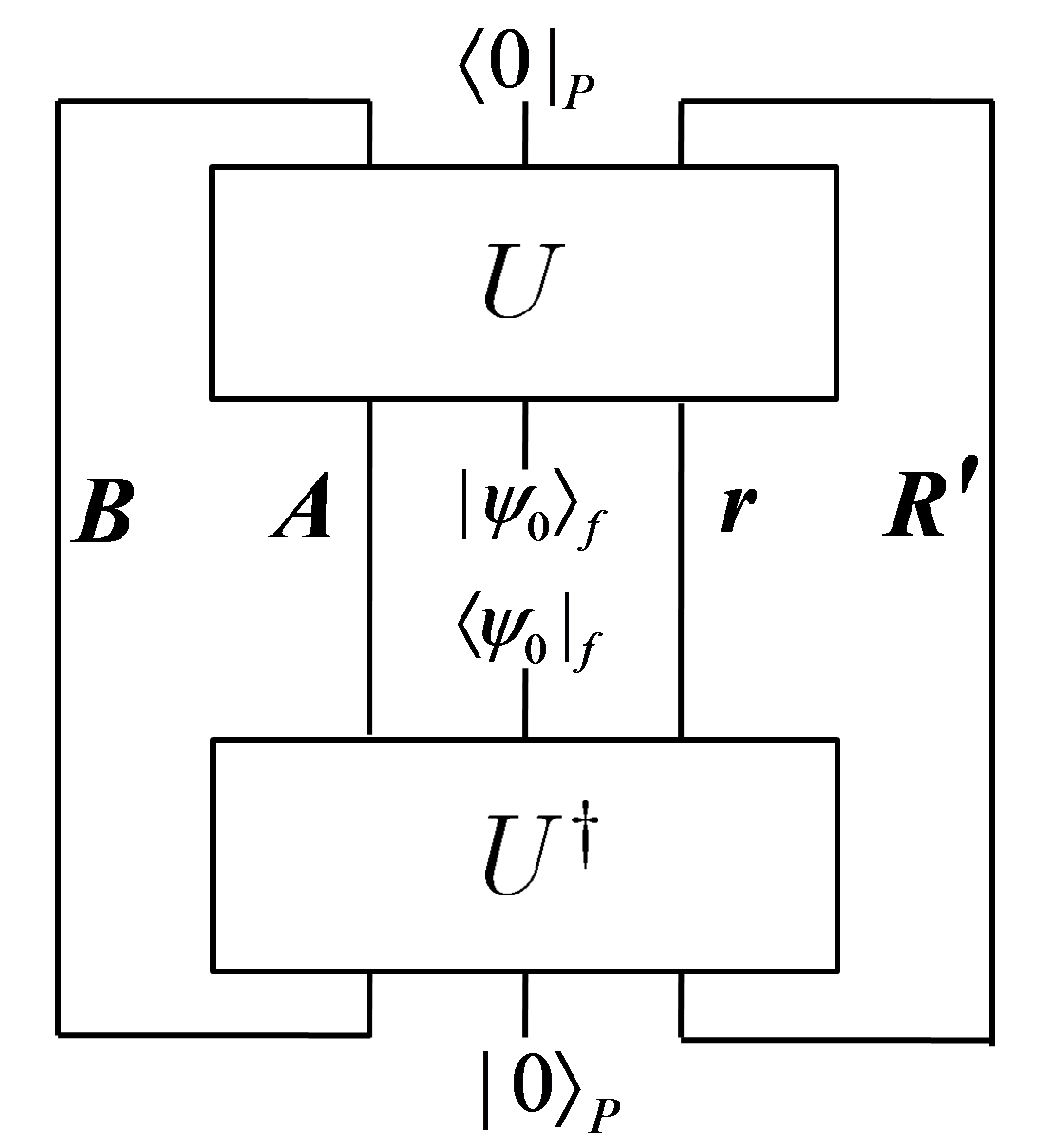}\;.
\end{eqnarray}
One can see 
\begin{eqnarray}
    \textrm{Tr}\rho_{A'B}\neq 1\;.
\end{eqnarray}
The reason is the same with $\langle\Psi_{\textrm{HP}}|\Psi_{\textrm{HP}}\rangle\neq 1$. The observation that the trace of $\rho_{A'B}$ is not equal to unity explains why we have put the double quotation marks to denote $\rho_{A'B}$.

However, for an observer in the effective description, the Haar average of $\textrm{Tr}\rho_{A'B}$ over the random unitary operator $U$ can be calculated by invoking the graphical representation in Eq.\eqref{2U_graph} 
\begin{eqnarray}
    \int dU \textrm{Tr}\rho_{A'B}=\frac{|P|}{d|A||r|}
    \figbox{0.2}{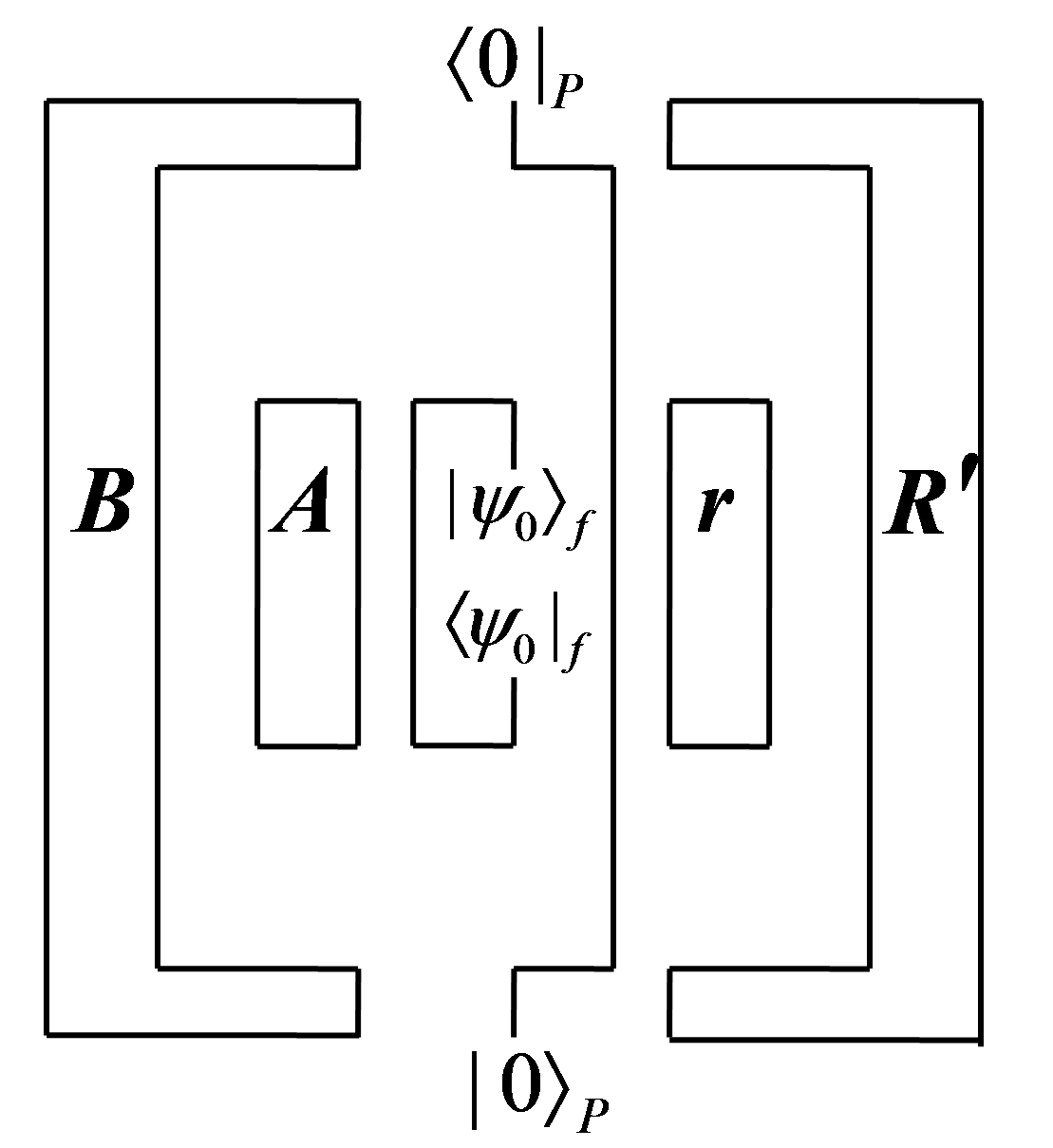}
    =\frac{|P||B||A||r||R'|}{d|A||r|}=1\;.
\end{eqnarray}
The technique in Eq.\eqref{HP_inner} is also used in the above derivation. This result shows that for an observer in the effective description, $\rho_{A'B}$ is a reduced density matrix on average over the random unitary operator.

Now we estimate the operator distance between the ``reduced density matrix" $\rho_{A'B}$ and the product state of the reference system $A'$ and the remnant black hole $B$ averaged over the random unitary operator $U$. We should consider the following quantity \cite{Harlow:2014yka} 
\begin{eqnarray}\label{quantity}
\left(\int dU \|\rho_{A'B}-\frac{1}{|A'||B|} I_{A'} \otimes I_B \|_1\right)^2\;,
\end{eqnarray}
where $\frac{1}{|A'|} I_{A'}$ and $\frac{1}{|B|} I_B$ are the maximally mixed density matrices of the system $A'$ and $B$. The operator trace norm $\|\cdot\|_1$ is the $L_1$ norm, defined for any operator $M$ as $\|M\|_1=\textrm{Tr}\sqrt{M^\dagger M}$. If the quantity in Eq.(\ref{quantity}) is small enough, the correlations between the reference system $A'$ and the remnant black hole $B$ can be ignored. Therefore, we try to estimate the upper bound of the quantity in Eq.(\ref{quantity}).

By defining the $L_2$ norm as $\|M\|_2=\sqrt{\textrm{Tr} M^\dagger M}$, and using the inequality $\|M\|_2\leq\|M\|_1\leq \sqrt{N} \|M\|_2$ with $N$ being the dimensionality of the Hilbert space, one can estimate
\begin{eqnarray}
\left(\int dU \|\rho_{A'B}-\frac{1}{|A'||B|} I_{A'} \otimes I_B \|_1\right)^2
&\leq& \int dU \|\rho_{A'B}-\frac{1}{|A'||B|} I_{A'} \otimes I_B \|_1^2\nonumber\\
&\leq& |A'||B| \int dU \|\rho_{A'B}-\frac{1}{|A'||B|} I_{A'} \otimes I_B \|_2^2\nonumber\\
&=&|A'||B| \int dU \textrm{Tr}\rho_{A'B}^2-1\;,
\end{eqnarray}
where we have used Jensen's inequality and the fact that $\int dU\textrm{Tr} \rho_{A'B}=1$.

To proceed, we calculate the average value of $\textrm{Tr}\rho_{A'B}^2$. Using the graphical representations of $\rho_{A'B}$ in Eq.\eqref{Eq:rho_a'b}, $\rho_{A'B}^2$ can be obtained by taking two copies of graphical representation of $\rho_{A'B}$ and connecting the legs of both the reference system $A'$ and the remnant black hole system $B$ in the middle. The trace of $\rho_{A'B}^2$ is obtained by connecting the remaining legs, which can be graphically expressed as
\begin{eqnarray}\label{Tr_rho_A'B_sq}
\textrm{Tr}\rho_{A'B}^2=\frac{|P|^2}{|A|^2|r|^2} ~~~\figbox{0.2}{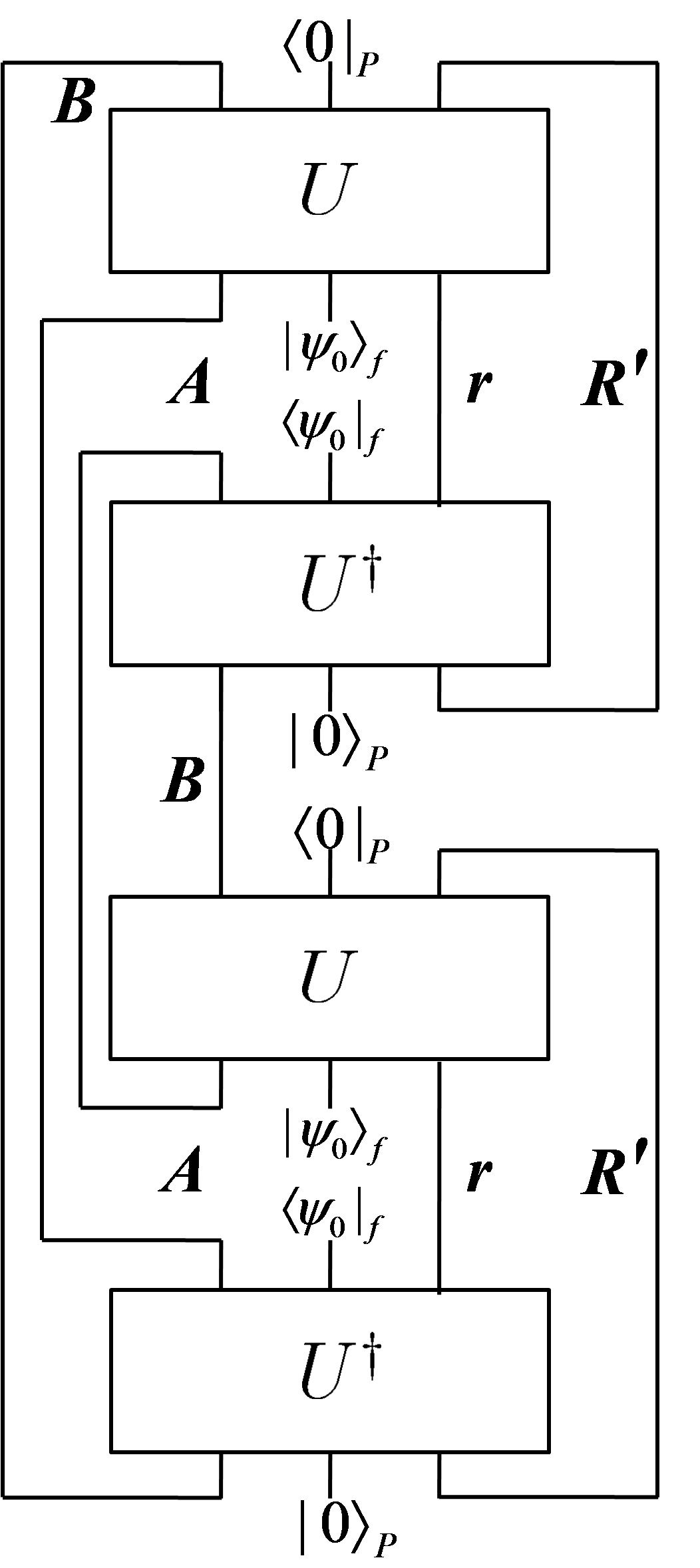}\;.
\end{eqnarray}
Computing the Haar average of $\textrm{Tr}\rho_{A'B}^2$ involves the following formula
\begin{eqnarray}\label{4U_integral}
\int dU\  U_{i_1j_1}U_{i_2j_2}U^\ast_{i_3j_3}U^\ast_{i_4j_4}
&=&\frac{\delta_{i_1i_3}\delta_{i_2i_4}\delta_{j_1j_3}\delta_{j_2j_4}
 + \delta_{i_1i_4}\delta_{i_2i_3}\delta_{j_1j_4}\delta_{j_2j_3}}{d^2-1}
 \nonumber\\
 &&-\frac{\delta_{i_1i_3}\delta_{i_2i_4}\delta_{j_1j_4}\delta_{j_2j_3}
 + \delta_{i_1i_4}\delta_{i_2i_3}\delta_{j_1j_3}\delta_{j_2j_4}}{d(d^2-1)}\,.
\end{eqnarray}
The details on evaluating such integrals are given in the Appendix A. With this in hand, the average of $\textrm{Tr}(\rho_{A'B})^2$ over the random unitary operator is given by
\begin{eqnarray}\label{Tr_rho_A'B_result}
    \int dU \textrm{Tr}\rho_{A'B}^2&=&\frac{|P|^2}{|A|^2|r|^2} \int dU U_{(a_1 f r_1)(b_1 0 r_1')}U_{(a_2 f r_2)(b_2 0 r_2')}U^*_{(a_2 f r_1)(b_2 0 r_1')}U^*_{(a_1 f r_2)(b_1 0 r_2')}\nonumber\\
    &=&\frac{|P|^2}{|A|^2|r|^2(d^2-1)} \left[ \delta_{a_1a_2}\delta_{r_1r_1}\delta_{a_2a_1}\delta_{r_2r_2}
    \delta_{b_1b_2}\delta_{r_1'r_1'}\delta_{b_2b_1}\delta_{r_2'r_2'}
    \right.
    \nonumber\\&&~~~~~~~~~~~~~~~~~~~
    +\delta_{a_1a_1}\delta_{r_1r_2}\delta_{a_2a_2}\delta_{r_2r_1}
    \delta_{b_1b_1}\delta_{r_1'r_2'}\delta_{b_2b_2}\delta_{r_2'r_1'}
    \nonumber\\&&~~~~~~~~~~~~~~~~~~~
    -\frac{1}{d}\delta_{a_1a_2}\delta_{r_1r_1}\delta_{a_2a_1}\delta_{r_2r_2}\delta_{b_1b_1}\delta_{r_1'r_2'}\delta_{b_2b_2}\delta_{r_2'r_1'}
    \nonumber\\&&~~~~~~~~~~~~~~~~~~~ \left.
     -\frac{1}{d}\delta_{a_1a_1}\delta_{r_1r_2}\delta_{a_2a_2}\delta_{r_2r_1}\delta_{b_1b_2}\delta_{r_1'r_1'}\delta_{b_2b_1}\delta_{r_2'r_2'}
    \right]\nonumber\\
    &=& \frac{|P|^2}{(d^2-1)}\left[\frac{|B||R'|^2}{|A|}+\frac{|B|^2|R'|}{|r|}-\frac{1}{d}\frac{|B|^2|R'|}{|A|}-\frac{1}{d}\frac{|B||R'|^2}{|r|}\right]\;.
\end{eqnarray}
Therefore, we have 
\begin{eqnarray}
|A'||B| \int dU \textrm{Tr}\rho_{A'B}^2-1&=&\frac{\left(|A|^2|f|-1\right)\left(d^2-|R'|^2|P|\right)}{\left(d^2-1\right)|R'|^2|P|}\nonumber\\
&\simeq& \frac{|A|^2|f|}{|R'|^2|P|}\left(1-\frac{1}{|B|^2|P|}\right)\nonumber\\
&\simeq& \frac{|A|^2|f|}{|R'|^2|P|}\;,
\end{eqnarray}
which gives the inequality of the operator distance between the  ``reduced density matrix" $\rho_{A'B}$ and the decoupled density matrix $\frac{1}{|A'||B|} I_{A'} \otimes I_B$ 
\begin{eqnarray}
   \int dU \|\rho_{A'B}-\frac{1}{|A'||B|} I_{A'} \otimes I_B \|_1\leq \sqrt{\frac{|f|}{|P|}} \frac{|A|}{|R'|}\;.
\end{eqnarray}
If the following condition is satisfied, i.e.,
\begin{eqnarray}\label{decoupling_cond}
   |R'| \gg \sqrt{\frac{|f|}{|P|}} |A|\;,
\end{eqnarray}
then we have 
\begin{eqnarray}
    \int dU \|\rho_{A'B}-\frac{1}{|A'||B|} I_{A'} \otimes I_B \|_1 \ll 1\;.
\end{eqnarray}
This equation implies that the operator distance between $\rho_{A'B}$ and the product state of the reference system $A'$ and the remnant black hole $B$ averaged over the random unitary operator is small enough. Therefore, the reference system $A'$ is decoupled from the remnant black hole $B$ and the entanglement between the reference system $A$ and the message system $A'$ is transferred to the entanglement between the reference system $A'$ and the newly-generated Hawking radiation $R'$. In this case, the information contained in the message system $A$ can be recovered by the outside observer who has the full access of the early Hawking radiation $R$ and the newly-generated Hawing radiation $R'$. In this sense, Eq.(\ref{decoupling_cond}) is the decoupling condition to guarantee the information can be retrieved from the black hole.

\section{Decoding Hawking radiation and the Page transition}\label{secV}

In this section, we consider how the observer who stays outside of the black hole can use the Hawking radiation that one collected and apply the Yoshida-Kitaev decoding strategy to recover the information thrown into the black hole. The strategy was firstly proposed by Yoshida and Kitaev in \cite{Yoshida:2017non} for the original model of Hayden-Preskill thought experiment. One can refer to \cite{Yoshida:2018vly,Bao:2020zdo,Cheng:2019yib,Li:2021mnl} for the decoding strategies with the quantum decoherence or noise. In these strategies, it is assumed that the outside observer has the full information about the information scrambling and black hole evaporating, which is usually represented by an random unitary operator.

\subsection{Probabilistic decoding and transition at Page time}

We have claimed that the dynamics of the information scrambling and the black hole evaporating are represented by the random unitary matrix. If the message system $A$, which is entangled with the reference system $A'$, is thrown into the black hole, after some time the quantum state of the whole system is described by the modified Hayden-Preskill state. For the outside observer, he has the full access to the early Hawking radiation $R$ and the newly-generated Hawking radiation $R'$. The observer wants to apply some operations to recover the information that is contained in the message system $A$. The probabilistic decoding strategy can be implemented as follows.

Firstly, we prepare one copy of $|\Psi_0\rangle_f$ and one copy of $|\textrm{EPR}\rangle_{A'A}$. The copy of $|\textrm{EPR}\rangle_{A'A}$ is denoted as $|\textrm{EPR}\rangle_{FF'}$. Then, with the modified Hayden-Preskill state in hand, apply the complex conjugate $U^*$ of the random unitary operator on the composed system of $R$, $f$ and $F$. The resultant state is denoted as $|\Psi\rangle_{in}$, which can be graphically expressed as  
\begin{eqnarray}
  |\Psi\rangle_{in}=|P|\ C~~~ \figbox{0.2}{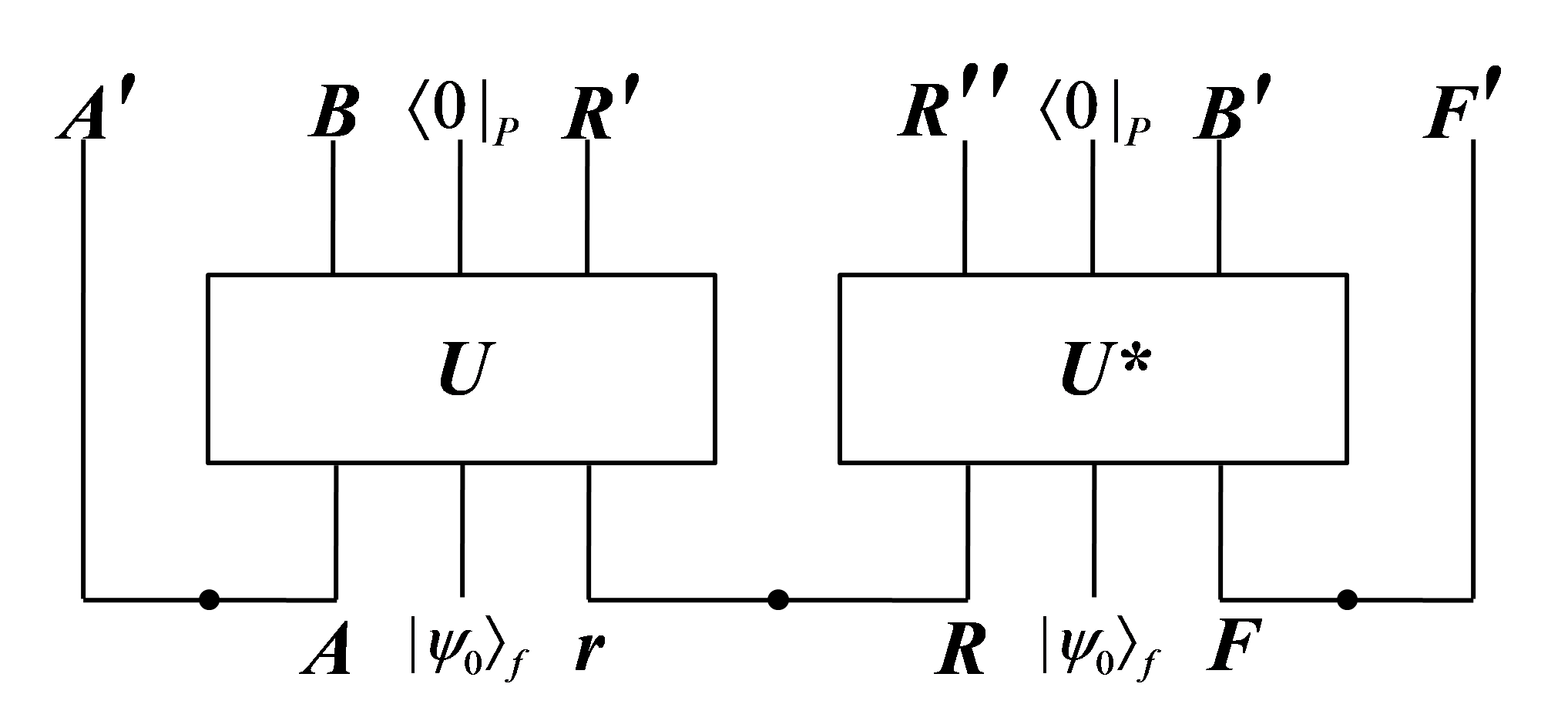}\;,
\end{eqnarray}
where $C$ is the normalization constant. The operator $U^*$ can be treated as the time reversal operator of the black hole dynamics. The output system of $U^*$ consists of a copy of newly-generated radiation $R''$, a copy of the remnant black hole $B'$ and another auxiliary system $P$. The output auxiliary system $P$ is post-selected or projected onto the fixed state $|0\rangle_P$. The normalization condition $\int dU ~_{in}\langle \Psi|\Psi\rangle_{in}=1$ gives the $C=\min\left(1,\sqrt{\frac{|A||B||R'|}{|r|}}\right)=\min\left(1,\sqrt{\frac{|f|}{|P|}}|A|\right)$. For small systems $A$ and $R'$ compared with the black hole size, the Page time is approximately when $|B|=|r|$. Therefore, roughly speaking $C=1$ before the Page time and $C=\sqrt{\frac{|A||B||R'|}{|r|}}$ starting from some period after the Page time. For generality of discussion and also the number of qubits used in the quantum simulation, we do not assume the above limit in our study.

Next, we project the system $R'$ and $R''$ onto the state $|\textrm{EPR}\rangle_{R'R''}$. This is to act the projecting operator $\Pi_{R'R''}=|\textrm{EPR}\rangle_{R'R''}\langle \textrm{EPR}|_{R'R''}$ on the system $R'$ and $R''$. The resulting state is denoted as $|\Psi\rangle_{out}$, which can be graphically expressed as 
\begin{eqnarray}
   |\Psi\rangle_{out}&=&\left(I_{A'B}\otimes\Pi_{R'R''}\otimes I_{B'F'} \right)|\Psi\rangle_{in}\nonumber\\
   &=&\frac{|P|C}{\sqrt{P_{\textrm{EPR}}}}~~~ 
   \figbox{0.2}{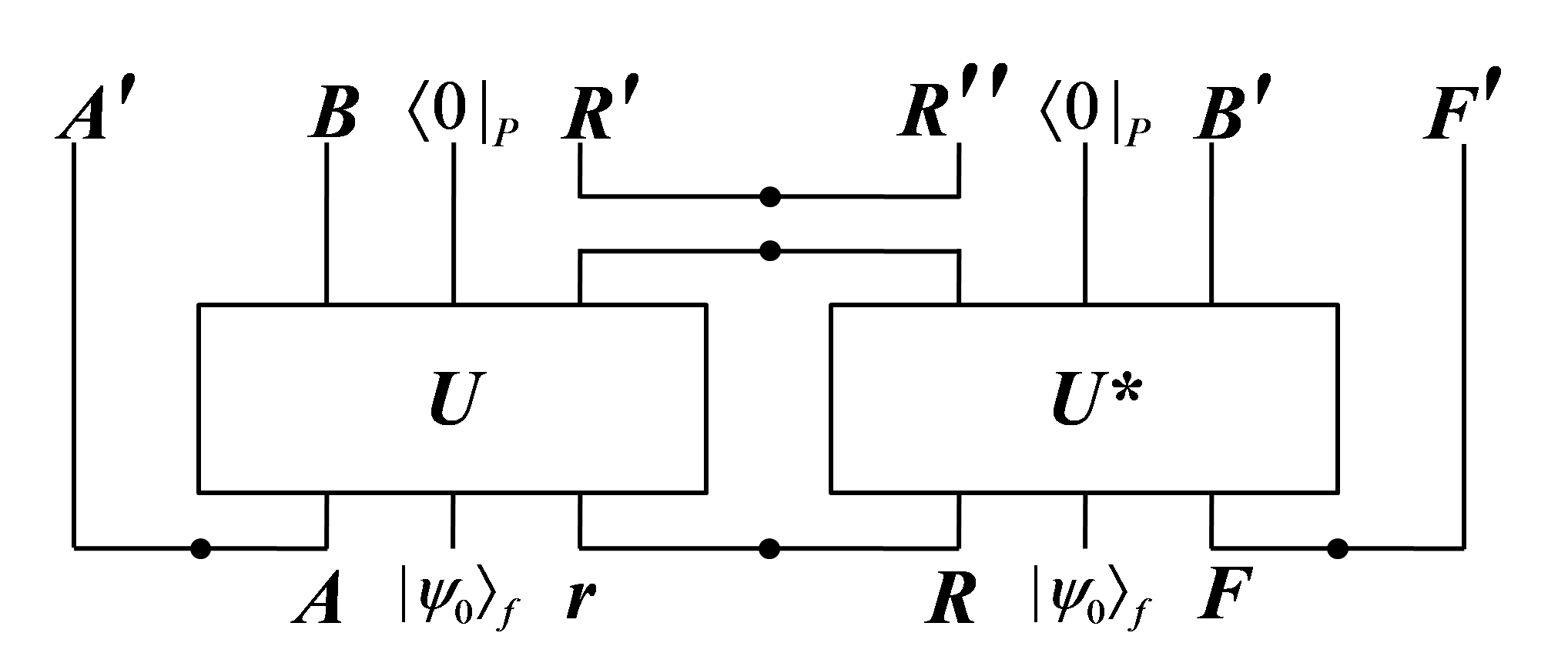}\;,
   \label{Eq:psiout}
\end{eqnarray}
where $P_{\textrm{EPR}}$ is the averaged projecting probability. The projecting operation $\Pi_{R'R''}$ serves to decouple the prepared system $F'$ from the remnant black holes $B$ and $B'$ and teleports the quantum state of the message system $A$ to the prepared system $F'$ owned by the outside decoder.

The factor $\frac{1}{\sqrt{P_{\textrm{EPR}}}}$ is introduced to preserve the normalization of the state $|\Psi\rangle_{out}$ on the Haar average of the random unitary operator. Therefore, the condition $\int dU ~_{out}\langle \Psi|\Psi\rangle_{out}=1$ gives the graphical representation of the projecting probability,
\begin{eqnarray}
  P_{\textrm{EPR}}=\frac{|P|^2 C^2}{|A|^2|r||R'|}\int dU \left( \figbox{0.2}{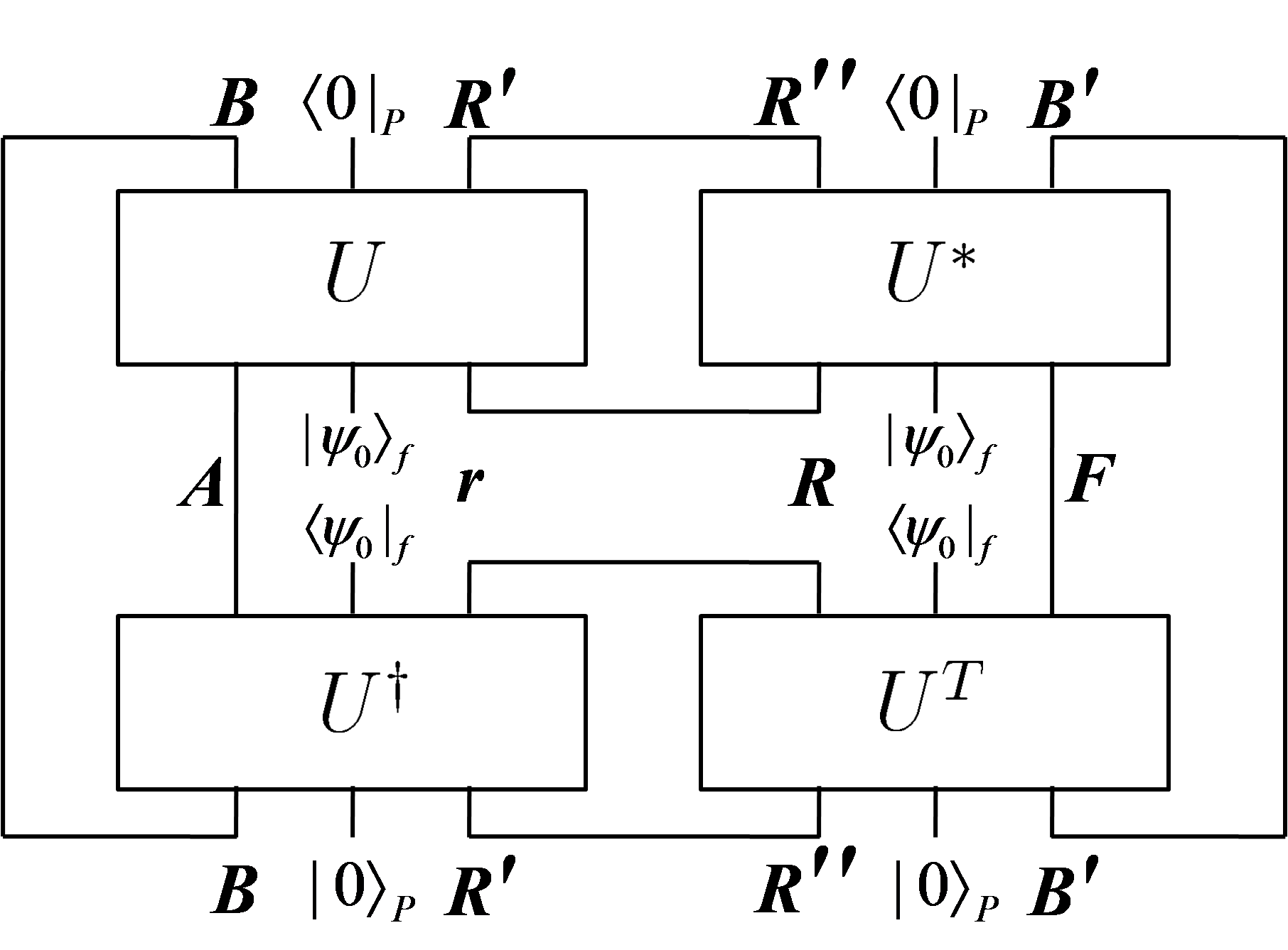}\right)\,,
\end{eqnarray}
where the inner product $~_{out}\langle \Psi|\Psi\rangle_{out}$ is represented by connecting the legs of $|\Psi\rangle_{out}$ in the upper half of the graph to the corresponding legs of $_{out}\langle \Psi|$ in the lower half. After a rearrangement of the unitary operators, this graph is equivalent to that of $\textrm{Tr}\rho_{A'B}^2$ in Eq.\eqref{Tr_rho_A'B_sq}, which results in the following relation
\begin{eqnarray}
    P_{\textrm{EPR}}=\frac{|r|C^2}{|R'|} \int dU \textrm{Tr}\rho_{A'B}^2\;.
\end{eqnarray}
By using the previous result of $\int dU \textrm{Tr}\rho_{A'B}^2$ given in Eq.\eqref{Tr_rho_A'B_result}, it can be shown that the projecting probability is given by 
\begin{eqnarray}
     P_{\textrm{EPR}}&=&\frac{|P|^2 C^2}{(d^2-1)}\left[\frac{|r||B||R'|}{|A|}+|B|^2-\frac{1}{d}\frac{|r||B|^2}{|A|}-\frac{1}{d}|B||R'|\right]\nonumber\\
    &\simeq& C^2 \left(\frac{|P|}{|f|}\frac{1}{|A|^2}+\frac{1}{|R'|^2}
    -\frac{1}{|f|}\frac{1}{|A|^2|R'|^2}-\frac{|P|}{d^2}\right)\;.
\end{eqnarray}
Under the decoupling condition \eqref{decoupling_cond}, the projecting probability can be further approximated as 
\begin{gather}
    P_{\textrm{EPR}}\simeq \min\left( 1,\frac{|P|}{|f|}\frac{1}{|A|^2}\right)\,.
\end{gather}

\begin{figure}[th]
  \centering
  \includegraphics[width=9cm]{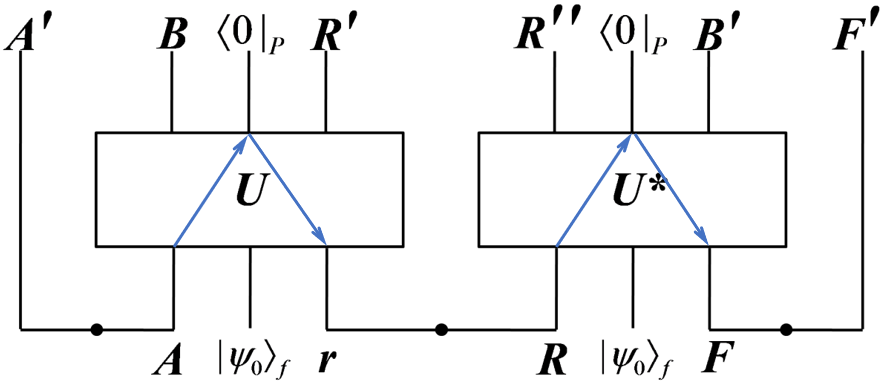}
  \caption{The flow of information through local projections on system $P$ after the transition around the Page time. Prior to the transition, the information is transmitted through the EPR projection on $R'$ and $R''$. The information initially stored in the entanglement between $A$ and $A'$ is transmitted to the entanglement between $A'$ and $F'$. The arrow indicates the direction of the information flow.}
  \label{Fig:psi_in_arrow}
\end{figure}

\begin{figure}
  \centering
  \includegraphics[width=.6 \columnwidth]{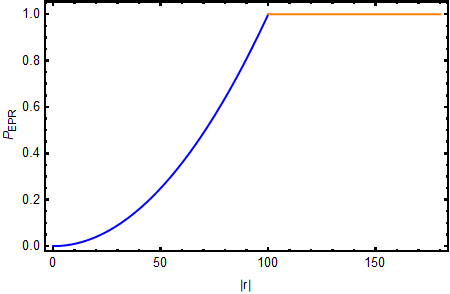}
  \caption{The probability of the EPR projection $P_{\mathrm{EPR}}$ varies with the dimension of early radiation $|r|$. The parameters used are: $|f|=10^4\,, |A|=1$. As shown, the transition emerges at the Page time defined by $|r|=\sqrt{|f|}=100$.}
  \label{Fig:pepr}
\end{figure}

Here only the leading order term is retained. This result shows that the projecting probability depends not only on the dimensionality of the message system $A$ but also depends on the ratio of the dimensionalities of the projecting system $P$ in the black hole interior and the initial collapsing matter's system $f$. For the case of $\frac{|P|}{|f|}\frac{1}{|A|^2}>1$, i.e., the late stage of the evaporation, the projection on $R'$ and $R''$ becomes unnecessary. From the decoupling condition, the new radiation $R'$ can be completely ignored or made arbitrarily small in dimension in the extreme case when $|P|\gg |f|$. This suggests that at the very late stage of the evaporation, the information is transmitted through a different channel shown in Fig.~\ref{Fig:psi_in_arrow}, while before the transition, information is transmitted through the EPR projection of $R'$ and $R''$. The transition can be shown through probability of EPR projections as shown in Figure~\ref{Fig:pepr}. Here, we have imposed the information constraint $|A||f|/|r|=|B||R'|$ and rewriten the probability of the EPR projection as $P_{\textrm{EPR}}\simeq \min\left( 1,\frac{|R|^2}{|f||A|^2}\right)$.

The decoding fidelity can be quantified by the derivation of the out state $|\Psi\rangle_{out}$ from $|\textrm{EPR}\rangle_{A'F'}$. The decoding fidelity is then defined and graphically expressed as
\begin{eqnarray}
F_{\textrm{EPR}}&=&\textrm{Tr}\left(\Pi_{A'F'}|\Psi\rangle_{out}~_{out}\langle\Psi| \right)\nonumber\\
&=&\frac{|P|^2C^2}{P_{\textrm{EPR}}|A|^3|r||R'| } ~~~\figbox{0.2}{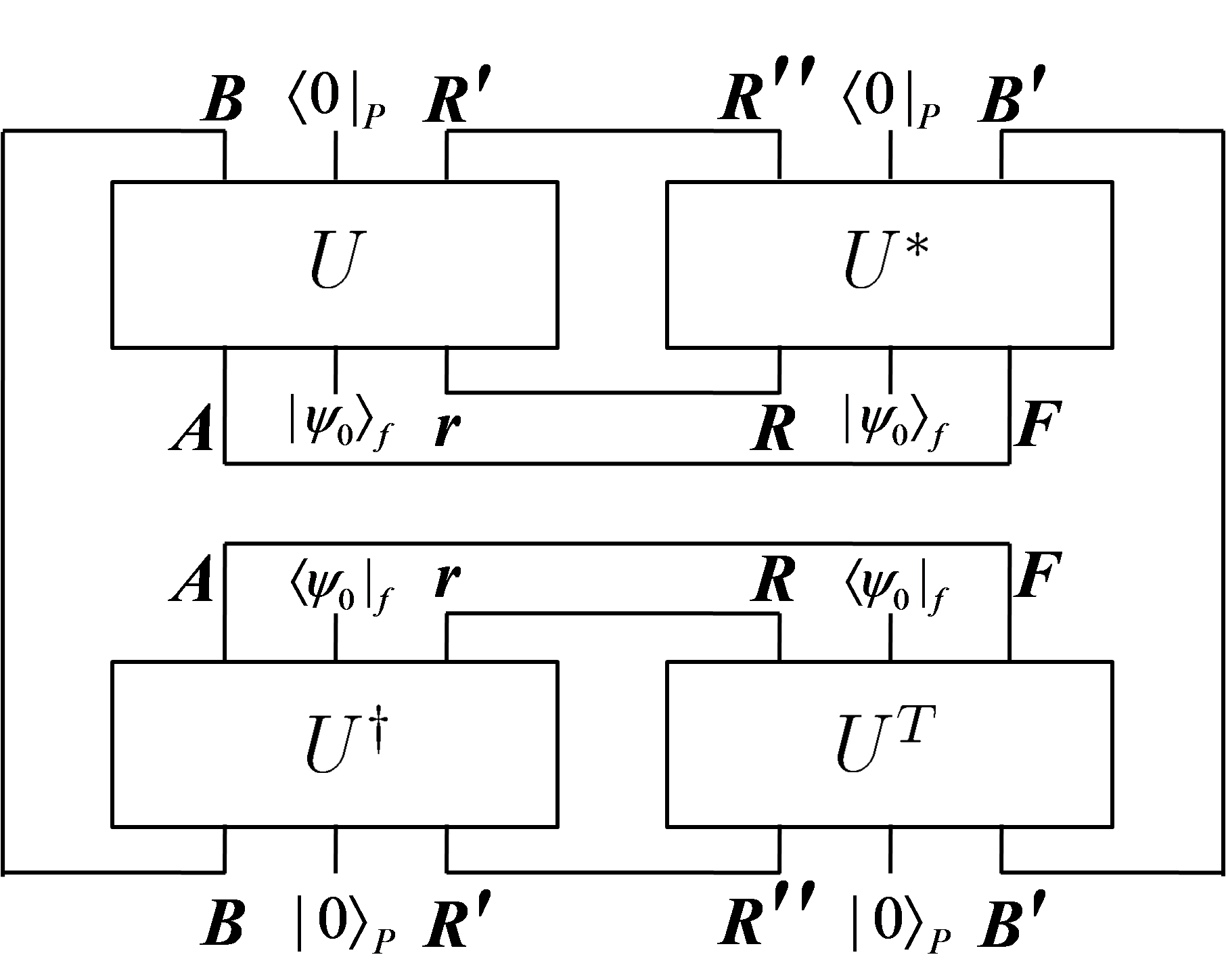}\;, 
\end{eqnarray}
where the upper half of the graph represents the state $|\Psi\rangle_{out}$ and the lower half represents $_{out}\langle \Psi|$. The techniques of the operators acting on the system and tracing out the system used in the previous calculations are also applied here.

The average of the decoding fidelity over the random unitary group can be calculated as 
\begin{eqnarray}
    \int dU F_{\textrm{EPR}} &=&\frac{|P|^2C^2}{P_{\textrm{EPR}}|A|^3|r||R'| } \int dU U_{(a_1 f r_1)(b_1 0 r_1')}U_{(a_2 f r_2)(b_2 0 r_2')}U^*_{(a_1 f r_1)(b_2 0 r_1')}U^*_{(a_2 f r_2)(b_1 0 r_2')}\nonumber\\
    &=&\frac{|P|^2C^2}{P_{\textrm{EPR}}|A|^3|r||R'|(d^2-1) } \left[ \delta_{a_1a_1}\delta_{r_1r_1}\delta_{a_2a_2}\delta_{r_2r_2}
    \delta_{b_1b_2}\delta_{r_1'r_1'}\delta_{b_2b_1}\delta_{r_2'r_2'}
    \right.
    \nonumber\\&&~~~~~~~~~~~~~~~~~~~
    +\delta_{a_1a_2}\delta_{r_1r_2}\delta_{a_2a_1}\delta_{r_2r_1}
    \delta_{b_1b_1}\delta_{r_1'r_2'}\delta_{b_2b_2}\delta_{r_2'r_1'}
    \nonumber\\&&~~~~~~~~~~~~~~~~~~~
    -\frac{1}{d}\delta_{a_1a_1}\delta_{r_1r_1}\delta_{a_2a_2}\delta_{r_2r_2}\delta_{b_1b_1}\delta_{r_1'r_2'}\delta_{b_2b_2}\delta_{r_2'r_1'}
    \nonumber\\&&~~~~~~~~~~~~~~~~~~~ \left.
     -\frac{1}{d}\delta_{a_1a_2}\delta_{r_1r_2}\delta_{a_2a_1}\delta_{r_2r_1}\delta_{b_1b_2}\delta_{r_1'r_1'}\delta_{b_2b_1}\delta_{r_2'r_2'}
    \right]\nonumber\\
    &=& \frac{d^2C^2}{P_{\textrm{EPR}}(d^2-1)|A|^2}\left[\frac{|P|}{|f|}+\frac{1}{|R'|^2}-\frac{1}{|f||R'|^2}-\frac{|P|}{d^2}\right]\nonumber\\
     &\simeq& \frac{|P|C^2}{P_{\textrm{EPR}}|A|^2|f|}\;.
   \end{eqnarray} 
    
If the decoupling condition \eqref{decoupling_cond} is satisfied, the projecting probability is approximated as $\frac{|P|}{|f|}\frac{1}{|A|^2}$, which implies that the decoding fidelity achieves the maximal decoding quality
\begin{eqnarray}
    F_{\textrm{EPR}}\simeq \frac{|P|C^2}{P_{\textrm{EPR}}|A|^2|f|}\simeq 1\;.
\end{eqnarray}
In summary, we have shown that the Yoshida-Kitaev probabilistic decoding strategy can be successfully employed in our modified Hayden-Preskill protocol to decode the Hawking radiation and recover the information falling into the black hole, and that the information is transferred to the outside of the black hole through two different channels switched at the Page time.

\subsection{Grover's search strategy}

We have shown that the probabilistic decoding strategy can be applied to recover the initial information. For $|P|\gg |f||A|^2$, no projection is necessary for the decoding procedure. For $|P|\lesssim |f|$, an additional projection probability is involved. In this subsection, we discuss a decoding strategy in reminiscent of the Grover's search algorithm for the modified Hayden-Preskill protocol, which can circumvent this probability \cite{Grover:1996rk}. 
 
Assuming $|P| < |f||A|^2$, we define the following operator 
\begin{eqnarray}
    \tilde{\Pi}_{R'BA}= |P|~~~\figbox{0.2}{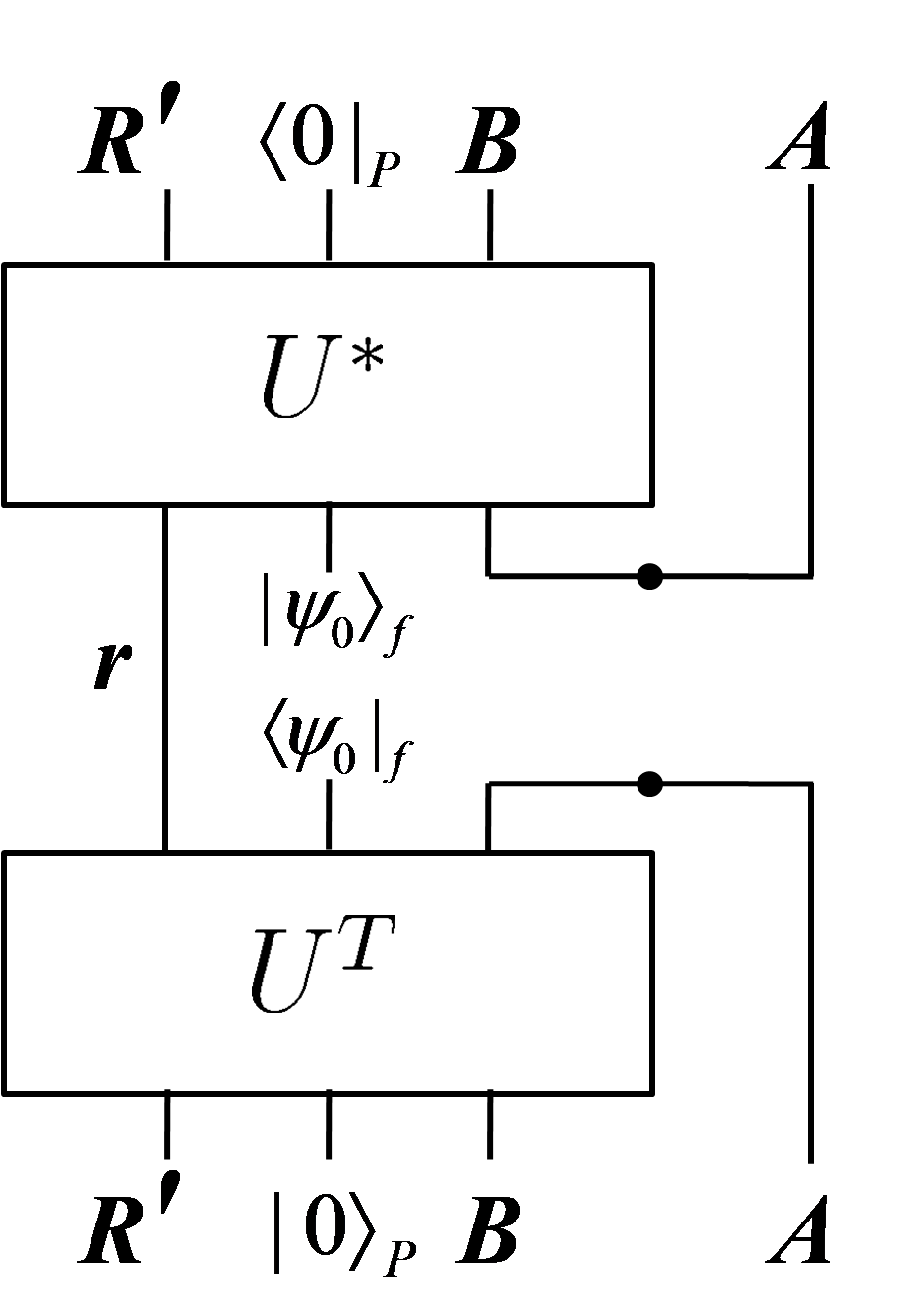}\;,
\end{eqnarray}
which operates on the newly-generated radiation $R'$, the remnant black hole $B$ and the message system $A$. In the ideal case, one needs to prove the following relations in order to apply the Grover's search algorithm
\begin{eqnarray}\label{G_eqs}
   \left(I_{A'B}\otimes\Pi_{R'R''}\otimes I_{B'F'}\right) |\Psi\rangle_{in}&=&\sqrt{\frac{|P|}{|f|}}\frac{1}{|A|}|\Psi\rangle_{out}\;,\nonumber\\
   \left(I_{A'B}\otimes\Pi_{R'R''}\otimes I_{B'F'}\right) |\Psi\rangle_{out}&=&|\Psi\rangle_{out}\;,\nonumber\\
  \left(I_{A'BR'}\otimes \tilde{\Pi}_{R''B'F'} \right) |\Psi\rangle_{in}&=&|\Psi\rangle_{in}\;,\nonumber\\
  \left(I_{A'BR'}\otimes \tilde{\Pi}_{R''B'F'} \right)|\Psi\rangle_{out}&=&\sqrt{\frac{|P|}{|f|}}\frac{1}{|A|}|\Psi\rangle_{in}\;.
\end{eqnarray}
The first two relations are apparent. The last two relations are satisfied only for the typical random unitary operator $U$ in the ideal case. The last two relations can be verified by showing that the distance of the density matrices for the states on the l.h.s and the r.h.s of the equations averaged over the random unitary group is small. A less rigorous verification of the last two relations is presented in the Appendix B and the proof is presented in the Appendix C.   

\begin{figure}
  \centering
  \includegraphics[width=9cm]{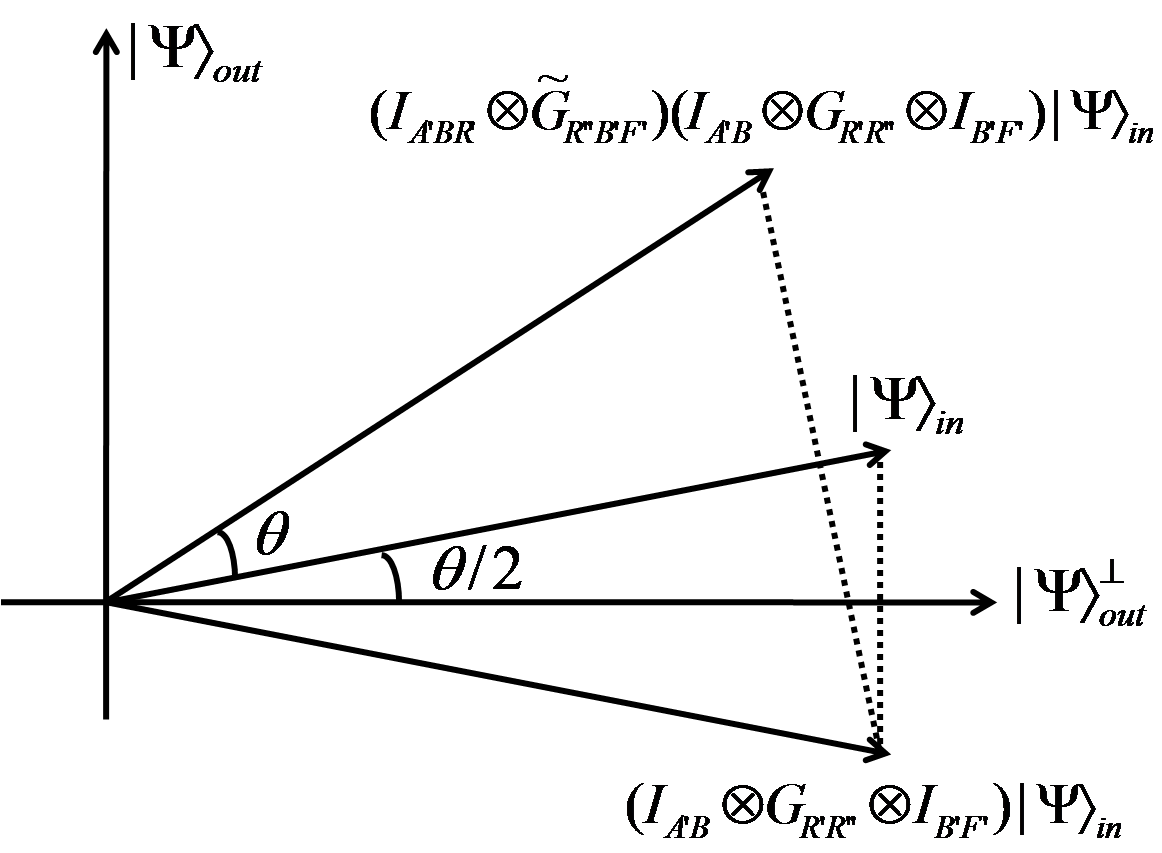}
  \caption{Grover rotation: an illustration of the application of the operators $G$ and $\tilde{G}$ on the state $|\Psi\rangle_{in}$. The two dimensional plane is spanned by $|\Psi\rangle_{in}$ and $|\Psi\rangle_{out}$. }
  \label{Grover_rotation}
\end{figure}

Consider the two dimensional plane spanned by $|\Psi\rangle_{in}$ and $|\Psi\rangle_{out}$. On this plane, we also introduce a state vector $|\Psi\rangle_{out}^{\perp}$ that is orthogonal to $|\Psi\rangle_{out}$. It is easy to check that 
\begin{eqnarray}
  |\Psi\rangle_{out}^{\perp}\propto \left(1-I_{A'B}\otimes\Pi_{R'R''}\otimes I_{B'F'}\right) |\Psi\rangle_{out}\;.
\end{eqnarray}
By defining the unitary operator $G$ as 
\begin{eqnarray}\label{G_def}
    G=1-2\Pi\;,
\end{eqnarray}
one can show that the inner product of $|\Psi\rangle_{in}$ and $|\Psi\rangle_{out}^{\perp}$ is equal to the inner product of $\left(I_{A'B}\otimes G_{R'R''}\otimes I_{B'F'}\right)|\Psi\rangle_{in}$ and $|\Psi\rangle_{out}^{\perp}$, i.e. the following relation holds
\begin{eqnarray}
    ~_{in}\langle\Psi|\Psi\rangle_{out}^{\perp}
    =~_{in}\langle\Psi|\left(I_{A'B}\otimes G_{R'R''}\otimes I_{B'F'}\right)|\Psi\rangle_{out}^{\perp}\;.
\end{eqnarray}
Therefore, the application of the operator $G$ on the state $|\Psi\rangle_{in}$ results in a reflection across the state $|\Psi\rangle_{out}^{\perp}$. The reflection angle $\theta$ is determined by the equation 
\begin{eqnarray}
   \sin\frac{\theta}{2}=\sqrt{\frac{|P|}{|f|}}\frac{1}{|A|}\;. 
\end{eqnarray}

Similarly, one can define the $\tilde{G}$ operator 
\begin{eqnarray}
    \tilde{G}=1-2\tilde{\Pi}\;.
\end{eqnarray}
The application of the operator $\tilde{G}$ on the state $\left(I_{A'B}\otimes G_{R'R''}\otimes I_{B'F'}\right)|\Psi\rangle_{in}$ means a reflection across the state $|\Psi\rangle_{in}$. The reflection angle is also given by $\theta$. Such a procedure is presented in Figure~\ref{Grover_rotation}, where the operation of $\tilde G$ is accomplished by $U^\ast G U^T$. Therefore, the application of the combined operator $\tilde{G} G$ on the state $|\Psi\rangle_{in}$ results in the rotation of this state on the two dimensional plane by the angle $\theta$. Such a procedure is similar to Grover's search algorithm. After $n$ times, we have 
\begin{eqnarray}
|\Psi(n)\rangle=\sin\left((n+1/2)\theta\right)|\Psi\rangle_{out}+\cos\left((n+1/2)\theta\right)|\Psi\rangle_{out}^{\perp}\;.
\end{eqnarray}

\begin{figure}
  \centering
  \includegraphics[width=8cm]{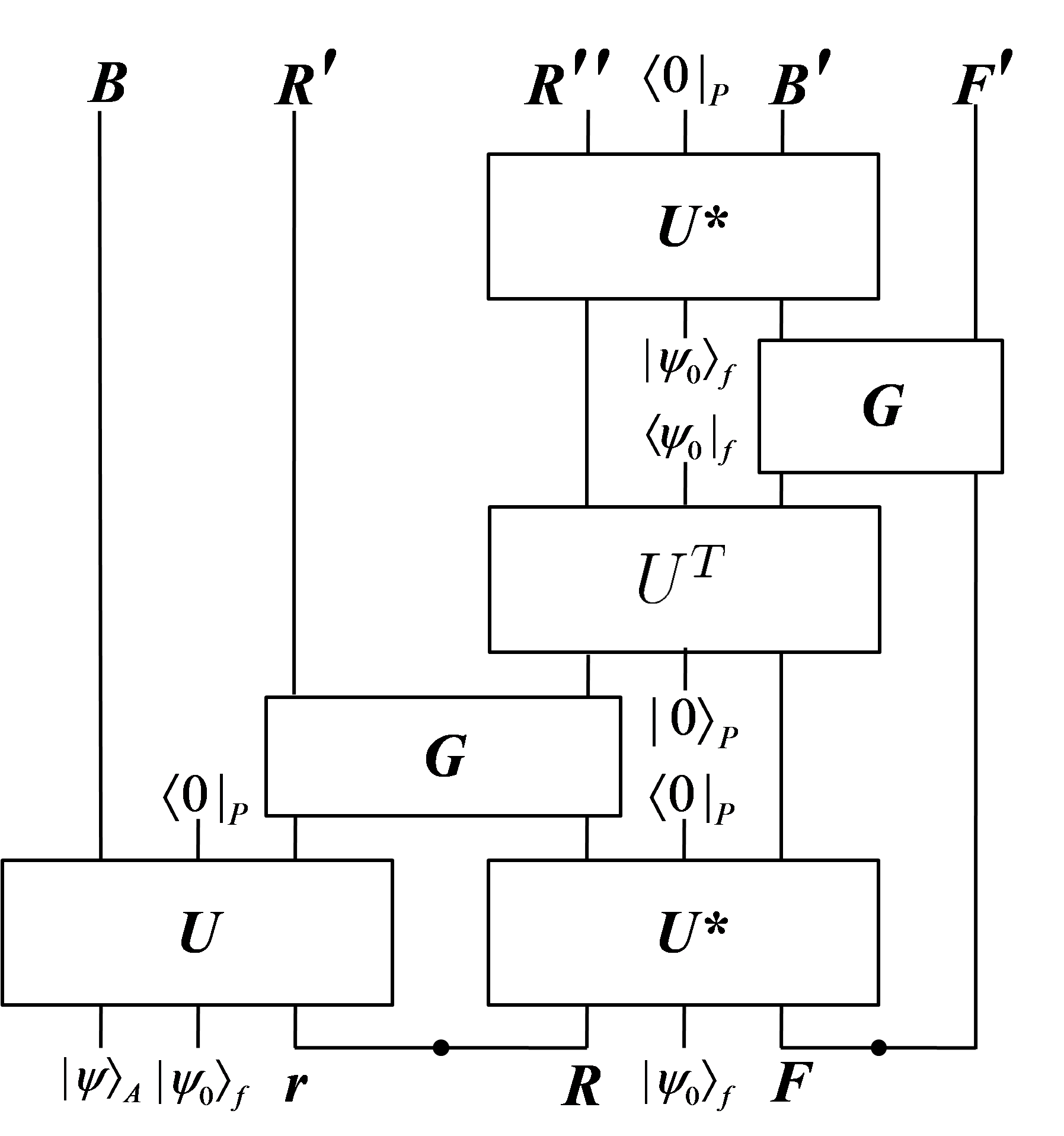}
  \caption{Grover's search decoding strategy of Hawking radiation in the non-isometric model of black hole interior. The message system $A$ is set to be in the quantum state $|\psi\rangle_{A}$. After applying the operators $G$, $U^{T}$, $G$ and $U^{\ast}$ successively, the state $|\psi\rangle_{A}$ is recovered by the prepared system $F'$.}
  \label{deterministic_decoder}
\end{figure}

For our quantum simulation that will be discussed in the following section, the message system $A$, the infalling matter system $f$ and the projecting system $P$ are represented by one qubit respectively. So we have $|P|=|f|=|A|=2$ and $\theta=\frac{\pi}{3}$. In this case, the initial quantum state of the message system $A$ can be successfully recovered by applying the combined operator $\tilde{G}G$ on the state $|\Psi\rangle_{in}$ only one time. Such a strategy is presented in Figure~\ref{deterministic_decoder}. Note that the operation of $\tilde G$ is accomplished by $U^\ast G U^T$. With the initial state $|\Psi\rangle_{in}$ in hand, the decoder should apply sequentially the reflection operator $G$ on the newly generated radiation $R'$ and its copy $R''$, the scrambling operator $U^{T}$ on the radiation copy $R''$ and the black hole copy $B'$, and again the reflection operator $G$ on the black hole copy $B'$ and the prepared system $F'$, and then $U^{\ast}$ on the radiation copy $R''$ and the black hole copy $B'$. In this way, the 
decoder can retrieve the information of the message system $A$, namely its quantum state $|\psi\rangle_{A}$, on the prepared system $F'$ outside the black hole.

\section{Quantum simulation of decoding Hawking radiation}\label{secVI}

Recently, works on the quantum processors have claimed to realize the equivalence of the ``traversable wormholes" on quantum chips and have attracted significant attentions \cite{Landsman:2018jpm,Brown:2019hmk,Nezami:2021yaq,Shapoval:2022xeo,Jafferis:2022crx,Shi:2021nkx}. These have stimulated researches on the simulations of quantum gravity in the laboratory. Benefited by the development of quantum computers, it is believed that certain essential quantum features of black holes can be simulated on the quantum computers, which will provide us a deeper understanding of the nature of quantum gravity.

In this section, we try to implement the probabilistic and the Grover's search decoding strategies for the Hawking radiation on the IBM quantum processors to verify the feasibility of the information recovery from the black hole. To this end, we experimentally realize the decoding strategies discussed in the last section on the 7-qubit IBM quantum processors using a 3-qubit scrambling unitary. The key is to realize the typical Haar scrambling unitary operator on the quantum processors. This is a difficult task especially in IBM quantum processors because the seven qubits on the IBM quantum processors are not fully connected. Some optimization schemes for the quantum circuit should be taken carefully.

\subsection{A typical scrambling unitary operator}

\begin{figure}
  \centering
  \includegraphics[width=9cm]{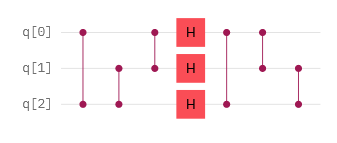}
  \caption{The scrambling unitary operator used to realize the quantum computation. The scrambling unitary is composed by six controlled-Z gates and three Hadamard gates. The ordering of the left or the right three controlled-Z gates does affect the scrambling unitary. }
  \label{U_scr}
\end{figure}

Firstly, we discuss how to realize the scrambling unitary operator on the IBM quantum processors. We consider the 3-qubit Clifford scrambler \cite{Yoshida:2018vly}. An ideal three-qubit Clifford scrambling unitary operator should transform single-qubit operations into three-qubit operations. An example of such scrambling unitaries satisfying Eq.~\eqref{Eq:scrambling} can be realized using the quantum circuit shown in Figure~\ref{U_scr}. Algebraically, the quantum circuit of the scrambling unitary in Figure~\ref{U_scr} can be expressed as 
\begin{eqnarray}
    U&=&\left(I \otimes I \otimes |0\rangle\langle0| + I \otimes \sigma_z \otimes |1\rangle\langle1|\right)
    \left(I \otimes |0\rangle\langle0| \otimes I + \sigma_z \otimes |1\rangle\langle1| \otimes I\right)\nonumber\\&&\times
    \left(I \otimes I \otimes |0\rangle\langle0|+\sigma_z \otimes I \otimes |1\rangle\langle1|\right)
    \left(H \otimes H \otimes H\right)
     \left(I \otimes |0\rangle\langle0| \otimes I + \sigma_z \otimes |1\rangle\langle1| \otimes I\right)\nonumber\\&&\times
    \left(I \otimes I \otimes |0\rangle\langle0| + I \otimes \sigma_z \otimes |1\rangle\langle1|\right)
    \left(I \otimes I \otimes |0\rangle\langle0|+\sigma_z \otimes I \otimes |1\rangle\langle1|\right) \nonumber\;,
\end{eqnarray}
where $\sigma_{x}$, $\sigma_y$, $\sigma_z$ are Pauli matrices and $I$ is the two dimensional identity matrix. Note that in Figure~\ref{U_scr}, the ordering of the left three controlled-Z gates or the right three controlled-Z gates does not affect the scrambling unitary. This unitary operator was used in \cite{Landsman:2018jpm} to realize the scrambling dynamics of the quantum information. It can be shown that the scrambling unitary in the computing basis can be expressed in the matrix form as 
\begin{eqnarray}
    U=\frac{1}{2\sqrt{2}} \begin{pmatrix}
1 & 1 & 1 & -1 & 1 & -1 & -1 & -1\\
1 & -1 & 1 & 1 & 1 & 1 & -1 & 1\\
1 & 1 & -1 & 1 & 1 & -1 & 1 & 1\\
-1 & 1 & 1 & 1 & -1 & -1 & -1 & 1\\
1 & 1 & 1 & -1 & -1 & 1 & 1 & 1\\
-1 & 1 & -1 & -1 & 1 & 1 & -1 & 1\\
-1 & -1 & 1 & -1 & 1 & -1 & 1 & 1\\
-1 & 1 & 1 & 1 & 1 & 1 & 1 & -1\\
\end{pmatrix}\;.
\end{eqnarray}

It is easy to check that the the scrambling unitary satisfies the following gate transformation identities
\begin{eqnarray}\label{Eq:scrambling}
U^\dagger \left( \sigma_x \otimes I \otimes I \right) U= \sigma_z \otimes \sigma_y \otimes \sigma_y\;,\nonumber\\
U^\dagger \left( I \otimes \sigma_x \otimes I \right) U= \sigma_y \otimes \sigma_z \otimes \sigma_y\;,\nonumber\\
U^\dagger \left( I \otimes I \otimes \sigma_x \right) U= \sigma_y \otimes \sigma_y \otimes \sigma_z\;,\nonumber\\
U^\dagger \left( \sigma_y \otimes I \otimes I \right) U= \sigma_y \otimes \sigma_x \otimes \sigma_x\;,\nonumber\\
U^\dagger \left( I \otimes \sigma_y \otimes I \right) U= \sigma_x \otimes \sigma_y \otimes \sigma_x\;,\\
U^\dagger \left( I \otimes I \otimes \sigma_y \right) U= \sigma_x \otimes \sigma_x \otimes \sigma_y\;,\nonumber\\
U^\dagger \left( \sigma_z \otimes I \otimes I \right) U= \sigma_x \otimes \sigma_z \otimes \sigma_z\;,\nonumber\\
U^\dagger \left( I \otimes \sigma_z \otimes I \right) U= \sigma_z \otimes \sigma_x \otimes \sigma_z\;,\nonumber\\
U^\dagger \left( I \otimes I \otimes \sigma_z \right) U= \sigma_z \otimes \sigma_z \otimes \sigma_x\;,\nonumber
\end{eqnarray}
which suggests that all single-qubit operators are dispersed into three-qubit operators after the operation of the scrambling unitary. This is the indication of its scrambling property. In the following, we will use the 3-qubit Clifford scrambler given in Figure~\ref{U_scr} to simulate the two decoding strategies.

\subsection{Simulation of the probabilistic decoding strategy}

\begin{figure}
  \centering
  \includegraphics[width=12cm]{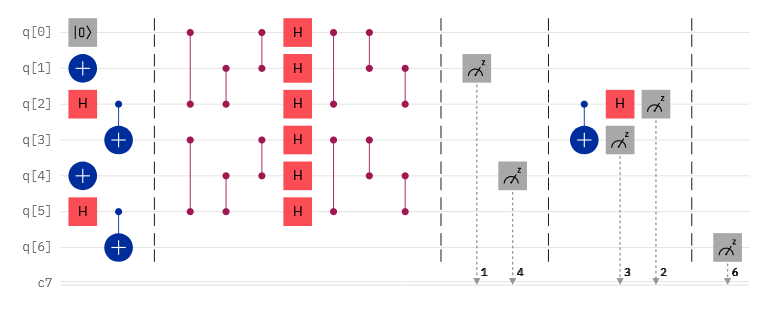}
  \caption{Quantum circuit for the probabilistic decoding strategy on the IBM quantum processor. This quantum circuit realizes the decoding strategy presented in Eq.~\eqref{Eq:psiout}. The message system $A$ is represented by the qubit $\textrm{q}[0]$. The infalling matter system $f$ is represented by the qubit $\textrm{q}[1]$. The entangled pair $\textrm{q}[2]$ and $\textrm{q}[3]$, which is prepared in the $\textrm{EPR}$ state, represents the interior mode $r$ and the exterior mode $R$ of the Hawking radiation. $\textrm{q}[4]$, $\textrm{q}[5]$ and $\textrm{q}[6]$ are the qubits owned by the decoder. We use seven classical bits to record the measurement results. The scrambling dynamics is realized by using the unitary $U$ depicted in figure~\ref{U_scr}. The vertical dashed lines represent the barriers, which are added just for the convenience of visualization. They will be deleted when executing the circuit on the quantum processor. In this diagram, the initial state of qubit $\textrm{q}[0]$ is set to be $|0\rangle$ as an example.}
  \label{quantum_circuit}
\end{figure}

The probabilistic decoding strategy is realized in the quantum circuit presented in Figure~\ref{quantum_circuit}. We use the first three qubits to represent $A$, $f$, and $r$, respectively. In addition, we use the next three qubits to represent $R$, $f$, and $F$, respectively. The last qubit represents the prepared system $F'$. The seven classical bits are used to record the measurement results. To simplify the model, we set the message system $A$ to be in a pure state without the reference system $A'$. The quantum circuit realizes the probabilistic decoding strategy in Eq.~\eqref{Eq:psiout}.

In Figure~\ref{quantum_circuit}, the vertical dashed lines represent the barriers, which are added just for the convenience of visualization. It is clear that the whole circuit is divided into five parts. In the first part, we prepare the entanglement state of the qubits of $\textrm{q}[2]$ and $\textrm{q}[3]$ and the entanglement state of the qubits of $\textrm{q}[5]$ and $\textrm{q}[6]$ and set the input states of $\textrm{q}1]$ and $\textrm{q}[4]$ to be $|1\rangle$. The qubits $\textrm{q}[1]$ and $\textrm{q}[4]$ represent the infalling matter system that collapses to the black hole. Without the loss of generality, the entanglement state is selected to be the $\textrm{EPR}$ state $|\textrm{EPR}\rangle=\frac{1}{\sqrt{2}}\left(|00\rangle+|11\rangle\right)$. The quantum state of $\textrm{q}[0]$, which is the state that we want to recover, can be set to be $|0\rangle$ or $|1\rangle$. Here, $|0\rangle$ and $|1\rangle$ represent the eigenstates of the spin operator $\sigma_z$. In figure~\ref{quantum_circuit}, the initial state of $\textrm{q}[0]$ is set to be $|0\rangle$. A $X$-gate that added on the $\textrm{q}[0]$ qubit can change this state to be $|1\rangle$. The first part prepares the initial setup of the quantum circuit.

In the second part, the first three qubits and the next three qubits are processed by the scrambling unitary operators $U$ and $U^*$, respectively. The scrambling unitary operator $U$ is given in Figure~\ref{U_scr}. Note that $U^*=U$, because the scrambling matrix is real. The second part realizes the scrambling dynamics in the black hole interior. In the third part, the qubits $\textrm{q}[1]$ and $\textrm{q}[4]$ are measured. The projection of a part of degrees of freedom in the black hole interior onto the system $P$ is realized by postselecting the measured value of $\textrm{q}[1]\textrm{q}[4]$ to be $00$ or $11$. In the forth part, we perform the $\textrm{EPR}$ projecting measurement on the qubits $\textrm{q}[2]$ and $\textrm{q}[3]$. The measured value of $\textrm{q}[2]\textrm{q}[3]$ being $00$ means that the success of the $\textrm{EPR}$ projection. In the last part, the qubit $\textrm{q}[6]$ is measured. If the input state of the first qubit is $|0\rangle$, the measured value of the last qubit is $0$ means the success of the decoding.

In this quantum circuit, the information contained in the qubit $\textrm{q}[0]$ is dispersed to the whole system by the scrambling unitary and is finally recovered in the qubit $\textrm{q}[6]$ by the projection operators. Without errors, the quantum circuit can be regarded as the realization of traversable wormhole on the quantum computer that teleports the information from the qubit $\textrm{q}[0]$ to $\textrm{q}[6]$. We implement the quantum circuit presented in figure~\ref{quantum_circuit} on the IBM quantum processor to verify the probabilistic decoding strategy. The circuit was run on IBM-nairobi processor, which is a 7-qubit quantum computer with quantum volume 32.

\begin{figure}
  \centering
  \includegraphics[width=12cm]{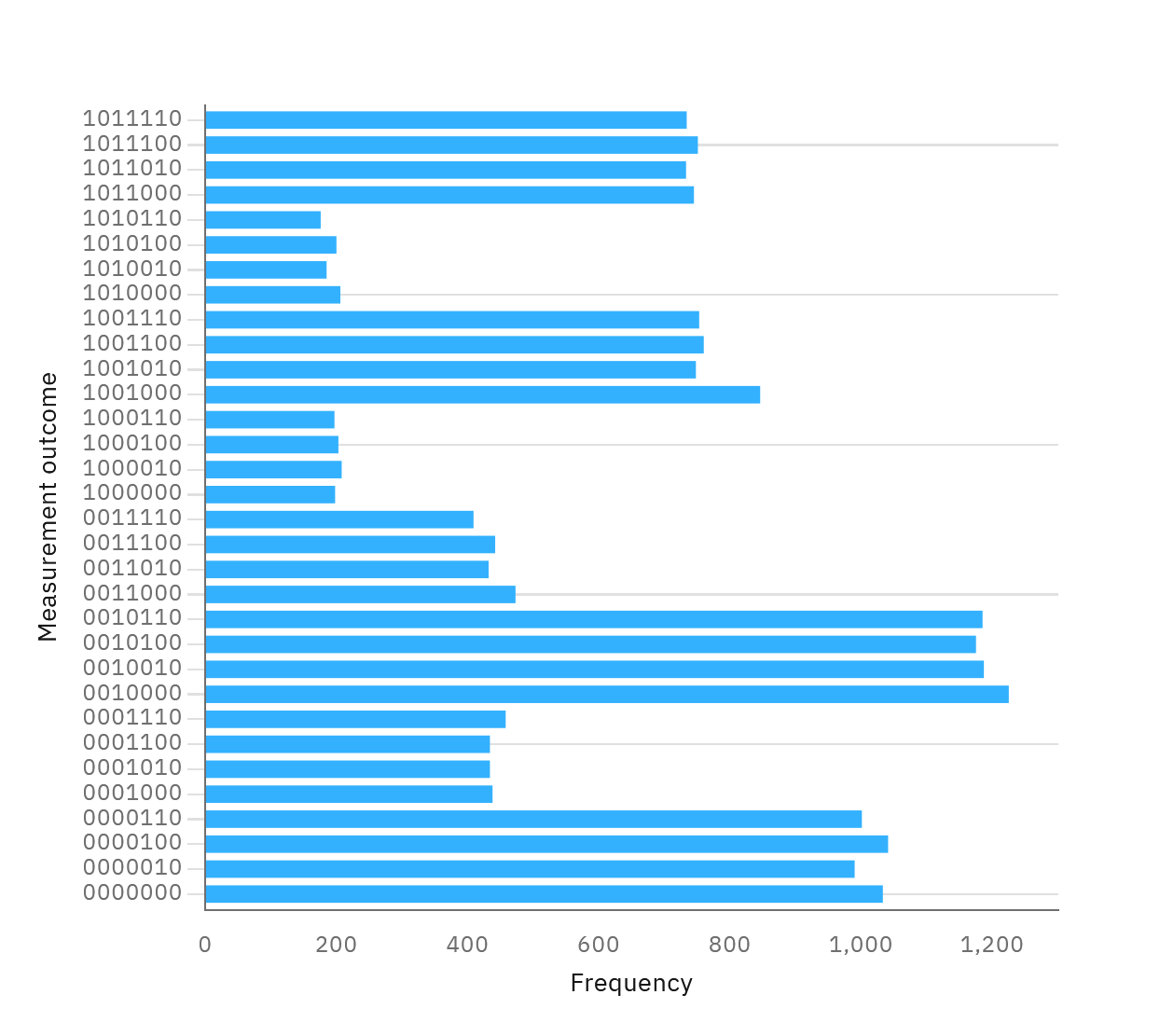}
  \caption{The original experimental results of running the quantum circuit with the initial input of $\textrm{q}[0]=|0\rangle$. The vertical axis labels the measurement outcome. For example, the numerical string $1010110$ represents the measurement outcome of the qubits $\textrm{q}[6]\textrm{q}[5]\textrm{q}[4]\textrm{q}[3]\textrm{q}[2]\textrm{q}[1]\textrm{q}[0]$. The horizontal axis labels the frequency of the specific measurement outcome. The circuit was run on the IBM-nairobi processor for $20000$ shots.} 
  \label{q0=0histogram}
\end{figure}

\begin{figure}
  \centering
  \includegraphics[width=7cm]{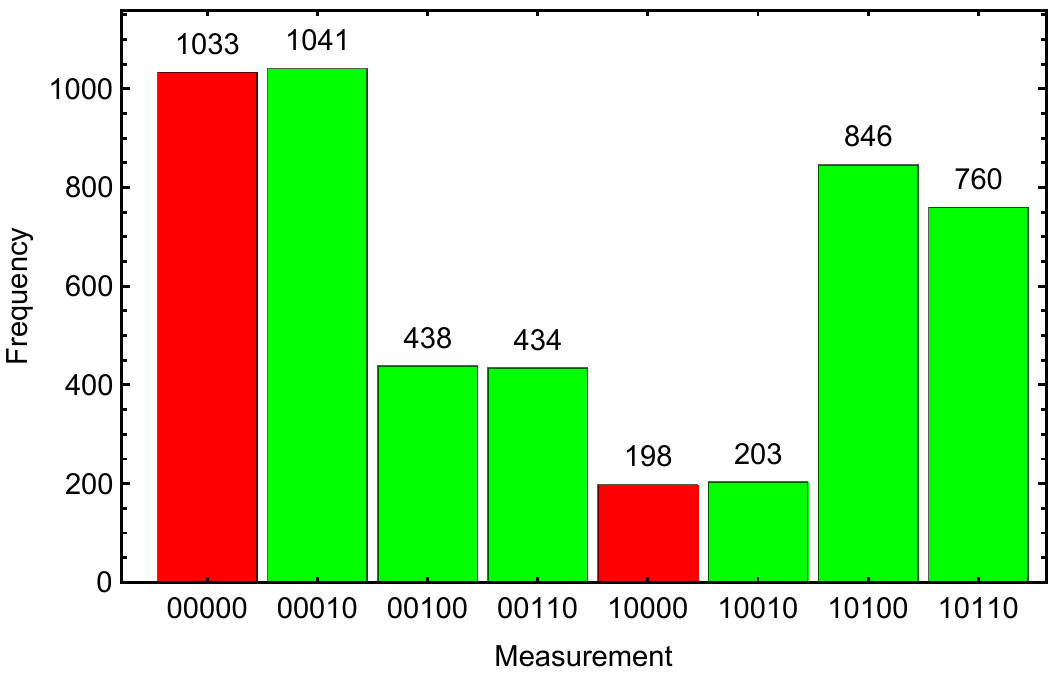}
    \includegraphics[width=7cm]{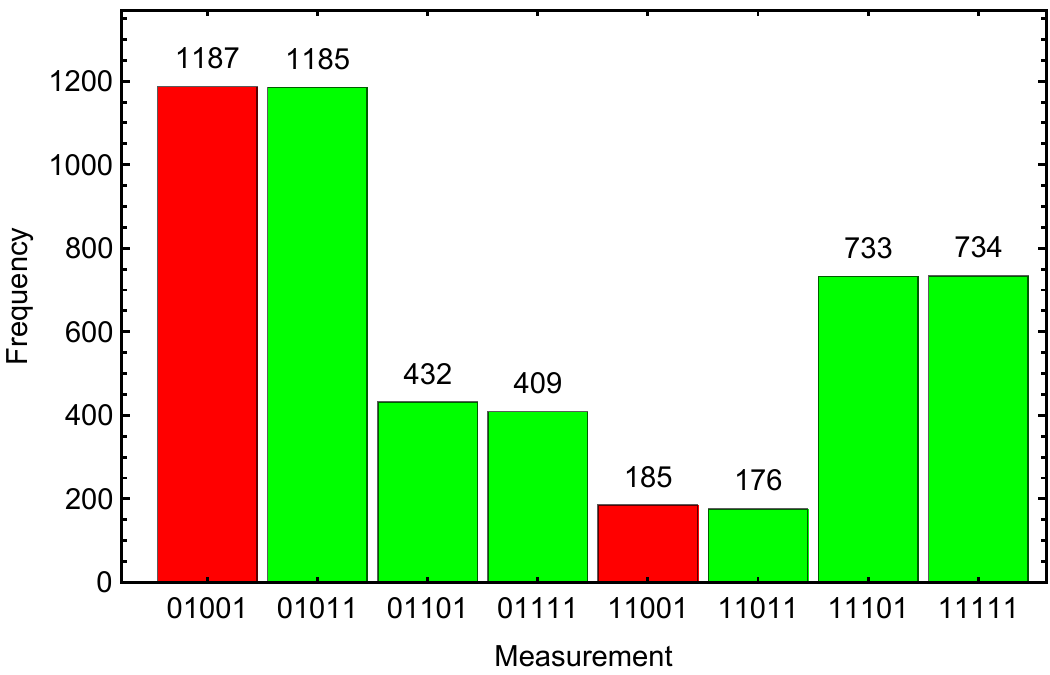}
  \caption{The statistics of the experimental results presented in Figure~\ref{q0=0histogram}. The horizontal axis represents the measurement results of $\textrm{q}[6]\textrm{q}[4]\textrm{q}[3]\textrm{q}[2]\textrm{q}[1]$ and the vertical axis represents the corresponding frequency. Left panel: The qubits $\textrm{q}[4]\textrm{q}[1]$ are projected to the state $|00\rangle$. Right panel: The qubits $\textrm{q}[4]\textrm{q}[1]$ are projected to the state $|11\rangle$. The red bars represent that the qubits $\textrm{q}[2]\textrm{q}[3]$ are successfully projected to the $\textrm{EPR}$ state $\frac{1}{\sqrt{2}}\left(|00\rangle+|11\rangle\right)$, and the green bars represent that the qubits $\textrm{q}[2]\textrm{q}[3]$ are projected to the other incorrect $\textrm{EPR}$ states.} 
  \label{q0=0barchart}
\end{figure}

In Figure~\ref{q0=0histogram}, we present the experimental results for the case that the initial input of $\textrm{q}[0]$ is $|0\rangle$. In this figure, we have presented all the measurement outcomes. The meaningful experimental results from figure~\ref{q0=0histogram} are presented in Figure~\ref{q0=0barchart}. In the left panel, the measurement outcome of the qubits $\textrm{q}[4]\textrm{q}[1]$ is selected to be $00$, which means that the qubits $\textrm{q}[4]\textrm{q}[1]$ are projected to the state $|00\rangle$. The red bars represent that the measurement outcome of the qubits $\textrm{q}[2]\textrm{q}[3]$ is $00$, which means that the qubits $\textrm{q}[2]\textrm{q}[3]$ are projected to the specific $\textrm{EPR}$ state $\frac{1}{\sqrt{2}}\left(|00\rangle+|11\rangle\right)$. The green bars represent that the qubits $\textrm{q}[2]\textrm{q}[3]$ are projected to other $\textrm{EPR}$ states, which means the failure of the $\textrm{EPR}$ projection. From the data presented in the left panel of Figure~\ref{q0=0barchart}, it can be calculated that the probability of projecting to the specific $\textrm{EPR}$ state $\frac{1}{\sqrt{2}}\left(|00\rangle+|11\rangle\right)$ is about $25\%$. In the ideal case, the projection on the $\textrm{EPR}$ state means the success of decoding radiation and recovering the information. However, due to the noise in the quantum processor, there are always errors in the circuit outcomes. In the left panel, the error is represented by the relatively low red bar where the output of the qubit $\textrm{q}[6]$ is $1$. The decoding efficiency in this case is about $84\%$. The decoding efficiency is defined as the ratio between the frequency of a successful decoding to the frequency of a successful $\textrm{EPR}$ projection. Therefore, there is the strong signal of recovering the information by executing the quantum circuit on the IBM quantum processor. In the right panel of figure~\ref{q0=0barchart}, the measurement outcome of the qubits $\textrm{q}[4]\textrm{q}[1]$ is selected to be $11$, which means that the qubits $\textrm{q}[4]\textrm{q}[1]$ are projected to the state $|11\rangle$. In this case, the $\textrm{EPR}$ projecting probability is estimated as $27\%$ and the decoding efficiency is about $87\%$. This result also implies the success of decoding the information.

\begin{figure}
  \centering
  \includegraphics[width=12cm]{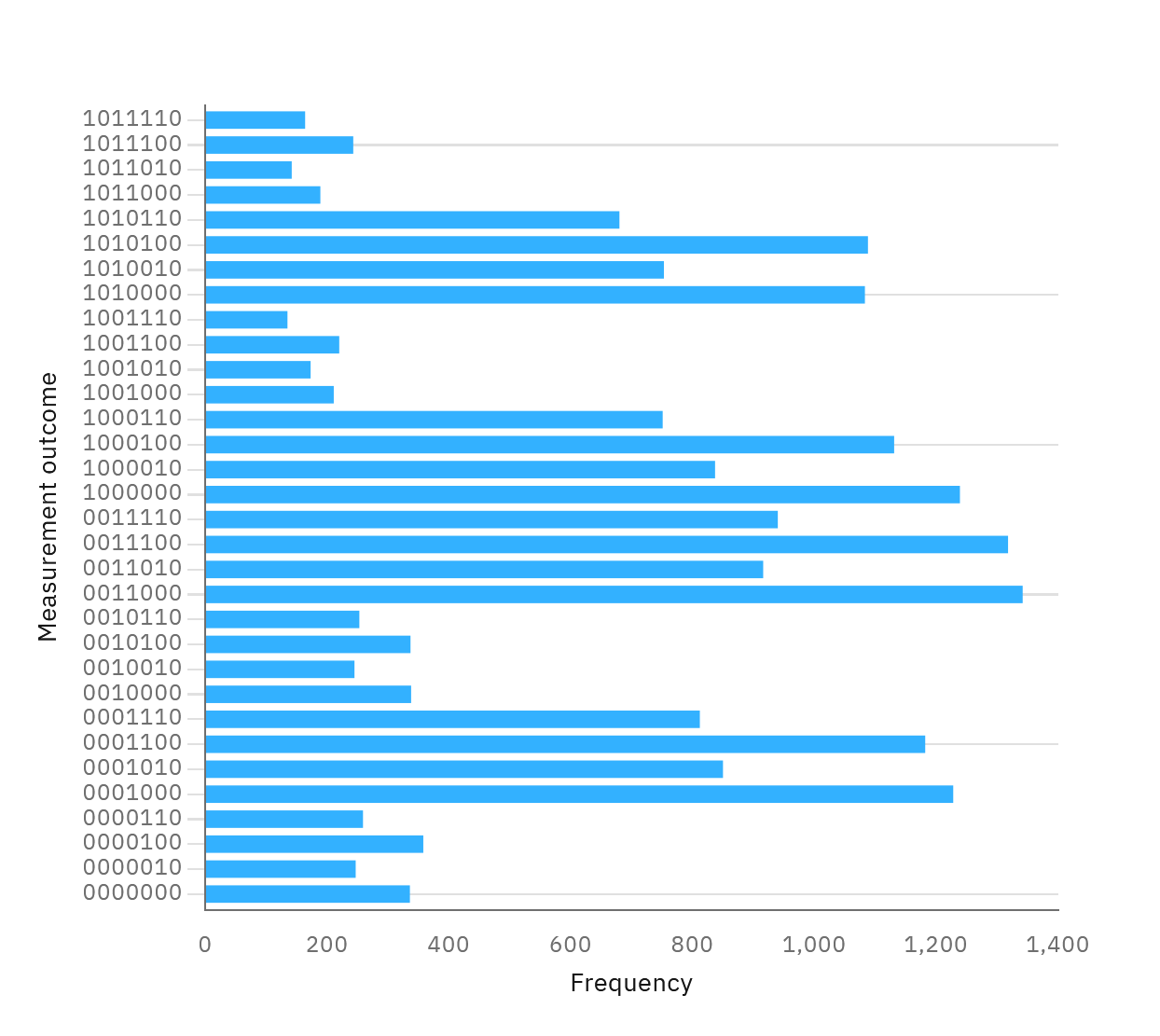}
  \caption{The original experimental results of running the quantum circuit with the initial input of $\textrm{q}[0]=|1\rangle$. The vertical axis labels the measurement outcome and the horizontal axis labels the frequency of the specific measurement outcome. The circuit was also run on the IBM-nairobi processor for $20000$ shots.} 
  \label{q0=1histogram}
\end{figure}

\begin{figure}
  \centering
  \includegraphics[width=7cm]{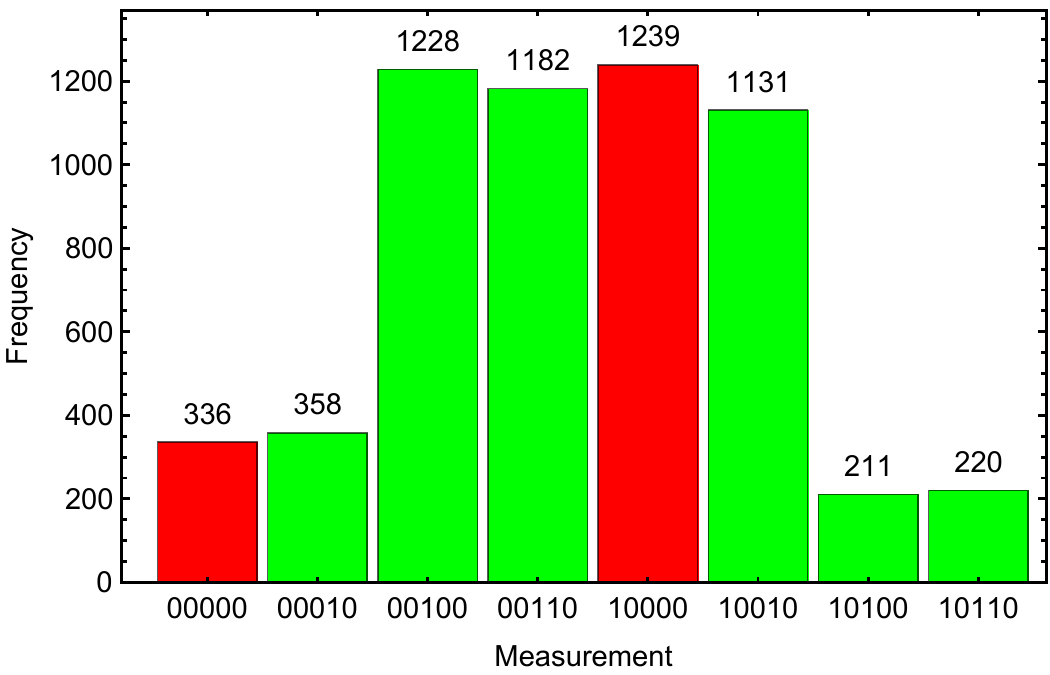}
    \includegraphics[width=7cm]{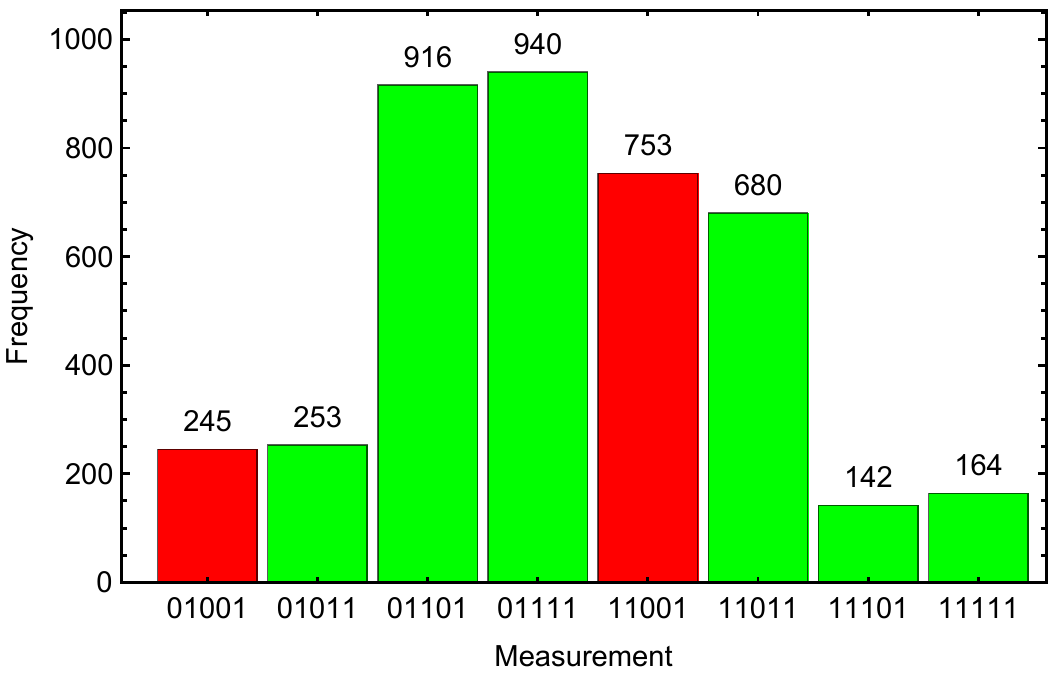}
  \caption{The statistics of the experimental results presented in Figure~\ref{q0=1histogram}. The horizontal axis represents the measurement results of $\textrm{q}[6]\textrm{q}[4]\textrm{q}[3]\textrm{q}[2]\textrm{q}[1]$ and the vertical axis represents the corresponding frequencies.} 
  \label{q0=1barchart}
\end{figure}

The original experimental results for the case that the initial input of $\textrm{q}[0]$ is $|1\rangle$ are presented in figure~\ref{q0=1histogram}. Similarly, we have plotted the meaningful experimental results in figure~\ref{q0=1barchart}. The red bars represent that the qubits $\textrm{q}[2]\textrm{q}[3]$ are projected to the correct $\textrm{EPR}$ state $\frac{1}{\sqrt{2}}\left(|00\rangle+|11\rangle\right)$ and the green bars represent that the qubits $\textrm{q}[2]\textrm{q}[3]$ are projected to other $\textrm{EPR}$ states. In the left panel, the measurement outcome of the qubits $\textrm{q}[4]\textrm{q}[1]$ is selected to be $00$, which means that the qubits $\textrm{q}[4]\textrm{q}[1]$ are projected to the state $|00\rangle$. In this case, the $\textrm{EPR}$ projecting probability is estimated as $27\%$ and the decoding efficiency is about $79\%$. In the right panel of figure~\ref{q0=1barchart}, the measurement outcome of the qubits $\textrm{q}[4]\textrm{q}[1]$ is selected to be $11$, which means that the qubits $\textrm{q}[4]\textrm{q}[1]$ are projected to the state $|11\rangle$. From these data, one can calculate the $\textrm{EPR}$ projecting probability is estimated as $24\%$ and the decoding efficiency is about $75\%$. The decoding efficiencies are smaller than those in the case that the initial input of $\textrm{q}[0]$ is $0$. This is caused by the fact that the qubit is more likely to decay to the state $|0\rangle$. These results indicate that the decoding strategy can also recover the information when the initial state of $\textrm{q}[0]$ is $|1\rangle$.

\subsection{Simulation of the Grover's search decoding strategy}

In this subsection, we discuss the experimental realization of the Grover's search decoding strategy of the Hawking radiation on the IBM-perth quantum processor. This processor is more suitable for conducting the task of the Grover's search decoding since the Grover's search decoding algorithm involves more gates operations and the decoherence time of the IBM-perth quantum processor is longer compared to the other machines available to us. In general, the efficiency of the Grover's search decoding strategy depends heavily on the quality of quantum processors and the IBM-perth processor performs better than the other IBM quantum processors available to us.

\begin{figure}[h]
  \centering
  \includegraphics[width=6cm]{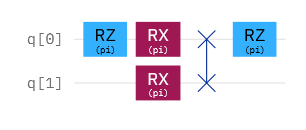}
  \caption{The circuit representation of the unitary operator $G$. }
  \label{G_operator}
\end{figure}

\begin{figure}[t]
  \centering
  \includegraphics[width=12cm]{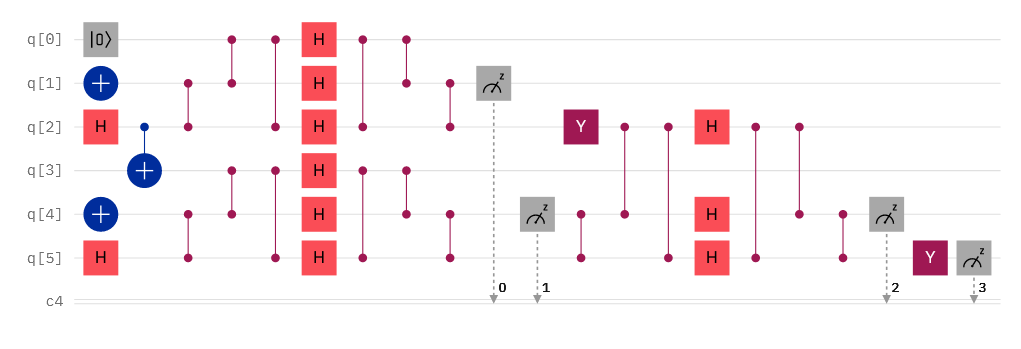}
  \caption{Quantum circuit that realizes the Grover's search decoding strategy. This simplified quantum circuit realizes the decoding protocol as shown in figure~\ref{deterministic_decoder}. In this diagram, the six qubits q$[0]$ to q$[5]$ are arranged the same way as in the case of probabilistic decoding. The message system $A$ represented by the qubit q$[0]$ is set to be $|0\rangle$ for the demonstration purpose. The measurement results are projected to the classical bits c$[0]$ to c$[3]$ which are abbreviated as c4 in the diagram. The information will be at the final state of qubit q$[5]$ before the measurement on q$[5]$.}
  \label{Grover_QC}
\end{figure}

The unitary operator $G$ in figure~\ref{deterministic_decoder} can be realized diagrammatically as shown in figure~\ref{G_operator}. It can be easily checked that the matrix representation of the operator $G$ in the computational basis coincides with that of the definition of $G$ operator in Eq.\eqref{G_def}. To realize the Grover's search decoding strategy with less operating gates, we simplify the quantum circuit in figure~\ref{deterministic_decoder} with the operator G to the following circuit shown in figure~\ref{Grover_QC}. Similar to the probabilistic decoding circuit, in this circuit the first three qubits represent $A\,,f$ and $r$ with the information to be recovered denoted by q$[0]$. The last three qubits represent $R\,,f$ and $F$, respectively. Ideally, the pre-measurement state coming out from the qubit q$[5]$ should recover the initial state of q$[0]$. In figure~\ref{Grover_QC}, the initial state is set to $|0\rangle$ as an example and it can be set to other states as well. The circuit for the operator $G$ is simplified to a single Y-gate for q$[2]$ after leaving out the swap-gate and rewiring the scrambling unitary $U^T$. Similarly, it is also simplified for q$[5]$ after we rewiring the measurement gate. For the Grover's search decoding protocol to work for the non-isometric holographic model, we need to post-select the measurement result on q$[4]$ to c$[2]$ to be the same as the initial state of q$[4]$, which is chosen to be $|1\rangle$ in this demonstration. The two measurements whose results are sent to bits c$[0]$ and c$[1]$ represent the projection onto the system $P$ and this projection can be realized by post-selecting the measurement results to be either $|00\rangle$ or $|11\rangle$. We test the decoding protocol of figure~\ref{Grover_QC} on the IBM-perth quantum processor with $20,000$ shots.

\begin{figure}[t]
  \centering
  \includegraphics[width=7cm]{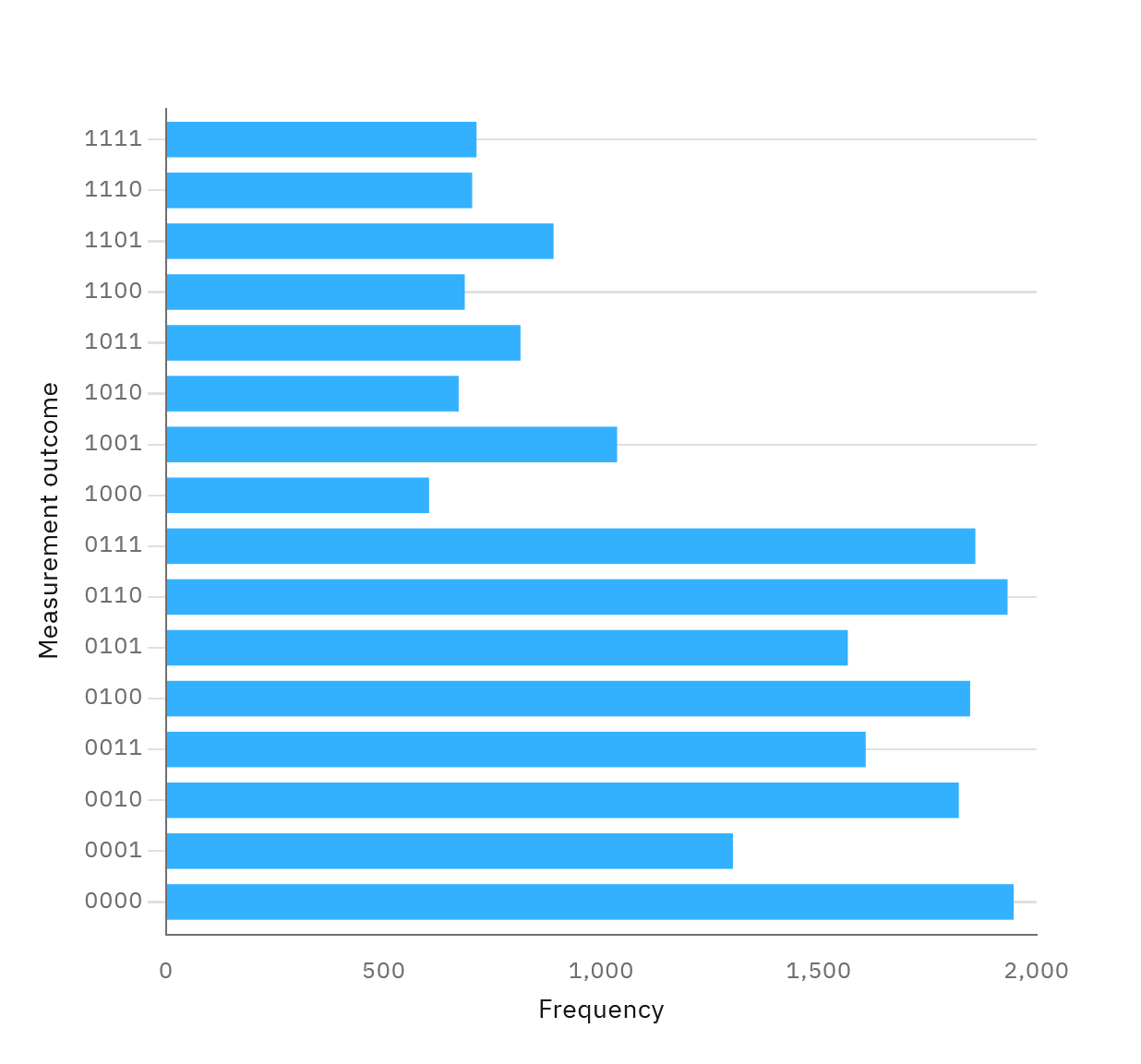}
    \includegraphics[width=7cm]{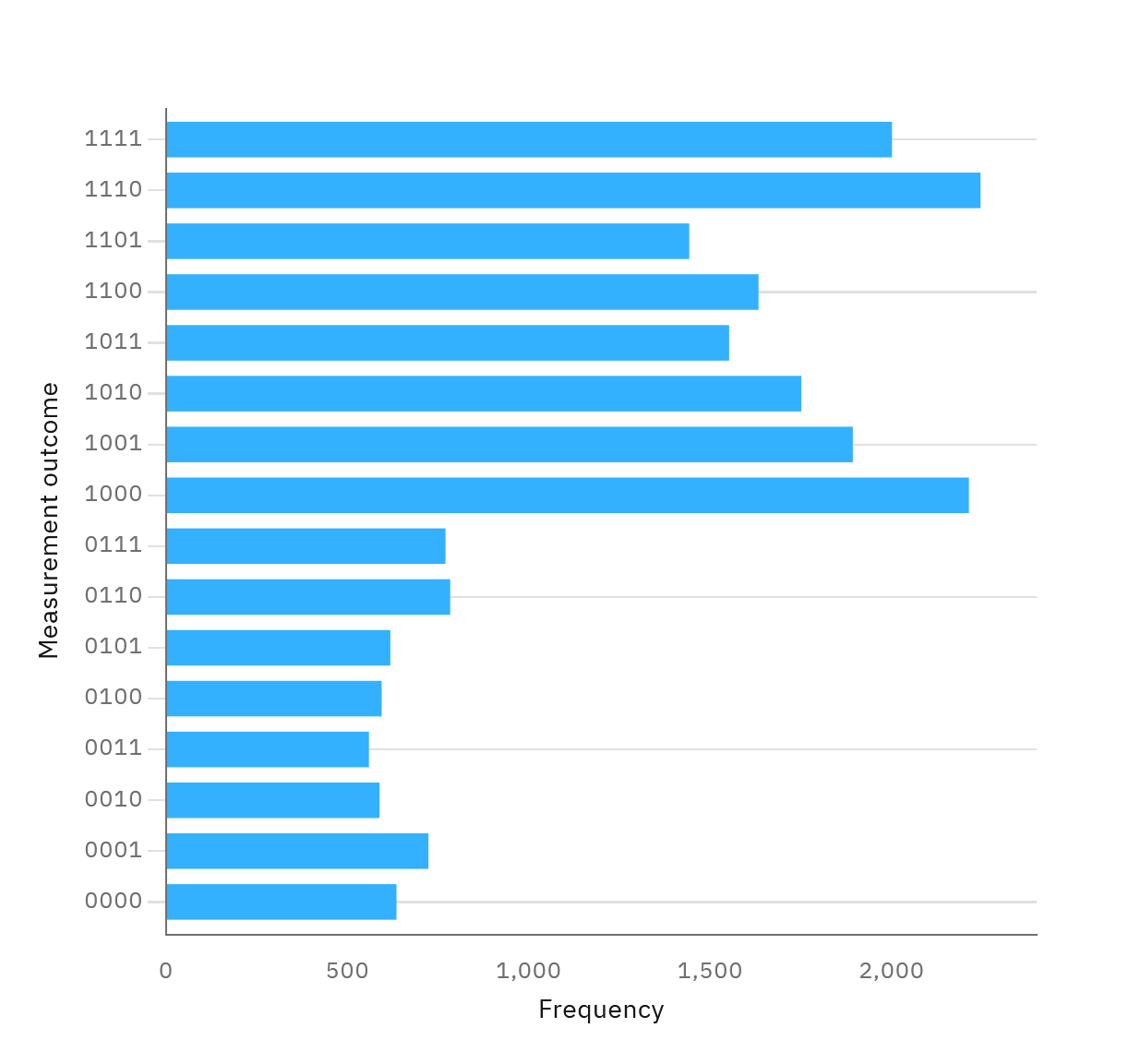}
  \caption{The original experimental data from executing the Grover's search decoding circuit on the IBM-perth quantum processor. Left: the initial input is $\textrm{q}[0]=0$. Right: the initial input is $\textrm{q}[0]=1$.}
  \label{grover_q0=0_histogram}
\end{figure}

\begin{figure}[h]
  \centering
  \includegraphics[width=7cm]{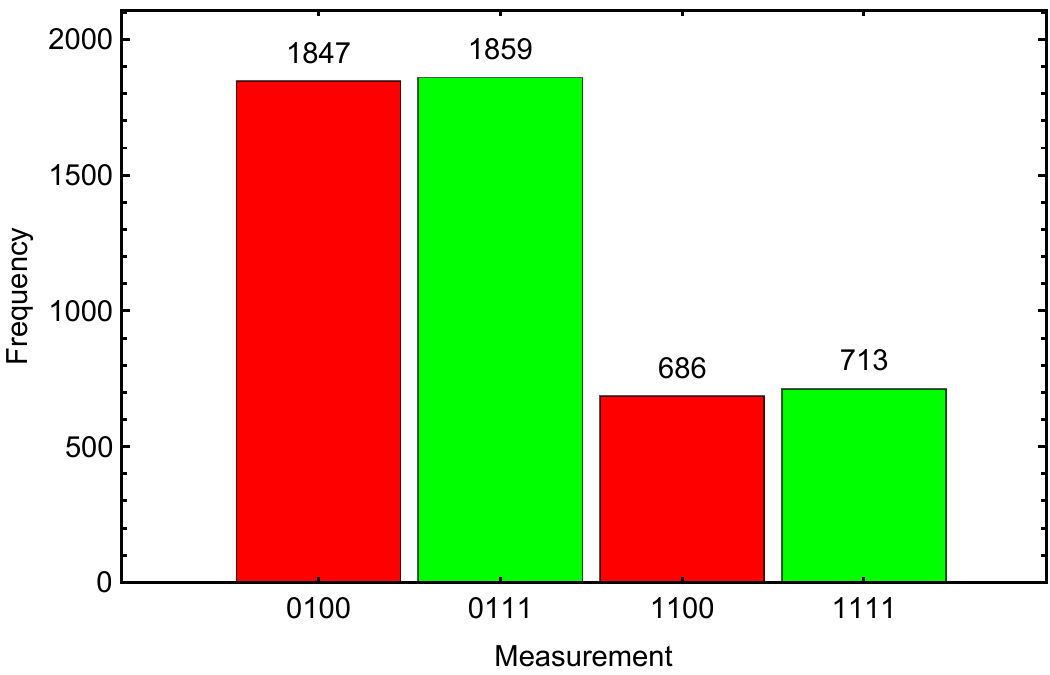}
    \includegraphics[width=7cm]{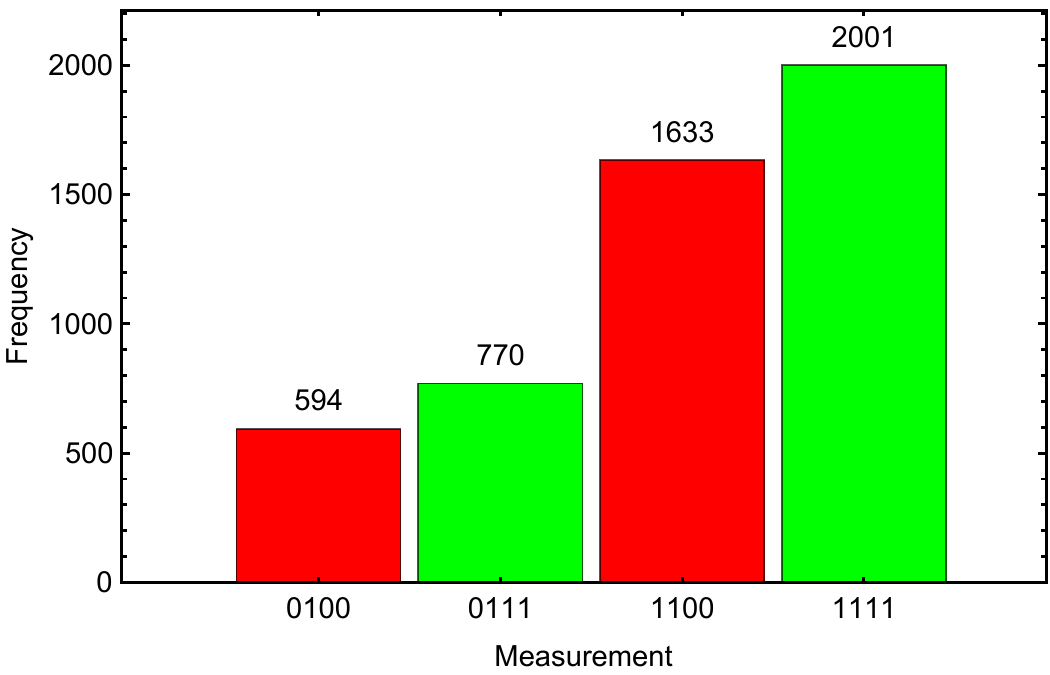}
  \caption{Left: the statistical results for the initial input of $\textrm{q}[0]=0$. Right: the statistical results for the initial input of $\textrm{q}[0]=1$.}
  \label{Grover_q0=0}
\end{figure}

In figure~\ref{grover_q0=0_histogram}, we present the original data of the test results. We remind that only the outcomes with the last three digits $100$ and $000$ in figure~\ref{grover_q0=0_histogram} are post-selected and other rows can be disregarded. The post-selected results are shown in figure~\ref{Grover_q0=0} where we include the decoding outcomes for projections onto both the $00$ states (the red bars) and the $11$ states (the green bars). For the initial input q$[0]=|0\rangle$, when the projection is onto the $00$ state the count for the successful decodings (labelled by $0100$ in figure~\ref{Grover_q0=0} (a)) is 1847. This corresponds to the successful decoding rate of approximately $73\%$. When the projection is onto the $11$ state, the decoding efficiency is about $72\%$. For the initial input q$[0]=|1\rangle$, the data are presented in figure~\ref{grover_q0=0_histogram} (b) and \ref{Grover_q0=0} (b). In this case, the decoding efficiency is about $73\%$ when the projection of system P is onto the $00$ state and the decoding efficiency is about $72\%$ when the projection is onto the $11$ state. It can be noticed that the decoding efficiencies for this decoding strategy are slightly compromised compared to the probabilistic decoding strategy due to the higher circuit complexity. However, in this strategy there is no additional probabilities of successful $\textrm{EPR}$ projections on which the probabilistic decoding efficiencies are conditioned on. Therefore, the overall decoding efficiencies of the Grover's search decoding strategy are much higher than the efficiencies in the probabilistic decoding.

In summary, we have experimentally verified the feasibility of the probabilistic and the Grover's search decoding strategies on the IBM quantum processors by using a typical scrambling unitary operator. it is shown that the initial quantum states can be recovered on the quantum circuits for the non-isometric model. Especially, for the probabilistic decoding, the quantum circuit can be viewed as a realization of quantum teleportation. On the other hand, the information recovery from decoding the Hawking radiation depends heavily on the scrambling dynamics in the black hole interior. Previous studies have relied on partial scrambling unitaries in accordance with the connecting configurations of the qubits on the IBM quantum processors to achieve the desired result \cite{Yan:2020fxu,Harris:2021mma}. The issue with such unitaries is that they do not satisfy the scrambling properties in Eq.~\eqref{Eq:scrambling} so that decoding information from such partial-scrambling unitaries is often impossible due to the information loss. In this study, we used a full scrambling unitary that satisfies Eq.~\eqref{Eq:scrambling}. Therefore, the successful simulation of the quantum circuits on IBM quantum processors indicates that high-quality three-qubit scrambling dynamics can be realized on IBM quantum processors even though the qubits are not fully connected. This requires extra effort in the simplification of the quantum circuits. This study may stimulate further investigations of black hole information problems on the IBM quantum processors and provide us more essential understandings of the nature of quantum gravity.

\section{On the interaction of infalling system with outside Hawking radiation} \label{secVII}

In our previous model, the interaction between the infalling message system $A$ and the right-moving mode $r$ of the radiation partner inside of the black hole was considered. But the interaction of the infalling message system $A$ with the outside right-going Hawking radiation $R$ is not taken into account, at least apparently. In this section, we will make a brief comment on the effect caused by this type of interaction \cite{Kim:2022pfp}.

In this case, the modified Hayden-Preskill state is graphically given by  
\begin{eqnarray}
|\Psi'\rangle_{\textrm{HP}}=  \figbox{0.2}{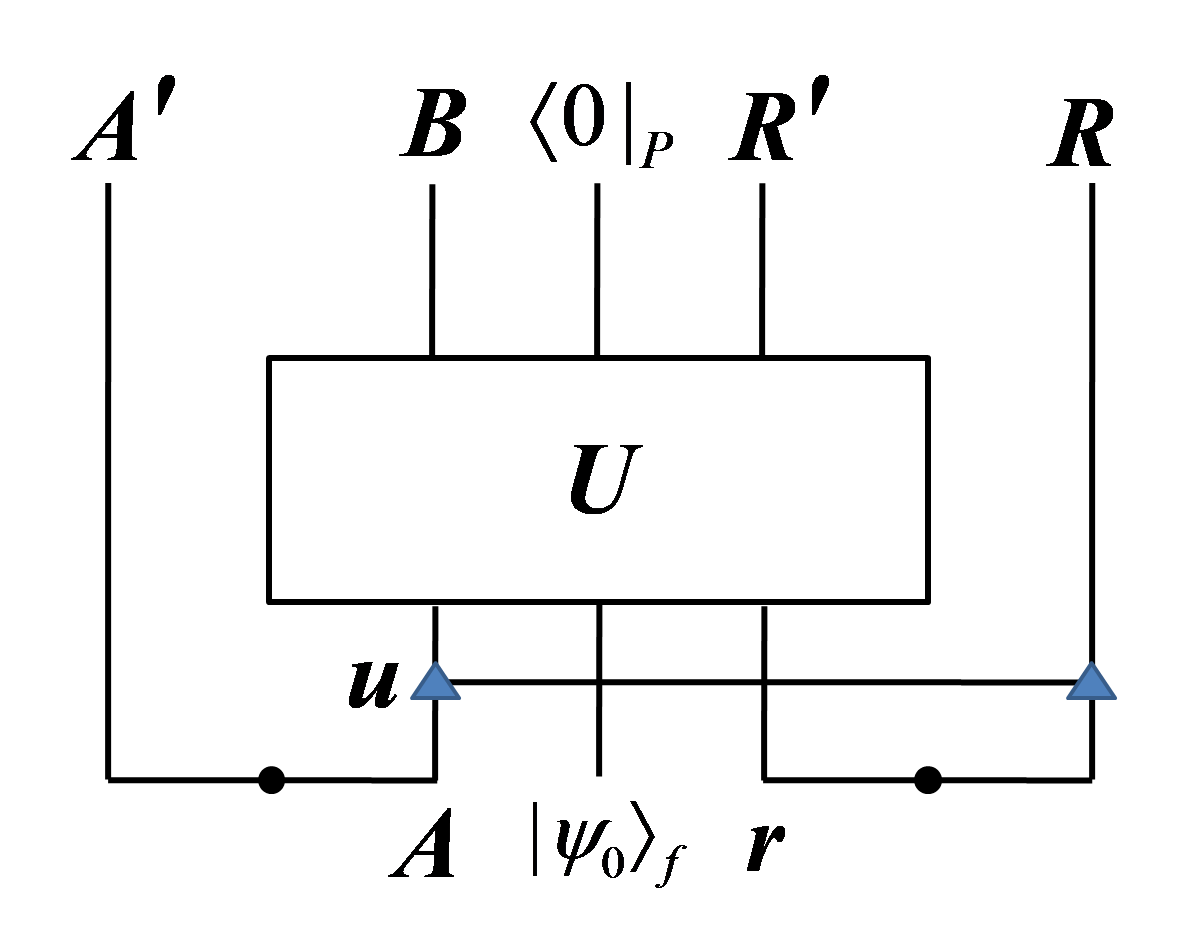}\;,
\end{eqnarray}
where $u$ represents the interaction between the message system $A$ and the Hawking radiation $R$. It is clear that the modified Hayden-Preskill state can be equivalently given by 
\begin{eqnarray}
|\Psi'\rangle_{\textrm{HP}}=   \figbox{0.2}{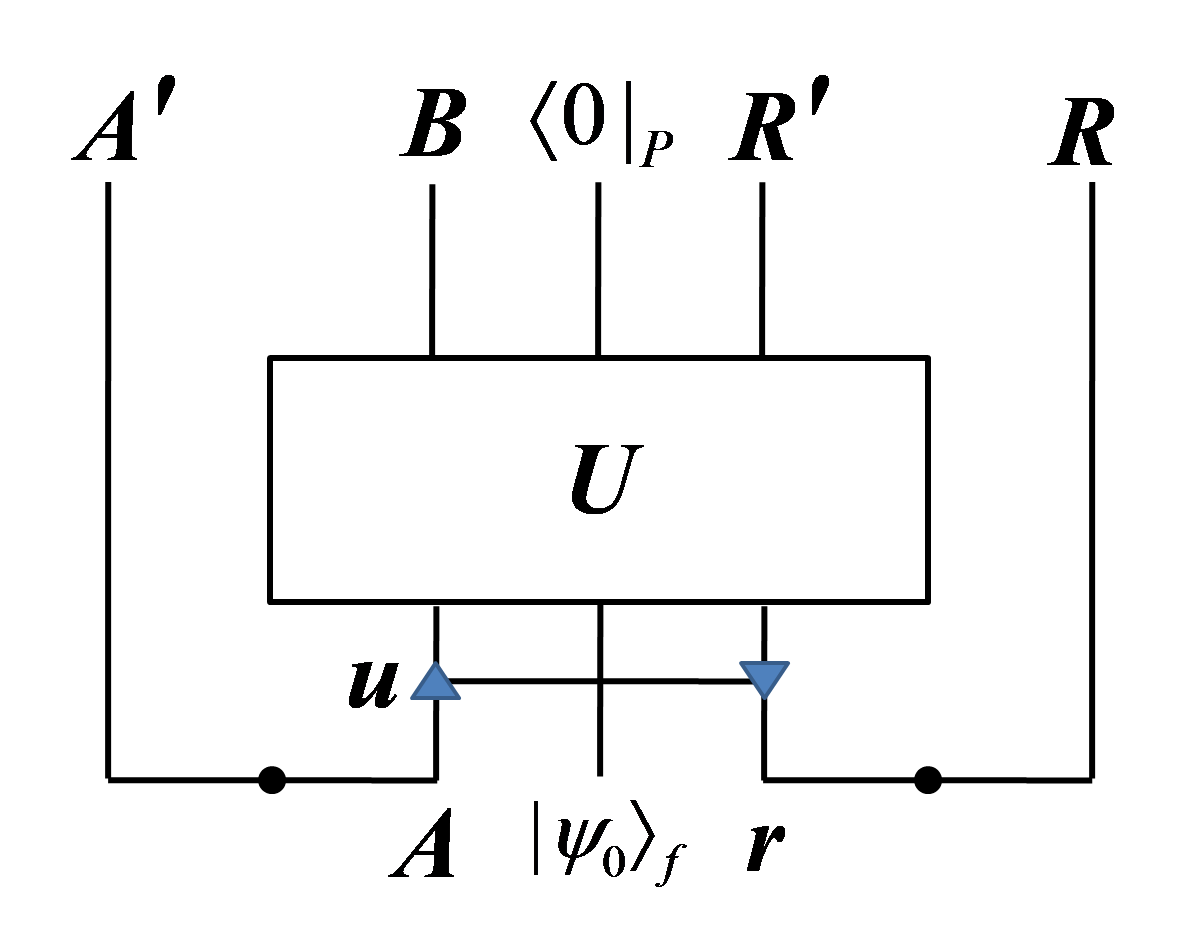}\;.
\end{eqnarray}
In this graphical representation, the interaction between the message system $A$ and the Hawking radiation $R$ is properly transferred into the interaction between the message system $A$ and the interior Hawking partner mode $r$. Therefore, we can further modify the scrambling unitary operator $U$ to be $U'=U_{(Afr)(BPR')}\cdot (v_{Ar}\otimes I_f)$ to take this type of the interaction into account. Finally, the modified Hayden-Preskill state can be represented by the original state without this type of interaction 
\begin{eqnarray}
|\Psi'\rangle_{\textrm{HP}}=   \figbox{0.2}{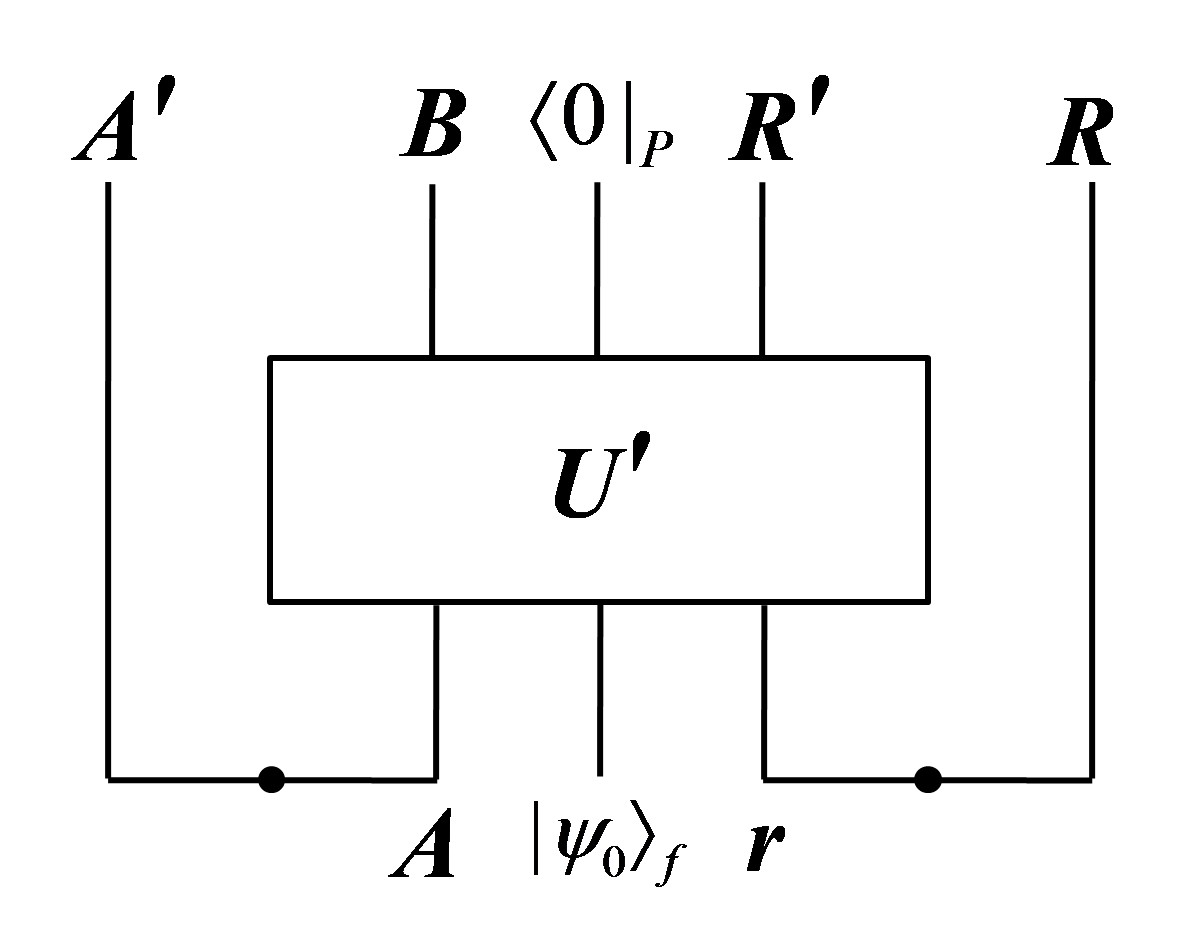}\;.
\end{eqnarray}
The discussion on the decoupling condition as well as the decoding strategy considered in Sec.\ref{secIII} and Sec.\ref{secIV} can be properly applied to study this case and the final conclusions do not change.

\section{Conclusion and discussion}\label{secVIII}

In the previous studies on Hayden-Preskill thought experiment of decoding the Hawking radiation \cite{Yoshida:2017non,Yoshida:2018vly,Bao:2020zdo,Cheng:2019yib,Li:2021mnl}, the full dynamics of the black hole evolution is assumed to be unitary and there is no question that under such assumption the information will come out from the black hole and can be decoded at late times. However, whether such decoding strategy can still be realized in the non-isometric model where the map from the fundamental to the effective descriptions involves nonunitary projections is still unclear. One may compare this model to the final-state proposal where information leaks through quantum teleportation only if the final state projection is finely-tuned. In this study, by reinterpreting and modifying the non-isometric holographic model, we investigated the possibility of decoding Hawking radiation and recovering information from a black hole when local projections inside the horizon are included. 

We firstly investigated the probabilistic decoding problem in the non-isometric model and presented the new decoupling condition under which the information can be retrieved by the outside observer. Under the assumption that the observer has a full access of the early-time and the late-time Hawking radiation as well as the full knowledge of the dynamics in the black hole interior, the Yoshida-Kitaev decoding strategy can be employed to decode the Hawking radiation and recover the information swallowed by the black hole. We showed that the new decoupling condition in this model is dependent on the size of the projected Hilbert space and is less stringent if a large effective degrees of freedom is projected out in the fundamental description. The projection operator in the map from the fundamental to effective descriptions can be realized by postselecting the measurement results in the quantum computer simulations. In the modified Hayden-Preskill protocol, the success of projection onto the $\textrm{EPR}$ state indicates the feasibility of recovering the information from the radiation. Importantly, we demonstrated that by locally projecting the annihilated states following the unitary scrambling, the channel for information transmission transitions from the EPR projections to the local projections at the Page time, reminiscent of a phase transition induced by local measurements. This offers a new perspective on the Page transition. A further improved Grover's search decoding algorithm can circumvent the issue of $\textrm{EPR}$ projections.

Furthermore, we implemented the decoding strategies through the quantum circuits of qubits and conducted tests of both decoding strategies on the IBM quantum computer using a full scrambling unitary circuit. The results from the quantum computers confirmed our analytical findings and demonstrated the feasibility of both probabilistic and Grover's search decoding strategies on the IBM quantum computer. At last, we also commented on the case where the infalling message system interacts with the outside Hawking radiation. We argued that this type of interaction causes no additional effect on the decoding or the recovery of the quantum information.

\section*{Acknowledgments}
We acknowledge the service of IBM Quantum for this work. 

\appendix

\section{Integral formulas over the Haar measure on random unitary group}

In this section, we present the general formula for evaluating the integral of the product of the $2n$ operators in the unitary group $U(d)$ with its normalized Haar measure $dU$. We consider the general $2n$-operator integral over the Haar measure, which is given by 
\begin{align}
    \int \ U_{i_1j_1} \dots U_{i_nj_n} U^\ast_{i_1'j_1'} \dots U^\ast_{i_n'j_n'} dU = \sum_{\sigma,\tau \in S_n}\delta_{i_1 i'_{\sigma(1)}}\dots \delta_{i_n i'_{\sigma(n)}}\delta_{j_1 j'_{\tau(1)}}\dots \delta_{j_n j'_{\tau(n)}}\mathrm{Wg}(\tau\sigma^{-1},n,d)\,,
    \label{Eq:Weingarten}
\end{align}
where $\sigma,\ \tau$ are the permutations of $n$ letters of the symmetric group $S_n$ and $\mathrm{Wg}(\rho,d)$ is the Weingarten function \cite{Collins2003,Collins2006}. In general, for a 2n-operator integral, there are $(n!)^2$ terms. For $d\ge n$, the Weingarten function takes the following form
\begin{gather}
    \mathrm{Wg}(\rho,n,d)=\frac{1}{(n!)^2}\sum_{\lambda \vdash n} \frac{\chi^\lambda (1)^2 \chi^\lambda(\rho)}{s_{\lambda,d}(1)}\,,
\end{gather}
where the sum is over all partitions $\lambda$ of $n$, $\chi^\lambda$ is the character of the symmetric group $S_n$, and $s$ is the Schur polynomial of $\lambda$.

Below are some explicit examples of the integrals used in this study. For the two-operator integral, the only relevant Weingarten function is 
\begin{eqnarray}
    \mathrm{Wg}([1],1,d)=\frac{1}{d}\;,
\end{eqnarray}
where $[1]$ is the identity map. Therefore, we have
\begin{gather}
    \int \  U_{i_1j_1}U_{i_1'j_1'}^\ast dU=\delta_{i_1i_1'}\delta_{j_ij_1'}\mathrm{Wg}([1],1,d)=\frac{\delta_{i_1i_1'}\delta_{j_ij_1'}}{d}\,.
\end{gather}
This is just the integral formula of Eq.\eqref{2U_integral}.

For the four-operator integral, 
\begin{align}
    \int \  U_{i_1j_1}U_{i_2j_2}U_{i_1'j_1'}^\ast U_{i_2'j_2'}^\ast dU=&(\delta_{i_1i_1'}\delta_{i_2i_2'}\delta_{j_1j_1'}\delta_{j_2j_2'}
 + \delta_{i_1i_2'}\delta_{i_2i_1'}\delta_{j_1j_2'}\delta_{j_2j_1'})\mathrm{Wg}([1,1],2,d)\nonumber \\
 &+(\delta_{i_1i_1'}\delta_{i_2i_2'}\delta_{j_1j_2'}\delta_{j_2j_1'}
 + \delta_{i_1i_2'}\delta_{i_2i_1'}\delta_{j_1j_1'}\delta_{j_2j_2'})\mathrm{Wg}([2],2,d)\,,
\end{align}
where $[2]$ denotes the permutation $(12)$ and
\begin{align}
    &\mathrm{Wg}([1,1],2,d)=\frac{1}{d^2-1}\,, 
    \nonumber\\
    &\mathrm{Wg}([2],2,d)=\frac{-1}{d(d^2-1)}\,.
\end{align}
This result is just the integral formula of Eq.\eqref{4U_integral}.

For the six-operator integral of our interest, the relevant Weingarten functions are
\begin{align}
    & \mathrm{Wg}([1,1,1],3,d) = \frac{d^2-2}{d(d^2-1)(d^2-4)}\,, \nonumber\\
    & \mathrm{Wg}([2,1],3,d) = \frac{-1}{(d^2-1)(d^2-4)}\,, \nonumber\\
    & \mathrm{Wg}([3],3,d) = \frac{2}{d(d^2-1)(d^2-4)} \,.
\end{align}
Eq.~\eqref{Eq:Weingarten} can be written out explicitly using the above functions as follows
\begin{align}
&\int \  U_{i_1j_1} U_{i_2j_2} U_{i_3j_3} U^\ast_{i_1'j_1'} U^\ast_{i_2'j_2'} U^\ast_{i_3'j_3'}\ dU \nonumber \\
=& \sum_{\sigma} \delta_{i_1 i'_{\sigma(1)}} \delta_{i_2 i'_{\sigma(2)}} \delta_{i_3 i'_{\sigma(3)}} \delta_{j_1 j'_{\sigma(1)}} \delta_{j_2 j'_{\sigma(2)}} \delta_{j_3 j'_{\sigma(3)}}\cdot \frac{(d^2-2)}{d(d^2-1)(d^2-4)}\nonumber\\
+& \sum_{\sigma} \left\{ \delta_{i_1 i'_{\sigma(1)}} \delta_{i_2 i'_{\sigma(2)}} \delta_{i_3 i'_{\sigma(3)}} \delta_{j_1 j'_{\sigma(2)}} \delta_{j_2 j'_{\sigma(1)}} \delta_{j_3 j'_{\sigma(3)}}\right.\nonumber\\
&\quad \quad +\delta_{i_1 i'_{\sigma(1)}} \delta_{i_2 i'_{\sigma(2)}} \delta_{i_3 i'_{\sigma(3)}} \delta_{j_1 j'_{\sigma(3)}} \delta_{j_2 j'_{\sigma(2)}} \delta_{j_3 j'_{\sigma(1)}}\nonumber\\
&\quad \quad \left.+\delta_{i_1 i'_{\sigma(1)}} \delta_{i_2 i'_{\sigma(2)}} \delta_{i_3 i'_{\sigma(3)}} \delta_{j_1 j'_{\sigma(1)}} \delta_{j_2 j'_{\sigma(3)}} \delta_{j_3 j'_{\sigma(2)}}\right\} \cdot \frac{(-1)}{(d^2-1)(d^2-4)}\nonumber\\
+& \sum_{\sigma} \left\{\delta_{i_1 i'_{\sigma(1)}} \delta_{i_2 i'_{\sigma(2)}} \delta_{i_3 i'_{\sigma(3)}} \delta_{j_1 j'_{\sigma(2)}} \delta_{j_2 j'_{\sigma(3)}} \delta_{j_{3} j'_{\sigma(1)}}\right. \nonumber\\
&\quad \quad +\left. \delta_{i_1 i'_{\sigma(1)}} \delta_{i_2 i'_{\sigma(2)}} \delta_{i_3 i'_{\sigma(3)}} \delta_{j_1 j'_{\sigma(3)}} \delta_{j_2 j'_{\sigma(1)}} \delta_{j_{3} j'_{\sigma(2)}} \right\} \cdot \frac{2}{d(d^2-1)(d^2-4)}\,.
\end{align}

For $d\gg 1$, the dominant contribution out of the 36 terms comes from the ones associated with $\mathrm{Wg}([1,1,1],3,d)$ which corresponds to identical permutations $\sigma=\tau$. Therefore, the leading order of the integral is given by
\begin{gather}
\int \  U_{i_1j_1} U_{i_2j_2} U_{i_3j_3} U^\ast_{i_1'j_1'} U^\ast_{i_2'j_2'} U^\ast_{i_3'j_3'}\ dU
\simeq \sum_{\sigma} \delta_{i_1 i'_{\sigma(1)}} \delta_{i_2 i'_{\sigma(2)}} \delta_{i_3 i'_{\sigma(3)}} \delta_{j_1 j'_{\sigma(1)}} \delta_{j_2 j'_{\sigma(2)}} \delta_{j_3 j'_{\sigma(3)}}\cdot \frac{1}{d^3}\,,
\label{Eq:6u}
\end{gather}
where $\sigma$ is the permutation on three letters and there are six choices of $\sigma$'s in this summation. 

For a general $2n$-operator integral with $n\ge 4$, direct computations of the Weingarten functions can be extremely involved. In this case, we can refer to the asymptotic behaviors of Weingarten functions in the limit $d\gg 1$,
\begin{gather}
    \mathrm{Wg}(\rho,n,d)\simeq d^{-n-|\rho|}\Pi_i (-1)^{|C_i|-1} c_{|C_i|-1}\,,
\end{gather}
where $\rho$ is a product of cycles of lengths $C_i$, $c_j=(2j)!/\left(j!(j+1)!\right)$ is the Catalan number, and $|\rho|$ is the smallest number of transpositions of the products. The leading order in $1/d$ of the Weingarten functions is obtained when $\rho=[1^n]$, which indicates that $|\rho|=0$ and the Catalan number $c_1=1$. Therefore, we have the asymptotic approximation
\begin{gather}
    \mathrm{Wg}([1^n],n,d)\simeq d^{-n}\,,
\end{gather}
and the $2n$-operator integral over the Haar measure can be approximated by
\begin{gather}
\int \  U_{i_1j_1} \dots U_{i_nj_n} U^\ast_{i_1'j_1'}\dots U^\ast_{i_n'j_n'}\ dU
\simeq \sum_{\sigma} \delta_{i_1 i'_{\sigma(1)}} \dots \delta_{i_n i'_{\sigma(n)}} \delta_{j_1 j'_{\sigma(1)}} \dots \delta_{j_n j'_{\sigma(n)}}\cdot \frac{1}{d^n}\,,
\end{gather}
where $\sigma$ is the permutation on $n$ letters. Given a particular diagram, usually only one term contributes dominantly in this study. The above formulae are exploited to evaluate the integrals in Appendix B and C below. 

\section{A quick check of the last two relations in Eq.(\ref{G_eqs})}

In this appendix, we show a not very rigorous demonstration of the last two relations in Eq.(\ref{G_eqs}) by using the integral formulas discussed in Appendix A. We have claimed that the two relations are satisfied only in the ideal case. This is to say that the unitary operator $U$ should be a typical one. 

A not-so rigorous check can be made by showing the following relations hold
\begin{eqnarray}
   && \int dU ~_{in}\langle \Psi| \left(I_{A'BR'}\otimes \tilde{\Pi}_{R''B'F'} \right) |\Psi\rangle_{in}=1\;,\nonumber\\
   && \int dU ~_{in}\langle \Psi| \left(I_{A'BR'}\otimes \tilde{\Pi}_{R''B'F'} \right) |\Psi\rangle_{out}=\sqrt{\frac{|P|}{|f|}}\frac{1}{|A|}\;.
   \label{Eq:grovercondition}
\end{eqnarray}
The two integrals involve the six-order unitary integral formula that is given in Eq.\eqref{Eq:6u}. In the ideal case, only one term contributes the final result. In the following, we will calculate the integral bu using the graphical representation.

Firstly, the first integral of Eq.~\eqref{Eq:grovercondition} can be graphically represented and approximately evaluated as
\begin{eqnarray}
  &&\int dU ~_{in}\langle \Psi| \left(I_{A'BR'}\otimes \tilde{\Pi}_{R''B'F'} \right) |\Psi\rangle_{in}\nonumber\\
  &=&   |P|^3 C^2\int dU~~~\left(\figbox{0.2}{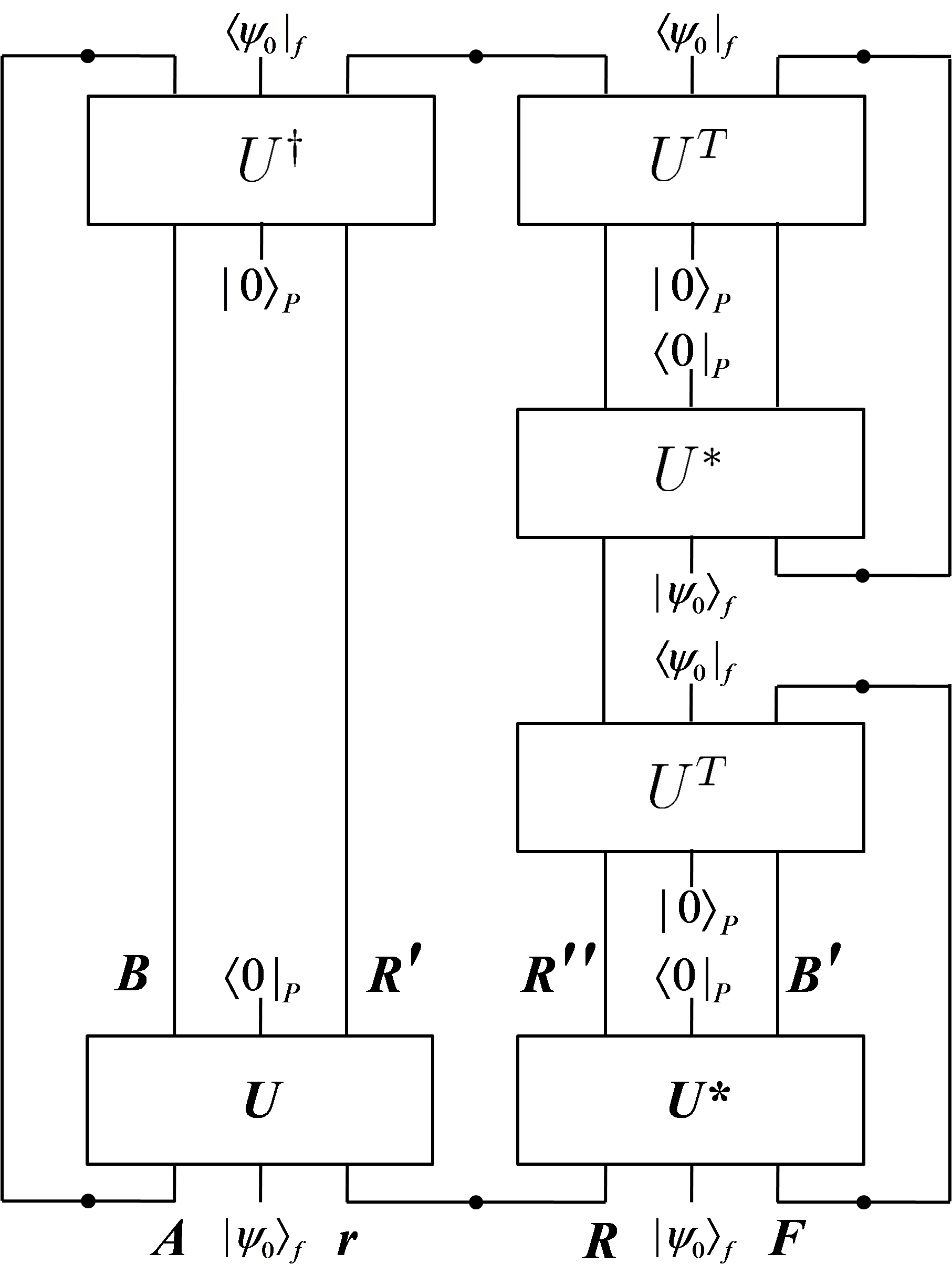}\right)\nonumber\\
  &\simeq& \frac{|P|^3 C^2}{d^3} \max (|B|^3|R'|^3,|r|^2|R'||B|/|A|^2) \nonumber\\
  &\simeq& \min\left(1,\frac{|f|}{|P|}|A|^2\right)\cdot \max\left(1,\frac{|P|^2}{|f|^2|A|^4} \right)\nonumber\\
  &=& \max\left(1,\frac{|P|}{|f||A|^2} \right) \nonumber\\
  &=&1\;,
\end{eqnarray}
where we have considered the ideal case of $d\gg 1$, $|P| < |f||A|^2$. In this calculation, only one particular choice of $\sigma$ in Eq.~\eqref{Eq:6u} returns the dominant contribution to $\int dU ~_{in}\langle \Psi| \left(I_{A'BR'}\otimes \tilde{\Pi}_{R''B'F'} \right) |\Psi\rangle_{in}$ of Eq.~\eqref{Eq:grovercondition}. In addition, we have omitted the lines that represent the normalization conditions $\langle \psi_0|\psi_0\rangle_f=1$ and $\langle 0|0\rangle_P=1$ in the graphical representation. Note that in deriving the above result, we also used the fact that $|A|=|F|$ and $|B|=|B'|$.

The second integral can be graphically represented and evaluated in the ideal case as 
\begin{eqnarray}
  &&\int dU ~_{in}\langle \Psi| \left(I_{A'BR'}\otimes \tilde{\Pi}_{R''B'F'} \right) |\Psi\rangle_{out}\nonumber\\
  &=& \frac{|P|^3}{\sqrt{P_\textrm{EPR}}} \int dU~~~\left(\figbox{0.2}{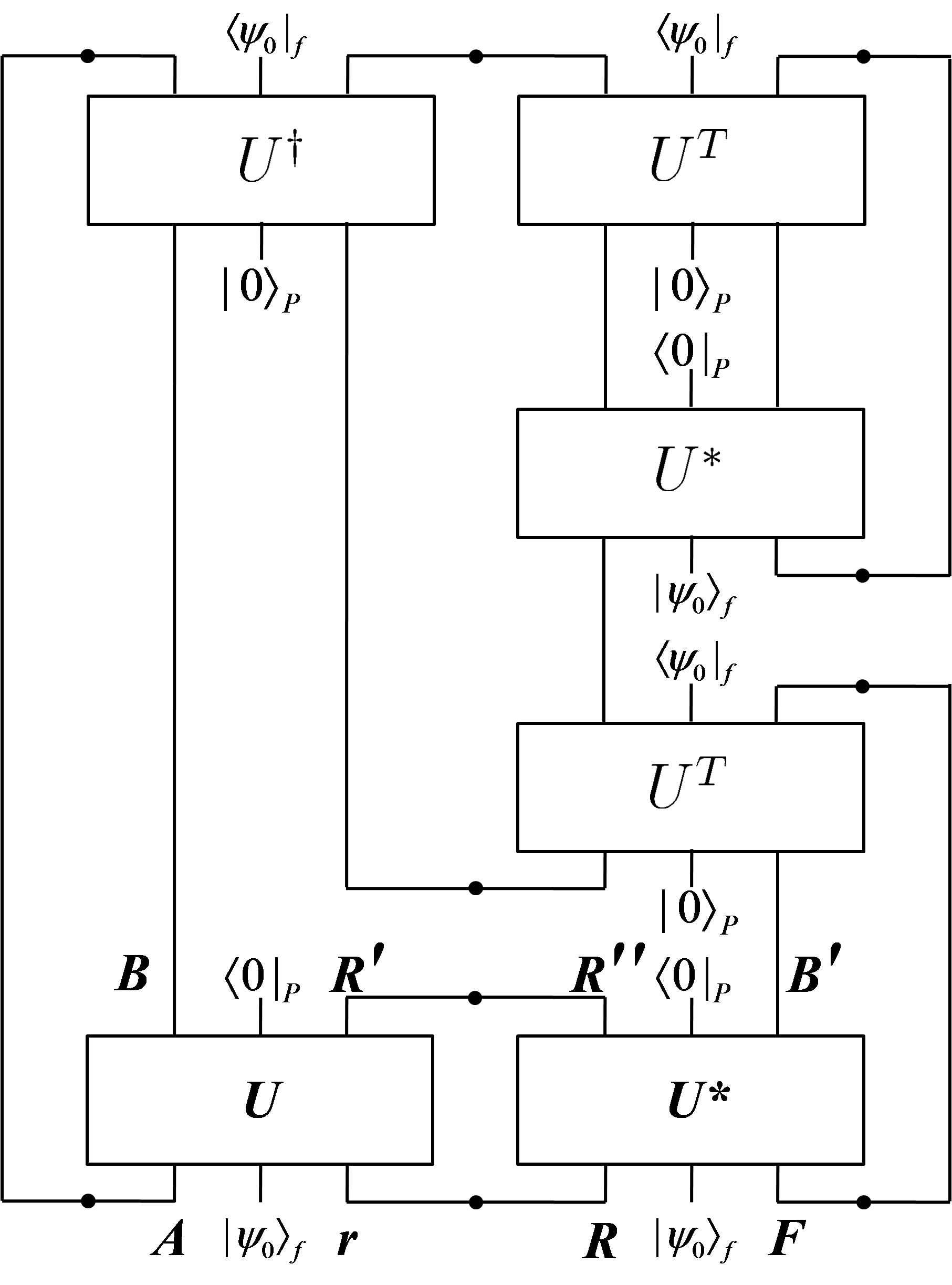}\right)\nonumber\\
  &\simeq& \frac{|P|^3}{\sqrt{P_\textrm{EPR}}}\frac{1}{d^3} ~~~\figbox{0.2}{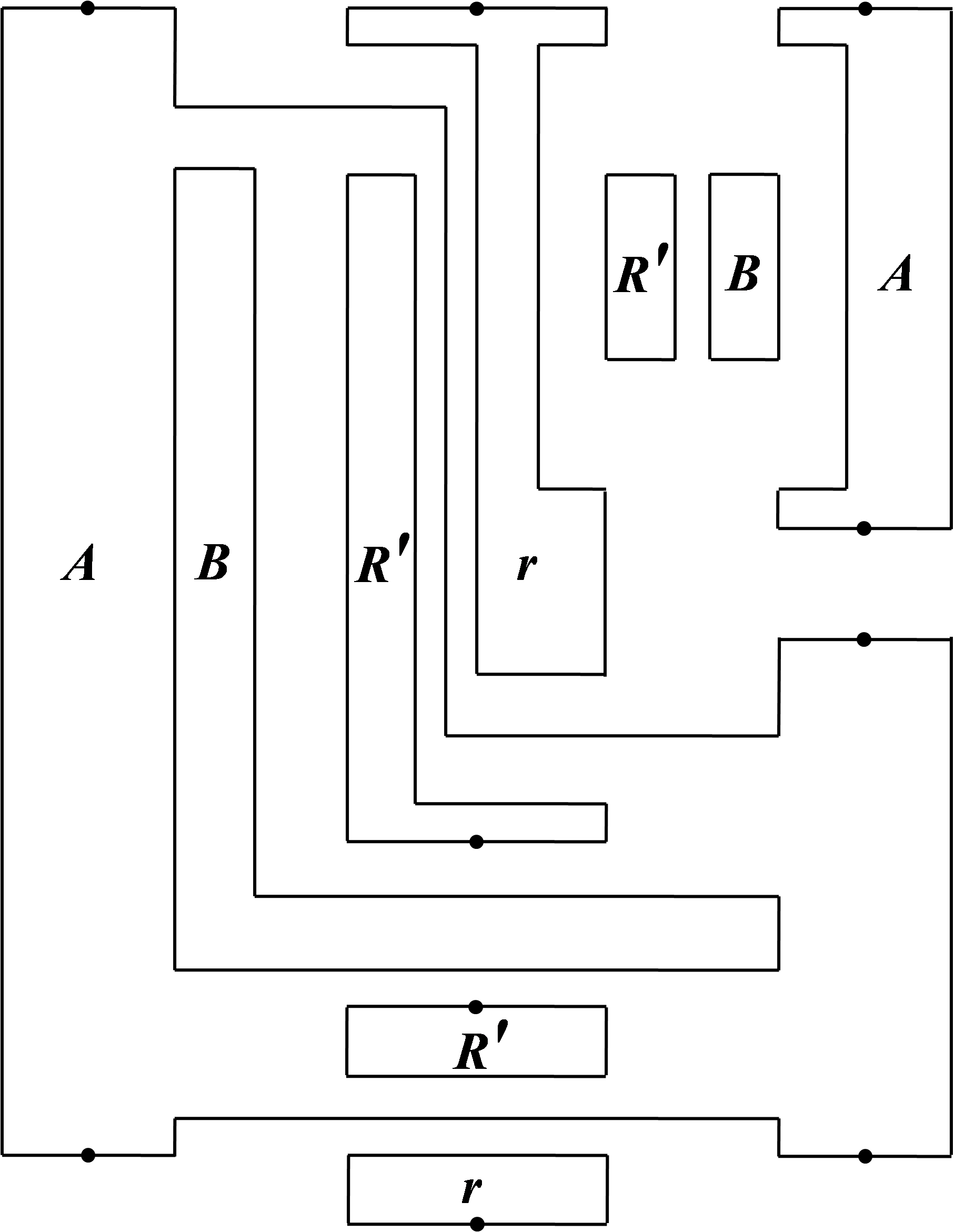}\nonumber\\
  &\simeq& \frac{|P|^3}{\sqrt{P_\textrm{EPR}}} \frac{|B|^2|r||R'|^2}{d^3 |A|} \nonumber\\
  &\simeq&\sqrt{\frac{|P|}{|f|}}\frac{1}{|A|}\;.
\end{eqnarray}

As we remarked, the above calculations serve as a simple demonstration that the last two relations in Eq.(\ref{G_eqs}) hold. However, a rigorous proof of them is much more tedious and should be carried out by showing that the operator distance of the density matrices for the states on the l.h.s and the r.h.s of the equations averaged over the random unitary group is small. This procedure involves the operator integrals of higher $n$ over the random unitary group, which will be presented in the Appendix C.

\section{Proof of the last two relations in Eq.(\ref{G_eqs})}

The last two relations in Eq.(\ref{G_eqs}) can be rewritten as  
\begin{eqnarray}\label{eq_pf}
  \left(I_{A'BR'}\otimes \tilde{\Pi}_{R''B'F'} \right) |\Psi\rangle_{in}&=&|\Psi\rangle_{in}\;,\nonumber\\
 \frac{1}{\sqrt{P_{\textrm{EPR}}}}\left(I_{A'BR'}\otimes \tilde{\Pi}_{R''B'F'} \right)|\Psi\rangle_{out}&=&|\Psi\rangle_{in}\;.
\end{eqnarray}
We now prove that the above relations are satisfied in the ideal case. Define the following three density matrices as 
\begin{eqnarray}
\rho_{in}&=&|\Psi\rangle_{in} ~_{in}\langle\Psi|\;,\nonumber\\
\rho_1&=&\tilde{\Pi}|\Psi\rangle_{in}~_{in}\langle\Psi|\tilde{\Pi}\;,\nonumber\\
\rho_2&=&\frac{1}{P_{\textrm{EPR}}}\tilde{\Pi}|\Psi\rangle_{out}~_{out}\langle\Psi|\tilde{\Pi}\;,
\end{eqnarray}
where we have omitted the identity operators and the subscripts for simplicity. The last two relations in Eq.(\ref{G_eqs}) can be proved by estimating the operator distances between $\rho_1$ and $\rho_{in}$ and between $\rho_1$ and $\rho_{in}$ are small enough in the ideal case. To our aim, only the dominant contribution from the unitary integral is considered.

Firstly, we evaluate the operator distance between $\rho_1$ and $\rho_{in}$. Let us consider 
\begin{eqnarray}
\left(\int dU ||\rho_1-\rho_{in}||_{1}\right)^2&\leq& 
\int dU ||\rho_1-\rho_{in}||_{1}^2\nonumber\\
&\leq& 2N \int dU ||\rho_1-\rho_{in}||_{2}^2\nonumber\\
&=& 2N \int dU \textrm{Tr}\left(\rho_1^2+\rho_{in}^2-2\rho_1\rho_{in} \right)\nonumber\\
&=& 2N \int dU \left( ~_{in}\langle\Psi|\tilde{\Pi}^2|\Psi\rangle_{in}^2 
+ ~_{in}\langle\Psi|\Psi\rangle_{in}^2 -2 ~_{in}\langle\Psi|\tilde{\Pi}|\Psi\rangle_{in}^2\right)\;,
\end{eqnarray}
where $N$ is the normalization factor. For a typical scrambling unitary which we consider or the normalized pure-state density matrices $\rho_1$ and $\rho_{in}$, the normalization factor is one. The factor of two in the second line comes from the operator inequality 
\begin{align}
    || X||_p\le \left(\mathrm{Rank}(X)\right)^{\frac{1}{p}-\frac{1}{q}} ||X||_q\,.
\end{align}

For simplicity,  we consider the ideal case and $|P| \ll |f||A|^2$. Then only the leading order's contribution to the integral on the right hand side of the inequality needs to be evaluated. Using the graphical representation, the first term can be calculated as 
\begin{eqnarray}
&&\int dU~_{in}\langle\Psi|\tilde{\Pi}^2|\Psi\rangle_{in}^2\nonumber\\
 &=&|P|^8 \int dU~~~\left(\figbox{0.2}{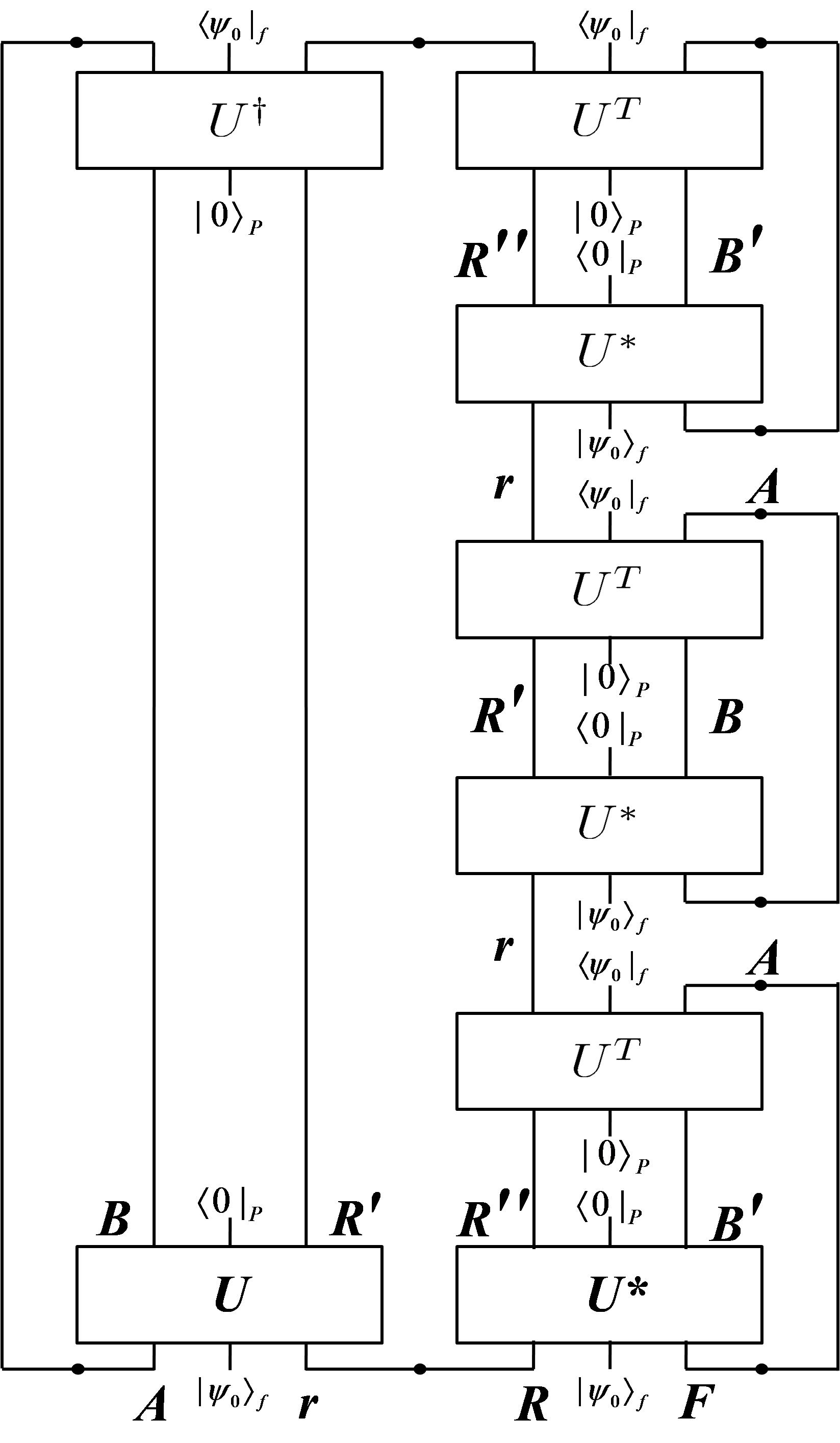}\right)^2  \nonumber\\
 &\simeq&\frac{|P|^8}{d^8} \int dU~~~\left(\figbox{0.2}{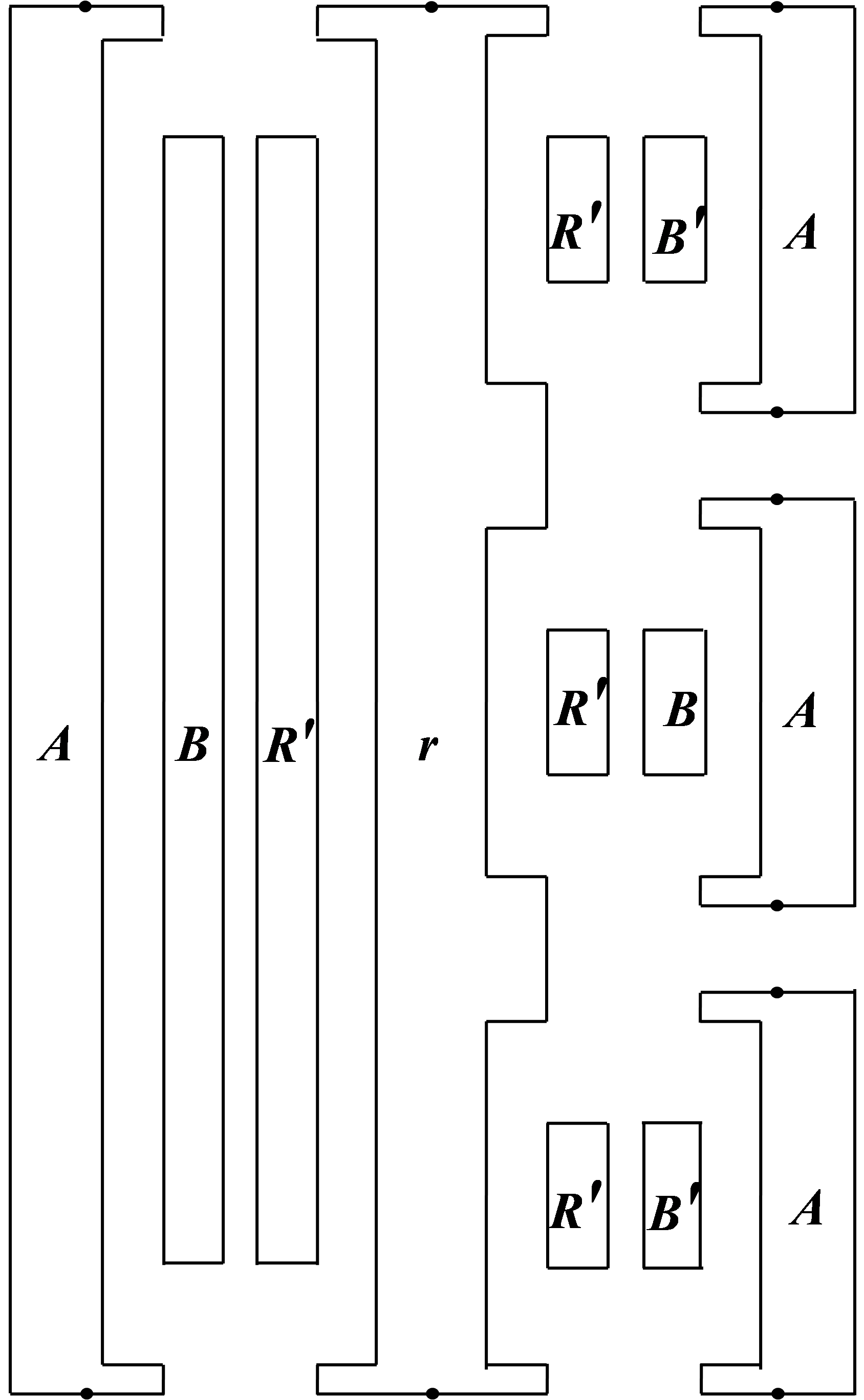}\right)^2  \nonumber\\
 &=&\frac{|P|^8 |B|^8 |R'|^8}{d^8}\nonumber\\
 &=&1\;.
\end{eqnarray}
Note that we have omitted the lines that represent the normalization conditions $\langle \psi_0|\psi_0\rangle_f=1$ and $\langle 0|0\rangle_P=1$ in this graphical representation. 

The second term can be calculated as 
\begin{eqnarray}
&&\int dU~_{in}\langle\Psi|\Psi\rangle_{in}^2\nonumber\\
 &=&|P|^4 \int dU~~~\left(\figbox{0.2}{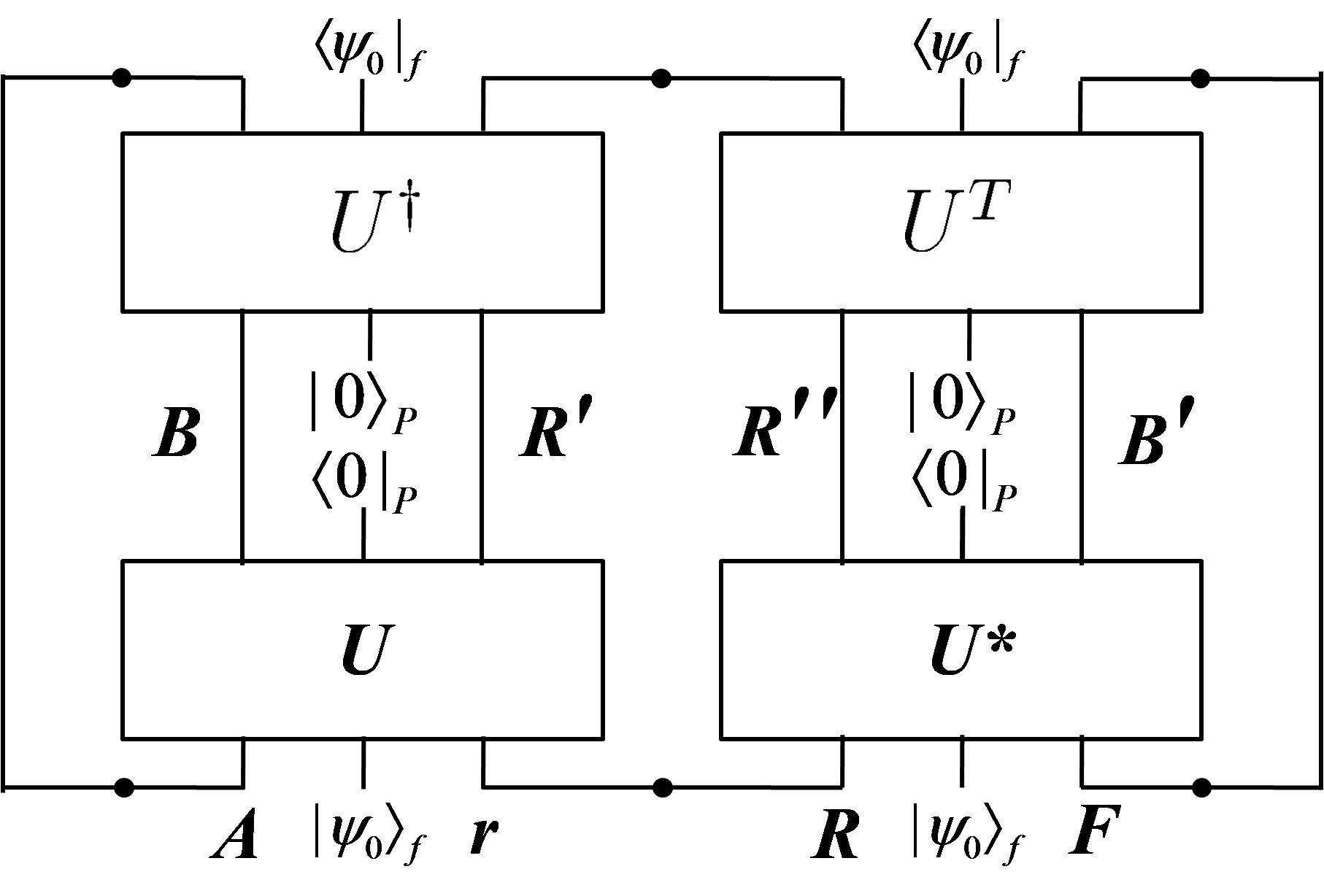}\right)^2  \nonumber\\
 &\simeq&\frac{|P|^4}{d^4} \int dU~~~\left(\figbox{0.2}{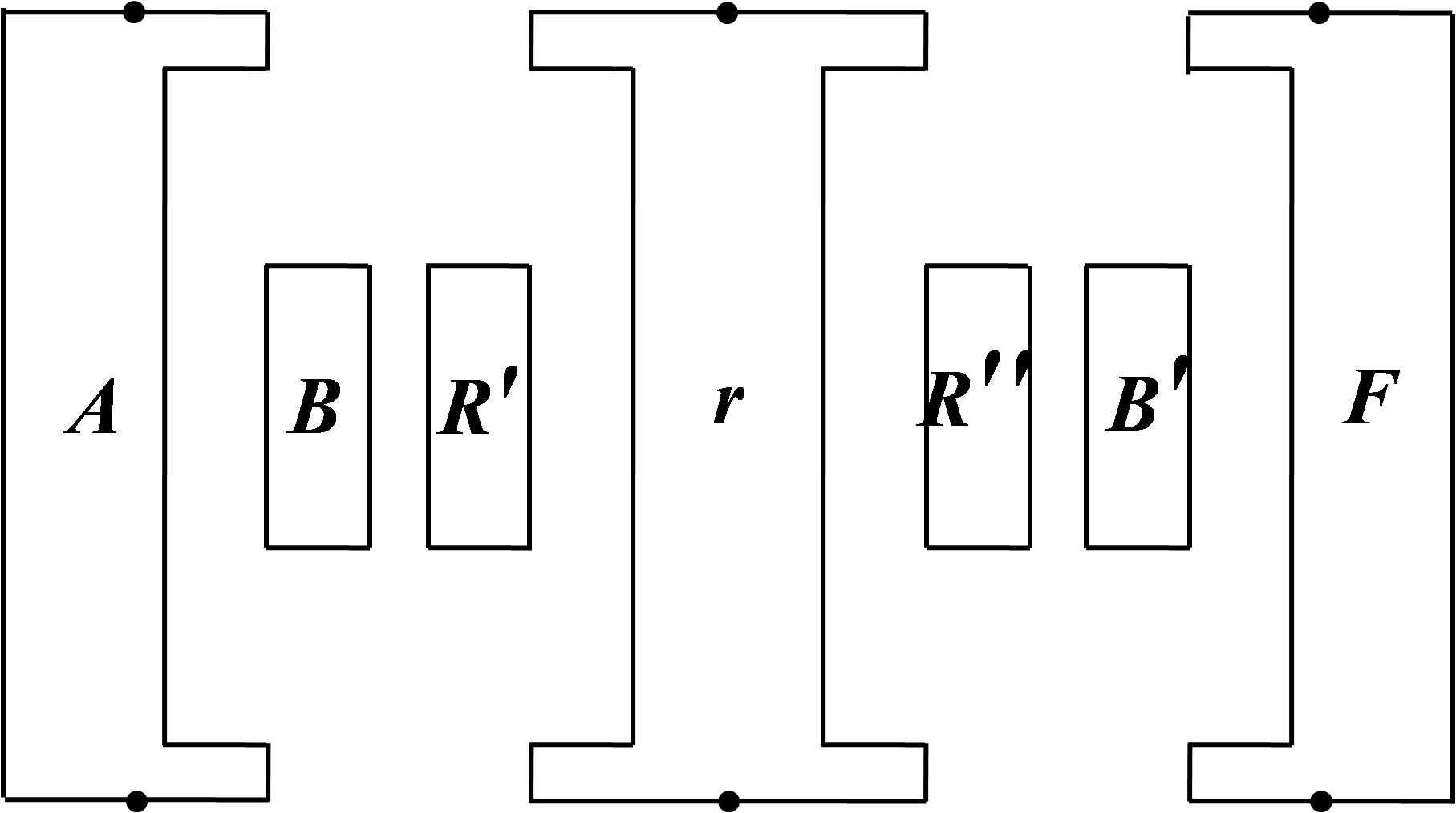}\right)^2  \nonumber\\
 &=&\frac{|P|^4 |B|^4 |R'|^4}{d^4}\nonumber\\
 &=&1\;.
\end{eqnarray}

The third term can be calculated as 
\begin{eqnarray}
&&\int dU~_{in}\langle\Psi|\tilde{\Pi}|\Psi\rangle_{in}^2\nonumber\\
 &=&|P|^6 \int dU~~~\left(\figbox{0.2}{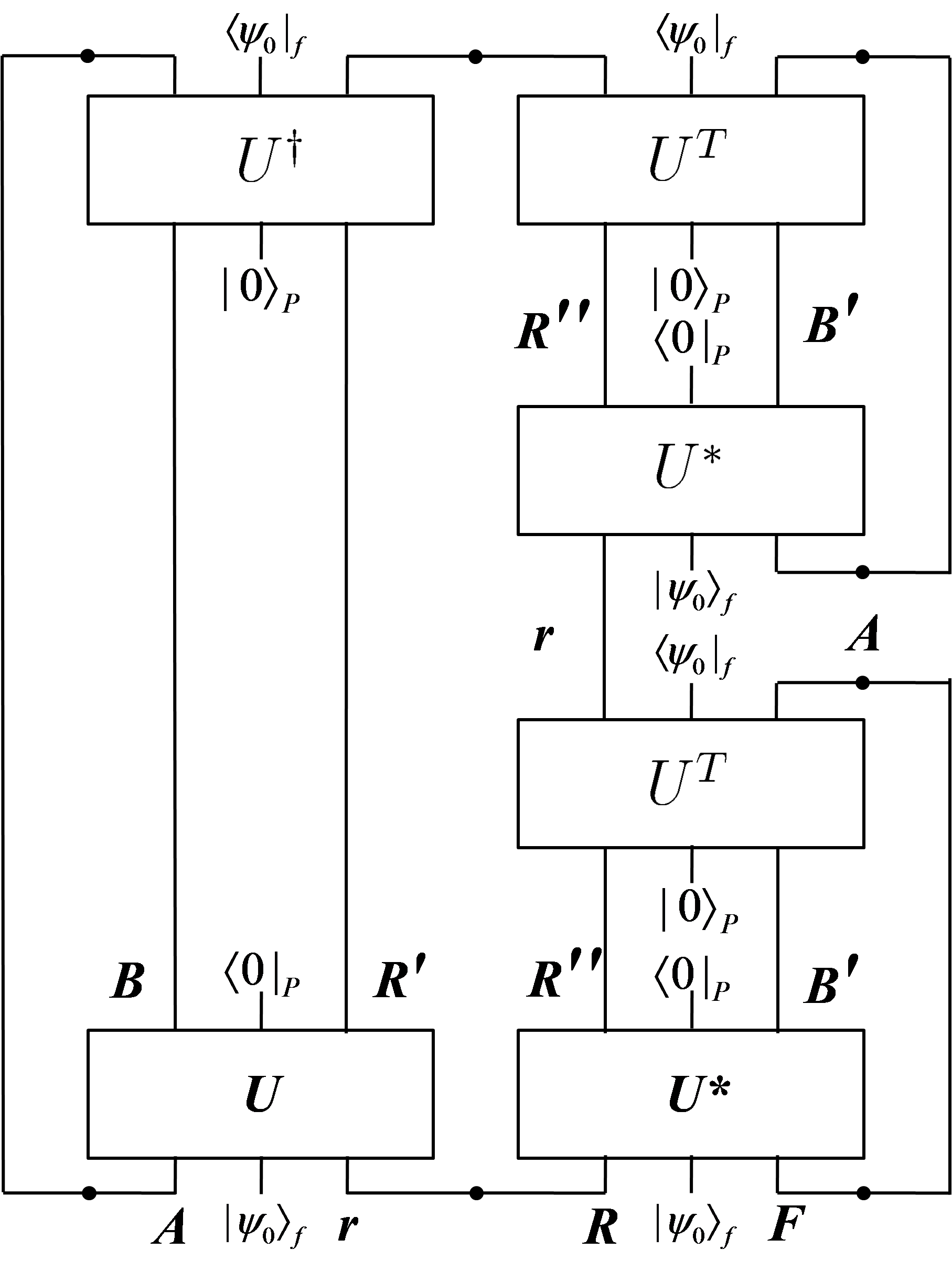}\right)^2  \nonumber\\
 &\simeq&\frac{|P|^6}{d^6} \int dU~~~\left(\figbox{0.2}{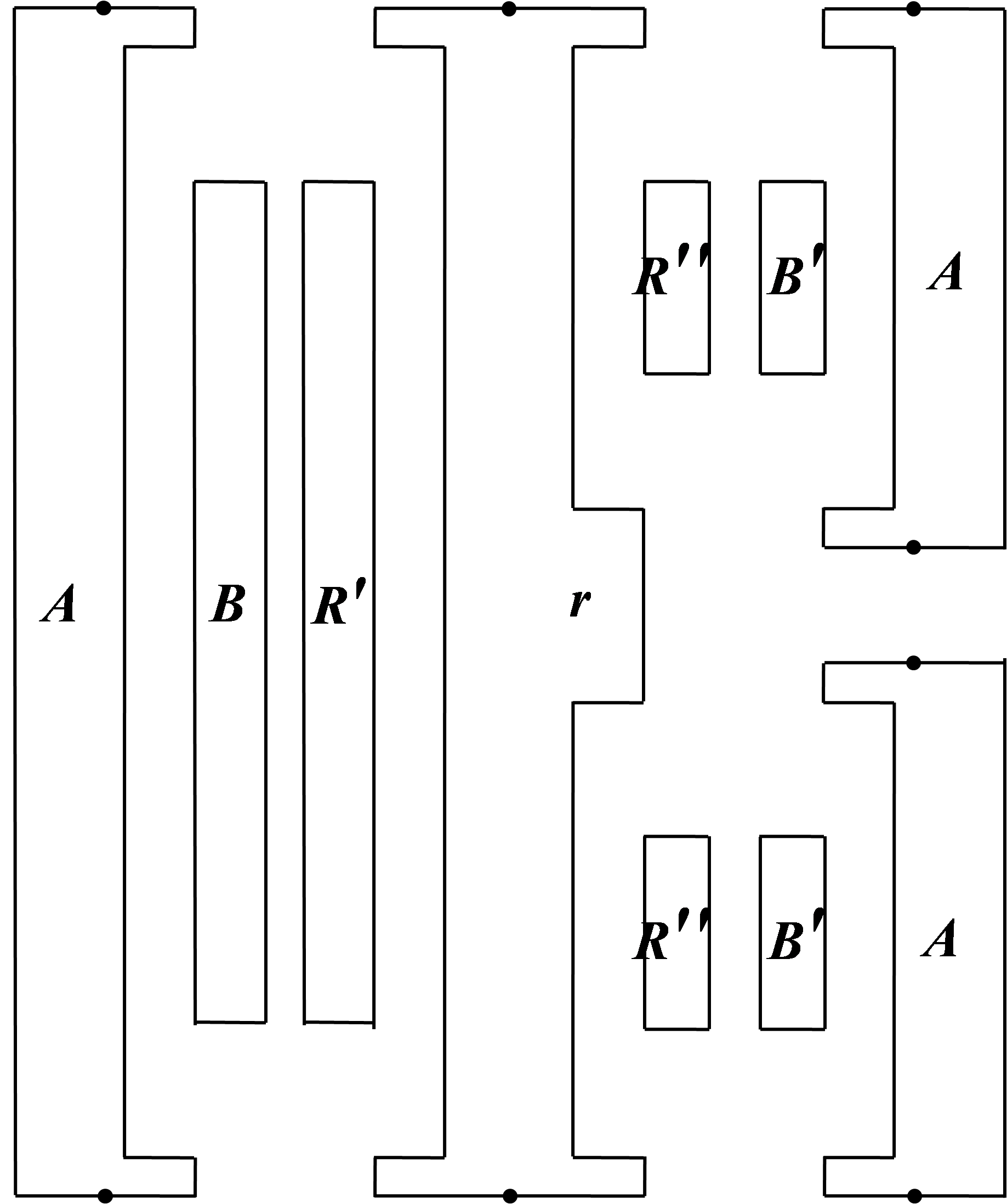}\right)^2  \nonumber\\
 &=&\frac{|P|^6 |B|^6 |R'|^6}{d^6}\nonumber\\
 &=&1\;.
\end{eqnarray}

Putting it together, the leading order contribution to the integral on the right hand side of the inequality is zero. Therefore, we have 
\begin{eqnarray}
    \int dU ||\rho_1-\rho_{in}||_{1}\leq\mathcal{O}(1)\;,
\end{eqnarray}
which implies that in the ideal case, the operator distance between $\rho_1$ and $\rho_{in}$ is small enough when averaged over the random unitary matrix. This gives that the first relation in Eq.(\ref{eq_pf}).

For the operator distance between $\rho_2$ and $\rho_{in}$, we have
\begin{eqnarray}
\left(\int dU ||\rho_2-\rho_{in}||_{1}\right)^2&\leq& 
\int dU ||\rho_2-\rho_{in}||_{1}^2\nonumber\\
&\leq& 2C \int dU ||\rho_2-\rho_{in}||_{2}^2\nonumber\\
&=&2C\int dU \textrm{Tr}\left(\rho_2^2+\rho_{in}^2-2\rho_2\rho_{in} \right)\nonumber\\
&=&2C \int dU \left( \frac{1}{P_{\textrm{EPR}}^2}~_{out}\langle\Psi|\tilde{\Pi}^2|\Psi\rangle_{out}^2 
+ ~_{in}\langle\Psi|\Psi\rangle_{in}^2 -\frac{2}{P_{\textrm{EPR}}} ~_{out}\langle\Psi|\tilde{\Pi}|\Psi\rangle_{in}^2\right)\;.\nonumber
\end{eqnarray}

We also evaluate the leading order contribution to the integral on the right hand side of the inequality. The first term can be calculated as 
\begin{eqnarray}
&&\int dU~_{out}\langle\Psi|\tilde{\Pi}^2|\Psi\rangle_{out}^2\nonumber\\
 &=&\frac{|P|^8}{P_{\textrm{EPR}}^2} \int dU~~~\left(\figbox{0.2}{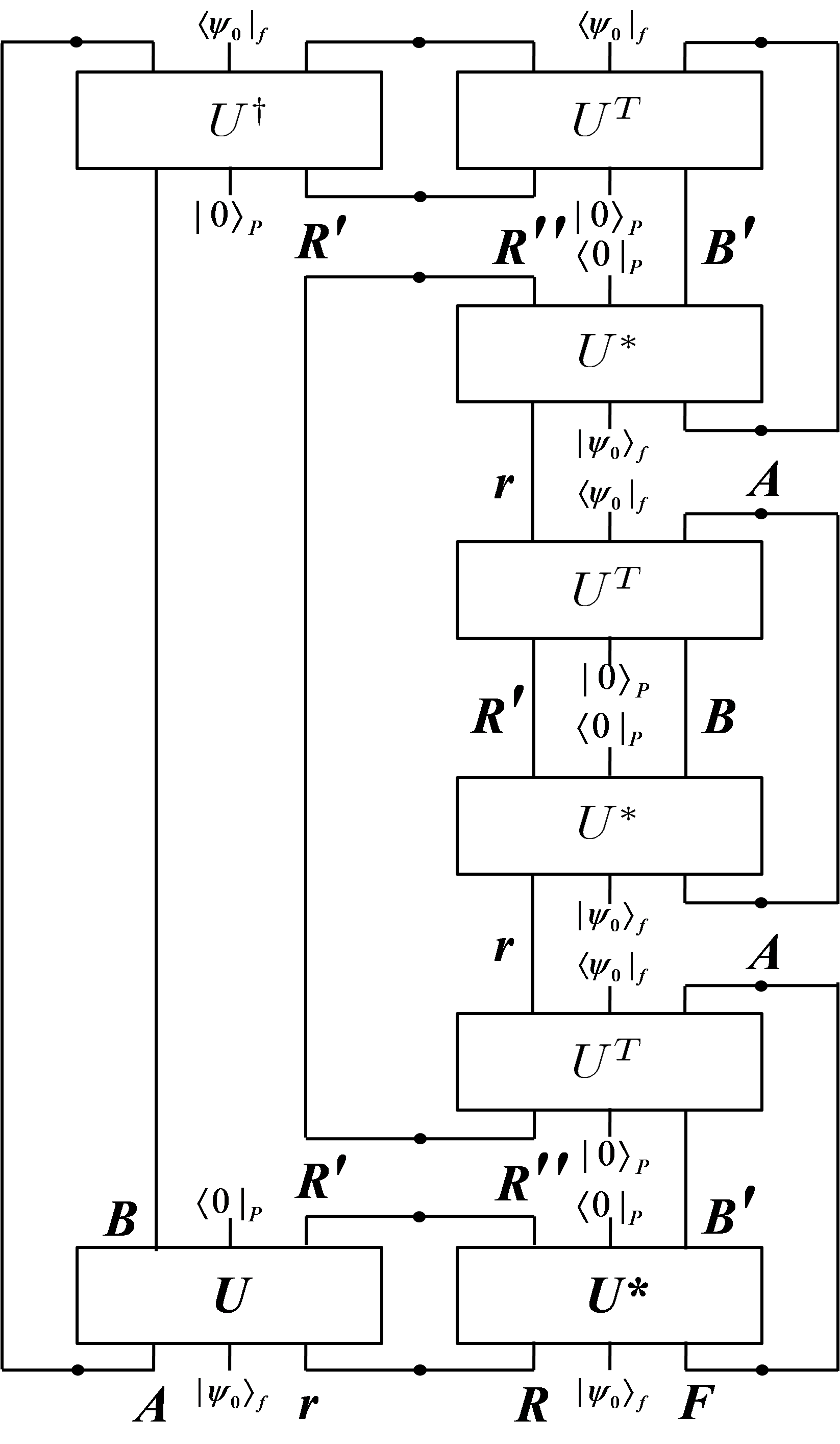}\right)^2  \nonumber\\
  &\simeq&\frac{|P|^8}{P_{\textrm{EPR}}^2}\frac{1}{d^8} \int dU~~~\left(\figbox{0.2}{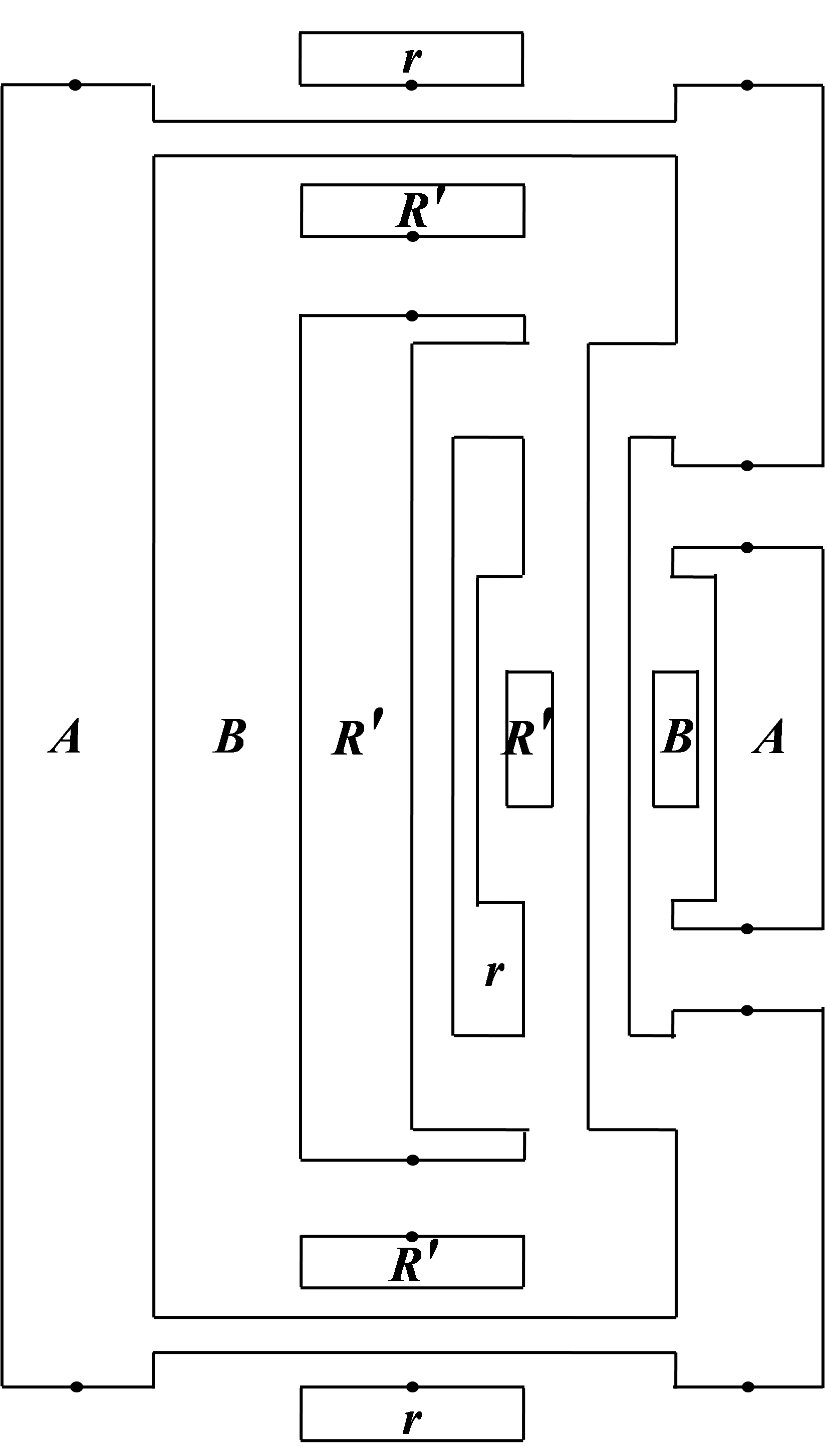}\right)^2  \nonumber\\
 &=&\frac{|P|^8}{P_{\textrm{EPR}}^2} \frac{|B|^4|r|^4|R'|^4}{d^8 |A|^4}\nonumber\\
 &\simeq&P_{\textrm{EPR}}^2\;.
\end{eqnarray}

The third term can be calculated as 
\begin{eqnarray}
&&\int dU~_{out}\langle\Psi|\tilde{\Pi}|\Psi\rangle_{in}^2\nonumber\\
 &=&\frac{|P|^6}{P_{\textrm{EPR}}} \int dU~~~\left(\figbox{0.2}{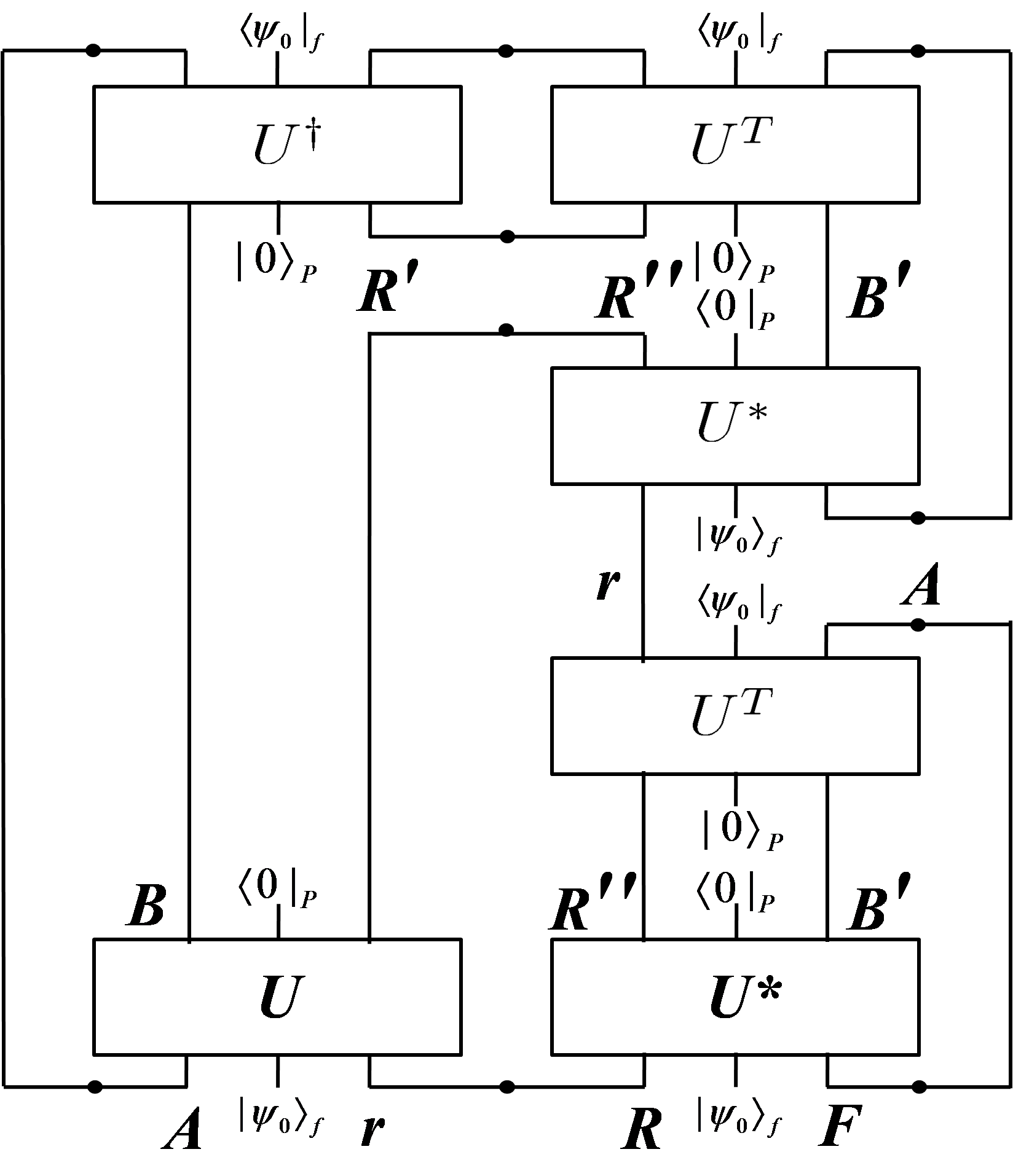}\right)^2  \nonumber\\
 &\simeq&\frac{|P|^6}{P_{\textrm{EPR}}} \frac{1}{d^6} \int dU~~~\left(\figbox{0.2}{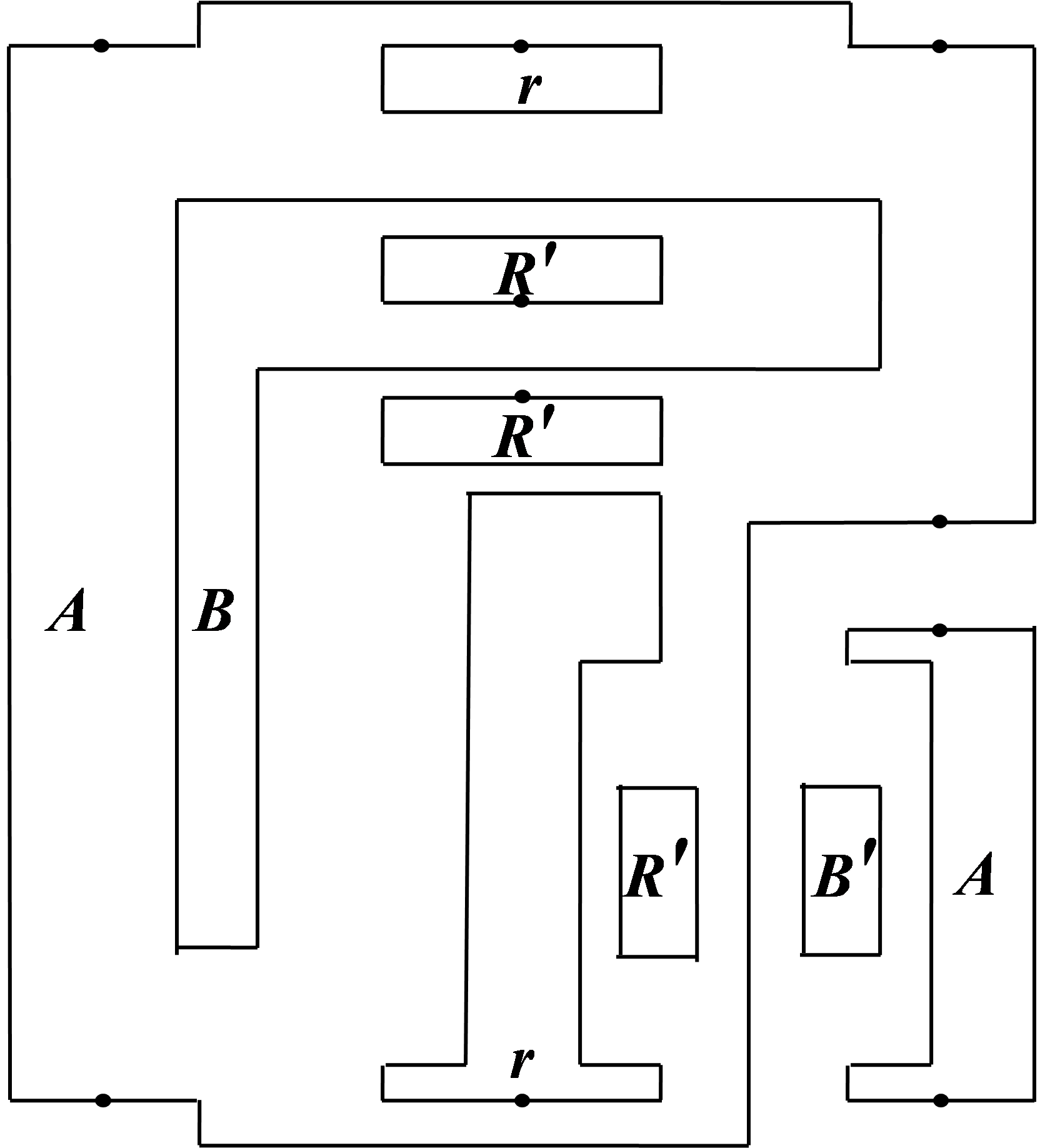}\right)^2  \nonumber\\
 &=&\frac{|P|^6}{P_{\textrm{EPR}}}\frac{|B|^4 |R'|^4 |r|^2}{d^6 |A|^2}\nonumber\\
 &\simeq&P_{\textrm{EPR}}\;.
\end{eqnarray}

Finally, we find that the leading order's contribution is also zero. Therefore, we have
\begin{eqnarray}
    \int dU ||\rho_2-\rho_{in}||_{1}\leq\mathcal{O}(1)\;.
\end{eqnarray}
We can conclude that the operator distance between $\rho_2$ and $\rho_{in}$ is also small enough in the ideal case. This gives that the second relation in Eq.(\ref{eq_pf}). In summary, we have proved the equations that used in the Grover's search decoding strategy.

\end{document}